\documentclass[aps,12pt,prd,showkeys,tightenlines,nofootinbib]{revtex4-1}

\usepackage{amsmath,amssymb, amsfonts}
\usepackage{graphicx}

\usepackage[pdftex]{hyperref}

\def\bra#1{\mathinner{\langle{#1}|}}
\def\ket#1{\mathinner{|{#1}\rangle}}
\def\braket#1{\mathinner{\langle{#1}\rangle}}
\newcommand{\sbraket}[1]{\lbrack #1\rbrack}

\DeclareMathOperator{\tr}{tr}

\makeatletter
\newcommand{\fmslash}[2][0mu]{%
  \mathchoice
    {\fmsl@sh\displaystyle{#1}{#2}}%
   {\fmsl@sh\textstyle{#1}{#2}}%
   {\fmsl@sh\scriptstyle{#1}{#2}}%
   {\fmsl@sh\scriptscriptstyle{#1}{#2}}}
\newcommand{\fmsl@sh}[3]{%
  \m@th\ooalign{$\hfil#1\mkern#2/\hfil$\crcr$#1#3$}}
\makeatother

\newcommand{\dalpha}{\dot{\alpha}}
\newcommand{\dbeta}{\dot{\beta}}

\begin{document}

\title{On-shell supersymmetry for massive multiplets} 
 
\author{Rutger H.  Boels} \affiliation{Universit\"at Hamburg,\\ II. Institut f\"ur Theoretische Physik,\\ Luruper Chaussee 149,\\ D-22761 Hamburg, Germany }
\author{Christian Schwinn} \affiliation{Albert-Ludwigs Universit\"at Freiburg, Physikalisches Institut, \\Hermann-Herder-Strasse 3,\\ D-79104 Freiburg, Germany\\}

\date{April 21, 2011\\}

\preprint{FR-PHENO-2010-007}

\begin{abstract}
\noindent The consequences of on-shell supersymmetry are studied for scattering amplitudes with massive particles in four dimensions. Using the massive version of the spinor helicity formalism the supersymmetry transformations relating products of on-shell states are derived directly from the on-shell supersymmetry algebra for any massive representation. Solutions to the resulting Ward identities can be constructed as functions on the on-shell superspaces that are obtained from the coherent state method. In simple cases it is shown that these superspaces allow one to construct explicitly supersymmetric scattering amplitudes. Supersymmetric on-shell recursion relations for tree-level superamplitudes with massive particles are introduced. As examples, simple supersymmetric amplitudes are constructed in Supersymmetric QCD (SQCD), the Abelian Higgs model, the Coulomb branch of $\mathcal{N}=4$, SQCD with an effective Higgs-gluon coupling and for massive vector boson currents.
\end{abstract}

\maketitle

\tableofcontents
\section{Introduction}

Recent developments inspired by insights into the twistor-space structure of scattering amplitudes in gauge theories~\cite{Witten:2003nn} led both to the discoveries of new symmetries and dualities of maximally supersymmetric Yang-Mills theory~\cite{Alday:2008yw,Drummond:2010ep} and to the development of new methods for the calculations of multi-particle processes relevant for physics at hadron colliders such as Tevatron or the LHC~\cite{Bern:2008ef}.  A prominent example of the simplicity of scattering amplitudes are the maximally helicity violating~(MHV) amplitudes in massless gauge theories~\cite{Parke:1986gb,Mangano:1990by,Dixon:1996wi}.  In supersymmetric theories, amplitudes with different field content are related by on-shell supersymmetry Ward identities~(SWIs)~\cite{Grisaru:1976vm,Grisaru:1977px} that allow to show for instance that the helicity equal amplitudes of massless particles in supersymmetric theories vanish to all orders in the coupling constant.  SWIs are also useful in order to obtain certain amplitudes in non-supersymmetric theories~\cite{Parke:1985pn,Kunszt:1985mg,Mangano:1990by,Dixon:1996wi}. In maximally supersymmetric $\mathcal{N}=4$ Yang Mills, the use of an on-shell superspace allows to solve the SWIs and combine the MHV amplitudes with different external field content into a ``supervertex''~\cite{Nair:1988bq}, nowadays known as a superamplitude. This on-shell superspace was the basis of the twistor-string theory of~\cite{Witten:2003nn} and also plays an important role in recent developments in $\mathcal{N}=4$ Yang-Mills theory such as dual superconformal symmetry~\cite{Drummond:2008vq,Alday:2008yw} as well as the supersymmetrized on-shell recursions relations~\cite{Brandhuber:2008pf, ArkaniHamed:2008gz}.
 At a practical level these are spaces with the coordinates given by the momentum eigenvalue and a number of Grassmann-valued parameters $\eta_i$. On these spaces a superfield admits an expansion in the fermionic variables
\begin{equation}
  \phi\left(k_{\mu}, \eta_i\right) = \phi\left(k_{\mu}\right) + \eta^i \phi_i\left(k_{\mu}\right) + \eta^i \eta^j \phi_{ij}\left(k_{\mu}\right) + \ldots
\end{equation}
such that typically the components are fields with a well-defined Lorentz quantum number (helicity in the massless case). A scattering amplitude becomes a function of the combined supersymmetric variable ${k_\mu, \eta_i}$ for each leg. Specific component amplitudes can be isolated by fermionic integration. As noted in~\cite{ArkaniHamed:2008gz}, the action of half of the supercharges can be diagonalized using a coherent-state representation, resulting in a useful method to simplify sums over the states of the supermultiplets in unitarity cuts and on-shell recursion relations.

It is an interesting question to which extent the simple structures of scattering amplitudes uncovered in massless theories survive once massive particles are included. In addition to being directly relevant for collider processes involving top quarks or electroweak gauge bosons for instance, amplitudes with massive particles also arise in $\mathcal{N}=4$ and $\mathcal{N}=2$ gauge theories away from the usually studied conformal point (see e.g. \cite{Schabinger:2008ah, Alday:2009zm} and references therein) or from compactifications of higher-dimensional field theories.  The extension of several of the new methods for the calculation of scattering amplitudes developed in the wake of~\cite{Witten:2003nn} to massive particles has been achieved. For instance on-shell recursion relations~\cite{Britto:2004ap,Britto:2005fq} have been generalized for amplitudes which involve massive scalars, quarks or vector bosons~\cite{Badger:2005zh,Badger:2005jv,Schwinn:2007ee} or to higher dimensional theories~\cite{ArkaniHamed:2008yf}. Furthermore the MHV vertex rules~\cite{Cachazo:2004kj} have been extended to amplitudes with external Higgs~\cite{Dixon:2004za} particles and electroweak gauge bosons~\cite{Bern:2004ba} as well as propagating massive scalars and quarks~\cite{Boels:2007pj,Boels:2008ef,Ettle:2008ey,Schwinn:2008fm} and spontaneously broken gauge theories~\cite{Buchta:2010qr}. More recently the `constructability' of amplitudes in theories with massive particles was studied in~\cite{Cohen:2010mi}. The extension of spinor-helicity methods to higher dimensions is discussed in~\cite{Cheung:2009dc, Boels:2009bv,Bern:2010qa,CaronHuot:2010rj}.

Following the example of the MHV amplitudes in massless theories an interesting starting point for the study of amplitudes with massive particles are the simplest non-vanishing scattering amplitudes which contain a pair of massive particles in the fundamental representation together with equal-helicity gluons. A closed all-multiplicity expression for  massive scalars and positive helicity gluons has been found first in~\cite{Forde:2005ue} while a particular compact form has been given in~\cite{Ferrario:2006np}:
\begin{equation}\label{eq:rodrigo}
A(\bar \phi_1^+,g_2^+,\dots,\phi_n^-)    = i 2^{n/2-1}m^2 \frac{  \braket{2+|\prod_{j=3}^{n-2} \left(y_{1,j}-\fmslash k_j\fmslash k_{1,j}\right)|(n-1)-}}{y_{1,2} y_{1,3}\dots y_{1,n-2} \braket{23}\braket{34}\dots\braket{(n-2)(n-1)}}, 
\end{equation}
where $ y_{1,j}=k_{1,j}^2-m^2$ with $k_{1,j}=k_1+\dots k_j$.  The analogous amplitudes with a pair of massive quarks can be obtained from~\eqref{eq:rodrigo} using SWIs~\cite{Schwinn:2006ca}. For amplitudes with a single negative helicity gluon~\cite{Forde:2005ue,Schwinn:2007ee} similar SWIs are also useful. Up to now no result of a similar simplicity as~\eqref{eq:rodrigo} is available for all-multiplicity amplitudes of massive vector bosons and general kinematics (some earlier works discussed amplitudes in the high-energy limit~\cite{Dunn:1990bp,Mahlon:1992cs} or for special kinematic configurations~\cite{Selivanov:1999ie}).  

The main aim of this article is to provide a complete discussion of on-shell supersymmetry for particles in massive supermultiplets, parallel to the discussion to massless particles. To this end a covariant form is derived of on-shell supersymmetry~(SUSY) transformations for any massive representation of the SUSY algebra in four dimensions in the framework of the spinor-helicity formalism. This follows the treatment of SUSY in higher dimensions in~\cite{Boels:2009bv} which highlighted a covariant construction of the representations of the SUSY algebra in a general Lorentz frame.  The results reproduce the expressions for the massive quark multiplet obtained in~\cite{Schwinn:2006ca} from an explicit analysis of the representation of the SUSY algebra on field operators and generalizes directly to all massive representations including the massive vector multiplet. As emphasized in~\cite{ArkaniHamed:2008gz} the on-shell superspaces for massless particles arise as fermionic and covariant coherent state representations of the on-shell SUSY algebra. Hence our analysis allows us to construct on-shell superspaces for massive particles which in turn lead directly to the formulation of supersymmetric amplitudes. This allows us for instance to demonstrate neatly the vanishing of some classes of amplitudes. Several explicit examples of superamplitudes in a variety of theories with massive particles will be provided. A general  method for solving the supersymmetric Ward identities is provided as well as a general analysis of supersymmetric on-shell recursion relations.  Note that a related discussion of BPS states in extended SUSY gauge theories in four dimensions has appeared in~\cite{Boels:2010mj} while coherent states for massless $\mathcal{N}=1$ and $\mathcal{N}=2$ supermultiplets have been discussed in~\cite{Boels:2006ir,Lal:2009gn,Elvang:2011fx}. In the $\mathcal{N}=4$ case there is a broad parallel to superspaces for the stress-energy multiplet as discussed  in \cite{Raju:2011ed, Raju:2011mp}. 

In detail, the article is structured as follows.  In section~\ref{sec:massivespinorhelictyform} we review the construction of polarization vectors for massive particles~\cite{Dittmaier:1998nn} within the spinor helicity formalism by defining the spin with respect to a fixed quantization axis. In section~\ref{sec:susyrepsmass} the SUSY transformations for the general massive $\mathcal{N}=1$ multiplet in an arbitrary Lorentz frame in the massive spinor helicity framework are derived. This  includes the transformation of the massive quark multiplet~\cite{Schwinn:2006ca} as a special case. Section~\ref{sec:coherent} employs the coherent state approach for constructing on-shell superspaces of massive SUSY representations and uses it to establish the vanishing of the analog of the all-helicity equal amplitude for four dimensional spontaneously broken Yang-Mills theories. Some applications of the superspace technology to specific examples of amplitudes in gauge theories with massive particles are investigated in section \ref{sec:exampamplis}. Theories studied there include super QCD with massive scalars and quarks, the Abelian Higgs model, effective Higgs-gluon couplings as well as vector boson currents. Section \ref{sec:susyrecur} contains a discussion of supersymmetric on-shell recursion relations, including an analysis of so-called large BCFW shifts in SQCD. Conclusions are reached and a discussion ensues. Appendices contain (\ref{app:notation}) an overview of our conventions, (\ref{app:details}) details of some of the models studied and (\ref{app:3-pt}) the calculation of a three point supervertex with arbitrary spin axes.

\section{Massive spinor helicity formalism}\label{sec:massivespinorhelictyform}
In this article massive spinor helicity methods will be used to treat the polarization states of massive vector bosons and massive quarks. While this material is treated in the literature~\cite{Dittmaier:1998nn} (see also e.g. \cite{Kleiss:1985yh,Kleiss:1986qc,Mahlon:1998jd,Chalmers:2001cy}), in this section a concise introduction is provided  to set up the notation and framework used in section~\ref{sec:susyrepsmass} for the construction of the SUSY multiplets.  The quantization of massive one-particle states with a choice of spin quantization axis is reviewed first. This will lead to massive quark spinors and polarization vectors of massive spin one bosons.
A summary of our spinor conventions is given in appendix~\ref{app:notation}.

\subsection{Massive one-particle states with a fixed spin axis}
\label{sec:massive-states}
Massive one-particle representations of the Poincar\'e group are specified by one vector and two half-integer numbers: the momentum $k_\mu$, total spin $s$ and projected spin quantum number $s_n$ for a spin-quantization axis $n$. These are defined by the conditions
\begin{align}
\label{eq:states}
P_\mu |k,s, s_n  \rangle & = k_\mu |k,s,s_n \rangle ,\\
W^2 |k,s, s_n  \rangle & =-m^2s(s+1)|k,s, s_n  \rangle,\\
R_{n} |k,s, s_n  \rangle & = s_n |k,s, s_n \rangle,
\end{align}
where $W^2$ is the length of the Pauli-Lubanski vector~\eqref{eq:pauli}. Further, the operator $R_{n}$ is given by 
\begin{equation}\label{eq:Lorentzrot}
R_{n} = -\frac{1}{m}n_{\mu} W^\mu
\end{equation}
with a space-like spin axis $n^\mu$ ($n^2=-1$) orthogonal to $k$. 

There are several possible choices of the spin axis $n$ for massive
particles. A massive two-component spinor formalism based on helicity
eigenstates is discussed in~\cite{Dittmaier:1998nn}.  Since helicity
is not a Lorentz-invariant concept for massive particles, in this
article the spin axis for a particle with momentum $k$ will be fixed instead 
in a covariant way following~\cite{Kleiss:1986qc} by introducing a
light-like vector $q$:
\begin{equation}\label{eq:spinaxiscov}
n_q^{\mu} = \frac{k^\mu}{m} - \frac{m \,q^{\mu}}{q \cdot k}\,.
\end{equation}
Here the reference vector $q$ is arbitrary up to the requirement $(q\cdot k) \neq 0$. Due to this constraint, some care has to be taken in the choice of $q$ in order to obtain a smooth massless limit. Note that scattering amplitudes defined in terms of the external states $\ket{k,s,s_{n_q}}$ will in general explicitly depend on the spin quantization axis~\eqref{eq:spinaxiscov} and hence on the reference vector $q$. The Lorentz-generator which implements rotations around the axis~\eqref{eq:spinaxiscov} has a manifestly well-defined massless limit. Acting on single-particle momentum eigenstates it reads
\begin{equation}\label{eq:Lorentzrotprop}
R_{n_q} =\frac{q_{\mu} W^{\mu}}{q \cdot P} =
-\frac{\epsilon^{\mu \nu \rho \sigma} q_{\mu} P_{\nu}M_{\rho
    \sigma} }{2(q \cdot P)}.
\end{equation}
In the following all legs of an amplitude are assumed to share the same polarization axis $q$, unless explicitly stated otherwise.

In order to incorporate massive particles into the spinor-helicity formalism, it is useful to decompose the massive momenta into two light-like vectors using the same reference vector $q$ used to define the spin axis~\cite{Dittmaier:1998nn}:
\begin{equation}\label{eq:decompmomentum}
  k^\mu =k^{\flat;\mu}+\frac{m^2 }{2(q \cdot k)}q^{\mu}.
\end{equation}
A massive momentum can therefore be expressed in terms of four
two-component Weyl spinors $ k^\flat_{\alpha}, k^\flat_{\dalpha}$ and
$ q_{\alpha}, q_{\dalpha}$, c.f.~\eqref{eq:bi-spinor}.  For spinor
products associated to massive momenta the notation
$\braket{ij}=\braket{k_i^\flat k_j^\flat}$ will be used.  For the
later treatment of the SUSY algebra note these spinors form a basis of
the dotted and undotted spinors. Hence any undotted
(dotted) spinor can be expanded into the basis spanned by
\begin{equation}\label{spin-basis}
\left\{k^\flat_\alpha\,,\;\tilde{q}_{\alpha} = \frac{m}{\braket{kq}} q_{\alpha} \right\}
\;,\qquad
\left\{ k^\flat_{\dalpha}\,,\; \tilde{q}_{\dalpha} \equiv \frac{m}{\sbraket{qk}} q_{\dalpha} \right\},
\end{equation}
respectively. This will be important below.

\subsubsection{Polarization spinors of massive spin one-half particles}
For spin one-half particles the generator of rotations around the spin axis~\eqref{eq:Lorentzrotprop} is represented as
\begin{equation}\label{eq:defrotspinrep}
R_{n_q}=\frac{1}{2m}\gamma^5\fmslash n_q \fmslash k
\end{equation}
(c.f.~\eqref{eq:gen-onehalf} and~\eqref{eq:SD-sigma}).  In the literature on helicity amplitudes, usually all momenta are treated as outgoing. Therefore consider anti-particle spinors and conjugate particle spinors labeled by their eigenvalues of the projectors
$P_{n_q}(\sigma)=\frac{1}{2}(1+\sigma \gamma^5\fmslash n_q)$:
\begin{align}
  \bar u(k,\sigma)P_{n_q}(\sigma)&=\bar u(k,\sigma),&
  P_{n_q}(\sigma)v(k,\sigma)&=v(k,\sigma).
\end{align}
Since on-shell spinors satisfy the Dirac equation, the eigenvalues of
the rotation generator in equation \eqref{eq:defrotspinrep} are given
by $\sigma/2$ for the $v$ spinors and $-\sigma/2$ for the $u$-spinors.
 Writing the Dirac spinors in terms of two-component
Weyl spinors as $v=(v_\alpha,v^{\dalpha})^T$ and $\bar
u=(u^\alpha,u_{\dalpha})$, these conditions reduce to
\begin{align}
  v_\alpha(k,\pm)&=\pm n_{q;\alpha\dalpha} v^{\dalpha}(k,\pm) ,&
  u^\alpha(k,\pm)&=\mp  u_{\dalpha}(k,\pm)n_q^{\dalpha\alpha}.
\end{align}
It is easily seen that this is satisfied by the massive spinors in the  conventions of~\cite{Schwinn:2006ca}:
\begin{align}
\label{eq:spinors} 
\bar u(k,+) &=
 \begin{pmatrix}  \frac{m}{\braket{qk}}  q^\alpha,   k^\flat_{\dalpha} \end{pmatrix},&
 \bar u(k,-) &=
 \begin{pmatrix}   k^{\flat;\alpha},  \frac{m}{\sbraket{qk}}  q_{\dalpha}  \end{pmatrix}\,,\\
 v(k,+) &=
 \begin{pmatrix}  -\frac{m}{\braket{kq}}  q_\alpha\\  k^{\flat;\dalpha}  \end{pmatrix},&
 v(k,-) &=
 \begin{pmatrix}   k^\flat_{\alpha}\\  -\frac{m}{\sbraket{kq}}  q^{\dalpha}  \end{pmatrix}.
\end{align}

\subsubsection{Polarization vectors of massive spin one-bosons}

Polarization vectors for massive vector bosons are defined by the condition
\begin{equation}
\label{eq:masseigenv}
(R_{n_q})^\mu_\nu\epsilon^\nu(k,-\sigma)=\sigma \epsilon^\mu(k,-\sigma)\,,
\end{equation}
where $\sigma\in\{0,\pm 1\}$. The reversed label arises because again all particles are considered to be outgoing.  The action of the generator $R_{n_q}$ on four-vectors is obtained using the vector representation of the Lorentz generators~$(M^{\mu\nu})_{\rho\sigma}=i (\delta^\mu_\rho\delta^\nu_\sigma-\delta^\mu_\sigma \delta^\nu_\rho)$. Translating the vector indices to bi-spinor notation using the identity~\eqref{eq:eps-spinor} for the totally antisymmetric tensor gives the explicit condition
\begin{equation}
\frac{1}{2}(R_{n_q})^{\dalpha\alpha}_{\beta\dbeta}\epsilon^{\dbeta \beta}(k,\sigma)  = \frac{ q^{\alpha} k^{\flat;\dalpha}  \left(k^{\flat,\beta}\epsilon_{\beta\dbeta}(k,\sigma)q^{\dbeta}\right)}{\sbraket{qk^\flat} \braket{k^\flat q}} - \frac{ k^{\flat;\alpha} q^{\dalpha} \left(q^\beta\epsilon_{\beta\dbeta}(k,\sigma)k^{\flat;\dbeta}\right)  }{\braket{k^\flat q}  \sbraket{qk^\flat}} =-\sigma \epsilon^{\dalpha \alpha}(k,\sigma) .
\end{equation}
This condition is satisfied by the polarization vectors~\cite{Dittmaier:1998nn}
\begin{equation}\label{eq:massivepolarization}
\epsilon_{\alpha\dalpha}(k,+) =
 \sqrt{2} \frac{q_{\alpha}k^\flat_{\dalpha}}{\braket{qk^\flat}}\,, \quad
 \epsilon_{\alpha\dalpha}(k,-) =
 \sqrt{2} \frac{k^\flat_\alpha q_{\dalpha}}{\sbraket{k^\flat q}} \,,
\quad \epsilon_{\alpha\dalpha}(k,0)  = \frac{1}{m}\left(k^\flat_{\alpha} k^\flat_{\dalpha} - \frac{m^2}{2 q\cdot k } q_{\alpha} q_{\dalpha}\right)
\end{equation}
that are transverse, $k\cdot \epsilon(k,\sigma)=0$, normalized according to $\epsilon(k,\sigma)\cdot \epsilon(k,-\sigma)=1$ and span the space perpendicular to $k$:
\begin{equation}
\sum_{\sigma\in\{0,\pm 1\}}\epsilon_{\mu}(k,\sigma) \epsilon_\nu(k,-\sigma)
 = -g_{\mu \nu} + \frac{k^\mu k^\nu}{k^2}\,.
\end{equation}
The positive and negative helicity polarization vectors are direct generalizations of the massless polarization vectors in the spinor helicity formalism. However, in the massless case the choice of the spinors $q$ corresponds to a physically irrelevant gauge choice whereas in the massive case it corresponds to a choice of the spin axis which is physical. 
The basis of polarization vectors just constructed obeys the simple but powerful equations
\begin{equation}\label{eq:simplebutpowerful}
\epsilon_{\mu}(k_i,+) \epsilon^{\mu}(k_j,+)= \epsilon_{\mu}(k_i,-) \epsilon^{\mu}(k_j,-) =0
\end{equation}
for polarization vectors of two different particles $i,j$ if the same reference vector $q$ is used to define the polarization axis for these fields.

\subsubsection{Special choice of frame}
\label{sec:frame} 
As a useful illustration, let us make the above considerations more concrete by fixing a special reference frame. There is always a frame such that the reference vector $q$ reads
\begin{equation}
q^{\mu}=\left(1,0,0,-1\right)\,.
\end{equation}
For a massive particle one could transform to the rest-frame of the particle. To allow for a smooth massless limit it is more useful however to boost to the frame such that the spacial component of the momentum is directed along the $z$-axis singled out by the choice of $q$,
\begin{equation}
k^{\mu}=\left(k^0,0,0,k^3 \right) \,, 
\end{equation}
with the mass-shell condition $k^2 =m^2$. We will assume $q \cdot k \neq 0$, so 
\begin{equation}\label{eq:constrgoodmasslim}
(q\cdot k)=k^0+k^3\,.
\end{equation}
In this frame, the light-cone projection of the massive momentum is given by
\begin{equation}
k^{\flat;\mu}=\frac{(k^0+k^3)}{2} (1,0,0,1)\,.
\end{equation}
The spin axis~\eqref{eq:spinaxiscov} agrees with the helicity axis $n_k$ mentioned above:
\begin{equation}
  n_q^\mu=\frac{1}{m}\left(k^3,0,0,k^0\right)\,.
\end{equation}
In the rest-frame the spin vector simply becomes the unit vector along the $z$-axis: $n_q^\mu=(0,\vec e_z)$ so the operator $R_{n_q}=J_z$ generates rotations around the $z$-axis as expected.

It is also easy to determine the basis spinors used in~\eqref{spin-basis} from the explicit expressions for $q$ and $k^\flat$ . Up to the usual scaling ambiguity, the spinors can be written as
\begin{equation}\label{eq:explicitspinors}
\begin{array}{ccc}
  k^\flat_{\alpha}  = \sqrt{k_0 + k_3} \left(\begin{array}{c} 1 \\ 0 \end{array}\right) ,& & 
  k^\flat_{\dalpha} =  \sqrt{k_0 + k_3} \left( \begin{array}{cc} 1 & 0 \end{array} \right),\\
  \tilde{q}_{\alpha} = \frac{m}{\braket{kq}} q_{\alpha}
  = \sqrt{k_0 -k_3}
  \left(\begin{array}{c} 0 \\ 1 \end{array}\right) ,& & 
  \tilde{q}_{\dalpha} = \frac{m}{\sbraket{ qk}} q_{\dalpha}
  = \sqrt{k_0 - k_3} \left(\begin{array}{cc} 0 & 1 \end{array}\right).
\end{array}
\end{equation}
The final expressions are well-defined both in the massless limit $k_0 \rightarrow k_3$ and in the rest-frame. Note the other natural such limit, $k_0 \rightarrow - k_3$, runs afoul of the constraint \eqref{eq:constrgoodmasslim}. 

In the special frame where the massive vector boson moves along the $z$-axis and the elements of the spinor basis are given by~\eqref{eq:explicitspinors} one recovers the familiar expressions
\begin{align}
  \epsilon^\mu(k,\pm)&=\frac{1}{\sqrt 2}  \begin{pmatrix}    0, 1, \mp i, 0  \end{pmatrix},&
  \epsilon^\mu(k,0)&=\frac{1}{m}   \begin{pmatrix}    k^3, 0,0,k^0  \end{pmatrix}.
\end{align}

\subsection{Lorentz-invariance constraints on amplitudes}

For helicity amplitudes of massless particles, Lorentz-invariance implies the constraint
\begin{equation}
\label{eq:helicity-constraint}
\left(k_i^\alpha\frac{\partial}{\partial k_i^\alpha} -
k_i^{\dot\alpha}\frac{\partial}{\partial k_i^{\dot \alpha}}\right)
A(,\dots \psi^{h_i}(k_i),\dots)=-2h_i
A(,\dots \psi^{h_i}(k_i),\dots)\,.
\end{equation}
To generalize this constraint to the massive spin
states~\eqref{eq:states}, consider Lorentz-rotations around the spin
quantization axis $n$ defined by the matrix
\begin{equation}
  \omega_{n,\mu\nu}= \frac{1}{m}\epsilon_{\mu\nu\rho\sigma}n^{\rho}k^{\sigma}.
\end{equation}
Under these transformations, the spin states acquire only a phase given
by the projection of the spin on the quantization axis:
\begin{equation}
\label{eq:lorentz-spin}
  e^{-i/2 \theta \omega_{n,\mu\nu}M^{\mu\nu}}\ket{k,s,s_n}=
 e^{i \theta R_n}|k,s,s_n \rangle = e^{i\theta s_n}\ket{k,s,s_n}\,.
\end{equation}
For the spin quantization axis $n_q$ defined in~\eqref{eq:spinaxiscov},
scattering amplitudes can be expressed entirely in terms of the basis
spinors $\ket{k_i^\flat\pm}$ and $\ket{q_i\pm}$ for each particle.
The action of the generator of rotations around the spin-axis in the
spin one-half representation~\eqref{eq:defrotspinrep} on these basis
spinors is given by
\begin{equation}
\begin{aligned}
R_{n_q}\ket{k^\flat\pm}&
=\frac{1}{4(q\cdot k)}\gamma^5 \slash\!\!\! k\slash\!\!\! q \ket{k^\flat\pm}
=\pm\frac{1}{2}\ket{k^\flat\pm},\\
R_{n_q}\ket{q\pm}&
=-\frac{1}{4(q\cdot k)}\gamma^5 \slash\!\!\! q\slash\!\!\! k \ket{q\pm}
=\mp\frac{1}{2}\ket{q\pm}.
\end{aligned}
  \end{equation}
so that the Lorentz-rotations around the spin-axis of the basis
spinors are given by
\begin{align}
  e^{-i \theta R_n}\ket{k^\flat\pm}&=
  e^{\pm i\theta/2}\ket{k^\flat\pm},&
  e^{-i \theta R_n}\ket{q\pm}&=
  e^{\mp i\theta/2}\ket{q,\pm}.
\end{align}
Consistency with the transformation law of the spin
states~\eqref{eq:lorentz-spin} then implies the generalization of the
constraint~\eqref{eq:helicity-constraint} to amplitudes with massive
particles (see also~\cite{Cohen:2010mi}) :
\begin{equation}
\label{eq:spin-constraint}
\left(k_i^{\flat,\alpha}\frac{\partial}{\partial k_i^{\flat,\alpha}}
-q_i^{\alpha}\frac{\partial}{\partial q_i^{\alpha}}
+q_i^{\dot\alpha}\frac{\partial}{\partial q_i^{\dot \alpha}} -
k_i^{\flat,\dot\alpha}\frac{\partial}{\partial k_i^{\flat,\dot \alpha}}
\right)
A(\dots,\psi^{s_{n,i}}_{k_i},\dots)=-2s_{n,i}
A(\dots, \psi^{s_{n,i}}_{k_i},\dots),
\end{equation}
where the minus sign on the right-hand side is due to the convention to treat all particles as outgoing. 
The only dependence in Feynman diagrams on the $q$ spinors can arise through the external polarization vectors and spinors and through the decomposition of momenta into light-like vectors~\eqref{eq:decompmomentum}. Since these expressions are homogeneous of degree zero in the spinors $q_\alpha$ and $q^{\dot\alpha}$ it follows that the same property is shared by scattering amplitudes so that the $q$-dependent terms in~\eqref{eq:spin-constraint} always drop out.

\section{Covariant representation theory of the $\mathcal{N}=1$, $D=4$ SUSY algebra} \label{sec:susyrepsmass}

Recall that in the textbook treatment the massive representations in the rest-frame are constructed by acting with specific components (e.g. $\bar Q_{\dot 1,\dot   2}/\sqrt{2m}$) of the SUSY generators on a ``Clifford vacuum'' state $\ket{\Omega}$ annihilated by the conjugate generators (see e.g.~\cite{Weinberg:2000cr} for a detailed textbook discussion). Since our objective is to formulate the representation theory of the algebra directly in an arbitrary Lorentz frame the supercharges $Q_\alpha$ and $\bar Q^{\dalpha}$ should be projected  on operators with a well defined quantum number of the operator $R_{n_q}$ used to define the states~\eqref{eq:states}. 
 In this section it will be shown that the massive spinor helicity methods from section~\ref{sec:massivespinorhelictyform} allow to accomplish this in a simple way. The approach used here is inspired by the treatment of  generic massless representations in higher dimensions by one of the present authors~\cite{Boels:2009bv}.  With this approach, the results for the SUSY transformation massive quark multiplet obtained in~\cite{Schwinn:2006ca} from LSZ reduction are easily recovered and generalized to any massive multiplet. Our conventions for the SUSY algebra are summarized in appendix~\ref{app:susy}.

\subsection{Spin decomposition of the SUSY generators}

The key observation in order to make contact with the massive spinor helicity methods discussed in section~\ref{sec:massivespinorhelictyform} is to note that the supercharges can be expanded in the basis spanned by the spinors $\tilde q$ and $k$ as in~\eqref{spin-basis}:
\begin{align}\label{eq:covariantizationofsusy}
Q_{\alpha} & = \frac{q_{\alpha}}{\braket{k^\flat q}} \braket{k^\flat Q} +
               \frac{k^\flat_{\alpha}}{\braket{qk^\flat}}  \braket{qQ}
             =\tilde q_\alpha Q_+ + k^\flat_{\alpha} Q_-\,,\\
\bar{Q}_{\dalpha} & = \frac{q_{\dalpha}}{\sbraket{q k^\flat}} 
\sbraket{\bar Q k^\flat} +
               \frac{k^\flat_{\dalpha}}{\sbraket{k^\flat q}}  \sbraket{\bar Qq}
=\tilde q_{\dalpha}\bar{Q}_- + k^\flat_{\dalpha} \bar{Q}_+\,.
\end{align}
Here  the components of the supercharges have been introduced,
\begin{equation}
\label{eq:covariant-generators}
\begin{array}{ccccccc}Q_+
  & \equiv & \frac{\braket{k^\flat Q}}{m}, & \quad \quad &
  \bar{Q}_- & \equiv & \frac{\sbraket{\bar Q k^\flat}}{m},\\[4pt]
   Q_- & \equiv & \frac{\braket{q Q}}{\braket{q k^\flat}}, &
  \quad \quad & \bar{Q}_+ & \equiv &
   \frac{\sbraket{\bar Qq}}{\sbraket{k^\flat q}} ,\end{array}
 \end{equation}
that have a well defined spin quantum-number as will be shown shortly. The convenience of the decomposition of the generators is seen when inserting the decompositions~\eqref{eq:covariantizationofsusy} into the SUSY algebra $\{Q_{\alpha}, \bar{Q}_{\dalpha} \} = -2 \sigma^{\mu}_{\alpha \dalpha} k_{\mu}$ and using~\eqref{eq:decompmomentum} for the right-hand side:
\begin{multline}
\tilde q_{\alpha} \tilde q_{\dalpha}  \{ Q_+, \bar{Q}_- \} +
k^\flat_{\alpha} \tilde q_{\dalpha}  \{ Q_-, \bar{Q}_- \}
+ \tilde q_{\alpha} k^\flat_{\dalpha}  \{ Q_+, \bar{Q}_+ \}
+ k^\flat_{\alpha} k^\flat_{\dalpha} \{ Q_-, \bar{Q}_+ \}
= -2 \left(k^\flat_{\alpha} k^\flat_{\dalpha} + \tilde q_{\alpha} \tilde q_{\dalpha} \right).
\end{multline}
Since the four light-like Lorentz vectors on the left hand side form a (`light-like') basis of four dimensional space, from this expression the anti-commutators for the generators $Q_\pm,\bar{Q}_\pm$ can be read off,
\begin{equation}
\label{eq:cov-algebra}
\left(\begin{array}{cc}
\{ Q_+, \bar{Q}_- \} & \{ Q_+, \bar{Q}_+ \} \\
\{ Q_-, \bar{Q}_- \} & \{ Q_-, \bar{Q}_+  \}
\end{array} \right) = -\left(\begin{array}{cc} 2  & 0 \\
0 & 2
\end{array} \right).
\end{equation}

The action of the SUSY charges $Q_{\pm}$ and $\bar Q_{\pm}$ on the eigenstates of the Lorentz-generator $R_{n_q}$ can be obtained in an analogous way starting from the  commutation relations of the SUSY charges with the generators of the Lorentz-transformations~\eqref{eq:m-q-comm} and inserting the decomposition~\eqref{eq:covariantizationofsusy}. Using the (anti-) self-duality relations~\eqref{eq:sd-sigma} the commutators of the SUSY charges with the generators of the rotation around the spin axis can be evaluated as
\begin{align}
  [R_{n_q},Q_\alpha]&=-\frac{1}{(k\cdot q)}
  q_\mu k_\nu(\sigma^{\mu\nu})^{\beta}_\alpha Q_\beta
=\frac{1}{2}\left(\tilde q_\alpha Q_+- k^\flat_\alpha Q_-   \right),
\\
 [R_{n_q},\bar Q^{\dot \alpha}]&=\frac{1}{(k\cdot q)}
  q_\mu k_\nu(\bar \sigma^{\mu\nu})_{\dot\beta}^{\dot\alpha} \bar Q^{\dot\beta}
=\frac{1}{2}\left(k^{\flat;\dot\alpha}\bar Q_+-\tilde q^{\dot\alpha}\bar Q_- 
\right).
\end{align}
Inserting the decomposition~\eqref{eq:covariantizationofsusy} on the left-hand side and comparing the coefficients of the basis spinors  can be read off from  the commutation relations
\begin{align}
\label{eq:komm-r-q}
  [R_{n_q},Q_\pm]&=\pm\frac{1}{2}Q_\pm\,,&
 [R_{n_q},\bar Q_\pm]&=\pm \frac{1}{2}\bar Q_\pm\,.
\end{align}
Therefore, as anticipated by the notation, $Q_+$ and $\bar Q_+$ raise the spin quantum number $s_n$ by $1/2$ while $Q_{-}$ and $\bar Q_{-}$ lower it by one-half. Therefore a  covariant definition of SUSY generators has been obtained that have a well-defined spin quantum number with respect to the spin-axis $n_q$ in~\eqref{eq:spinaxiscov}. By the close relation to the usual rest-frame analysis it should come as no surprise that these generators can be used to construct the massive representations, which will be done in section~\ref{sec:mass-reps}.

A general SUSY transformation parameterized by two Grassmann valued spinors $\theta^\alpha$ and $\theta_{\dalpha}$ is expressed in terms of the supercharges~\eqref{eq:covariant-generators} as
\begin{equation}
\label{eq:decompose-susy}
  Q(\theta)=\theta^\alpha Q_\alpha+\theta_{\dot\alpha}\bar Q^{\dot \alpha}
= \braket{\theta k^\flat} Q_- +\sbraket{\theta k^\flat} \bar{Q}_+
+ m \frac{\braket{\theta q}}{\braket{k^\flat q}} Q_+ 
+m \frac{\sbraket{\theta q}}{\sbraket{q k^\flat}} \bar{Q}_-\,.
\end{equation}
This operator has a well-defined massless limit. Note that if the supersymmetry variation is directed along the light-cone vector $q$ used to define the polarization axis
\begin{equation}
  \label{eq:susy-q}
 \theta_\alpha=\theta q_\alpha\,,\qquad
\theta^{\dalpha}=\theta q^{\dalpha},
\end{equation}  with a Grassmann number
$\theta$, 
 the form of the algebra is exactly equivalent to the massless case, generalizing the findings of~\cite{Schwinn:2006ca} to arbitrary representations.

\subsection{General massive representations in an arbitrary frame}
\label{sec:mass-reps}
After the covariant analysis of the SUSY algebra its representations can be studied on generic massive one-particle states which have definite quantum numbers $s_n$ under the $R_{n_q}$ operator defined above in~\eqref{eq:states}. Since the operators $Q_\pm$ and $\bar{Q}_\pm$ raise and lower the value of $s_n$ by one half, their action on the basis states of the multiplets can be determined along the lines of the usual analysis of the rest-frame algebra. The action of a general supersymmetry transformation on a state in an arbitrary frame is then obtained from the operator~\eqref{eq:decompose-susy}.

\begin{figure}[t]
 \begin{center}
\includegraphics[width=0.6\textwidth]{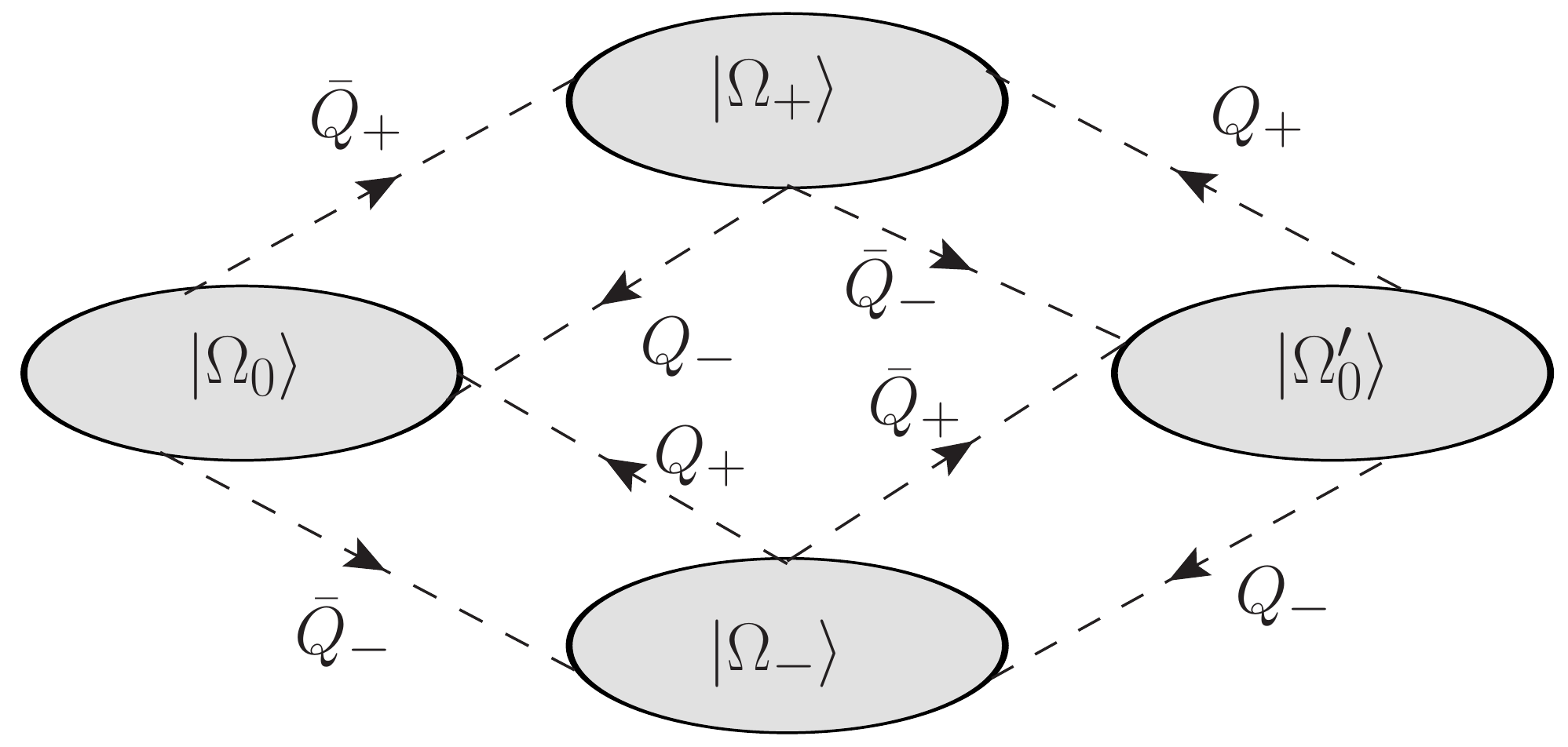}
\end{center}
  \caption{Sketch of the states in the massive multiplet
   and the
 action of the supercharges.}
  \label{fig:mass-rep}
\end{figure}

Following the usual treatment in the rest-frame a set of $2s^0+1$ states $\ket{\Omega_0, s_n^0}\equiv\ket{k,s^0,s^0_n}$ can be defined (the ``Clifford vacuum'') annihilated by $Q_-$ and $Q_+$:
\begin{equation}
  Q_-\ket{\Omega_0,s_n^0}=Q_+\ket{\Omega_0,s_n^0}=0.
\end{equation}
Such a state can be constructed from any state with $R_{n_q}$ eigenvalue $s_n^0$ by multiplying with $Q_-Q_+$ (or e.g. by multiplying by $Q_-$ in case that the state is already annihilated by $Q_+$). Using the fact that $\bar Q_+$ raises $s_n$ by one half and $\bar Q_-$ lowers it by one half  the following states can be defined (up to a phase choice)
\begin{align}
\label{eq:ket-states}
\ket{\Omega_\pm,s_n^0\pm\tfrac{1}{2}}&\equiv \frac{1}{\sqrt 2}
\bar Q_\pm\ket{\Omega_0,s_n^0},&
\ket{\Omega_0',s_n^0}&\equiv \frac{1}{2}\bar Q_-\bar Q_+\ket{\Omega_0,s_n^0}
\end{align}
which are eigenstates of the generator of rotations around the spin axis with quantum numbers $R_{n_q}\ket{\Omega_\pm}=(s_n^0\pm\tfrac{1}{2})\ket{\Omega_\pm}$ and $R_{n_q}\ket{\Omega'_0}=s_n^0\ket{\Omega'_0} $. 
In the following the spin labels on the states will be suppressed when no confusion can arise. 
 The action of the remaining generators on the ket-states is completely fixed by the SUSY algebra:
\begin{subequations}
\label{eq:susy-ket}
\begin{align}
\frac{1}{\sqrt 2} Q_\pm\ket{\Omega_\mp}&=-\ket{\Omega_0},
\\
\frac{1}{\sqrt 2} \bar Q_\pm\ket{\Omega_\mp}&=
\mp\ket{\Omega_0'},
\\
\frac{1}{\sqrt 2} Q_\pm\ket{\Omega_0'}&
=\mp \ket{\Omega_\pm },
\end{align}
\end{subequations}
where all other combinations of generators and states vanish. Therefore, as sketched in figure~\ref{fig:mass-rep}, acting with the generators $Q_\pm$ and $\bar Q_\pm$ moves around within the states $(\Omega_+,\Omega_0,\Omega_0',\Omega_-)$ so they form an irreducible representation of the SUSY algebra.
The states $\ket{\Omega_+}$ and
$\ket{\Omega_-}$ are in general superpositions of
states with spin $s^0\pm\tfrac{1}{2}$
\begin{equation}
\label{eq:clebsch-states}
\begin{aligned}
\ket{\Omega_+}&=
c_{++}
\ket{k,s^0+\tfrac{1}{2},s_n^0+\tfrac{1}{2}}+
c_{-+}\ket{k,s^0-\tfrac{1}{2},s_n^0+\tfrac{1}{2}},      \\
\ket{\Omega_-}
&=c_{--}
\ket{k,s^0-\tfrac{1}{2},s_n^0-\tfrac{1}{2}}+
c_{+-}
\ket{k,s^0+\tfrac{1}{2},s_n^0-\tfrac{1}{2}},
\end{aligned}
\end{equation}
with the Clebsch-Gordan coefficients
\begin{equation}
c_{\lambda\lambda_n}=\braket{s^0+\tfrac{\lambda}{2},s_n^0+\tfrac{\lambda_n}{2}|s^0,s_n^0,\tfrac{1}{2},\tfrac{\lambda_n}{2}}.
\end{equation} 
The spin decomposition of these states is identical to that in the rest-frame since for a Lorentz boost from the rest frame the states transform as
$  U(\Lambda)\ket{k,s,s_z}=\ket{\Lambda k,s,s_z'}$ without a  Wigner rotation. Explicit examples for the states $\Omega_\pm$ for $s_n^0=0$ and $s_n^0=\pm\frac{1}{2}$ will be given below.

The action of a general SUSY transformation~\eqref{eq:decompose-susy} parameterized by the Grassmann valued spinors $\theta_\alpha$ and $\theta^{\dalpha}$ on the four states of the massive multiplet can now be worked out. 
Since the same results can be obtained in the superfield formalism to be introduced below (see~\eqref{eq:mass-multiplet}), the results will not be recorded here .

It is also useful to record the definitions of the out-states
explicitly.  The conjugate
states satisfying the conditions
$\bra{\Omega_\pm}R_{n_q}=(s_n^0\pm\tfrac{1}{2})\bra{\Omega_\pm}$ and $
\bra{\Omega_0'}R_{n_q}=s_n^0\bra{\Omega_0'}$ are defined as
\begin{align}
\label{eq:bra-states}
\bra{\Omega_\pm,s_n^0\pm\tfrac{1}{2}} & = -\frac{1}{\sqrt 2}
\bra{\Omega_0,s_n^0} Q_\mp\,,       & \bra{\Omega_0',s_n^0}&= \frac{1}{2} \bra{\Omega_0,s_n^0}Q_-Q_+\,,
\end{align}
which is consistent with the algebra~\eqref{eq:cov-algebra} and the definition of the conjugate Clifford vacuum satisfying
$\bra{\Omega_0}\bar Q_\pm=0$.  The action of the SUSY charges on the bra-spinors~\eqref{eq:bra-states} is obtained from~\eqref{eq:susy-ket} using the complex conjugation of the supercharges $Q_\pm^\dagger\equiv \bar Q_\mp$.

\subsubsection{Massive quark multiplet}

The simplest massive representation is based on a scalar Clifford
vacuum $\ket{\Omega_0,0}=\ket{k,0,0}$. In this case the
superpositions~\eqref{eq:clebsch-states} in the definitions of the
states $\ket{\Omega_\pm,\pm\tfrac{1}{2}}$ collapse to a single state with spin
$s^0\pm\tfrac{1}{2}=\pm\tfrac{1}{2}$.  Therefore the multiplet includes two
spin states of a massive Majorana fermion
$\ket{\Omega_\pm}=\ket{k,\frac{1}{2},\pm \frac{1}{2}}$ and the states 
 $\ket{\Omega_0}$ and $\ket{\Omega_0'}$ of a complex scalar.  
For a massive multiplet in the fundamental representation of $SU(N)$ the fermion necessarily must be a Dirac fermion so one has to double the number of states and include two complex scalar fields.
The on-shell SUSY
transformations of the spin states with the external
spinors~\eqref{eq:spinors} have been obtained in~\cite{Schwinn:2006ca}
from the transformations of the field operator. In order to make contact
with the conventions used in that reference,  the following identifications 
for the out states must be made:
\begin{equation}
\begin{aligned}
\bra{\Omega_0}&=\bra{k,0,0}\equiv \bra{\bar\phi^{+}_k}, &
\bra{\Omega_0'}&=\bra{k,0',0}\equiv \bra{\bar\phi^-_k},\\
\bra{\Omega_+}&=\bra{k,\tfrac{1}{2},\tfrac{1}{2}}\equiv \bra{\bar Q^+_k},
 &
 \bra{\Omega_-}&=\bra{k,\tfrac{1}{2},-\tfrac{1}{2}}
\equiv \bra{\bar Q^-_k},
\end{aligned}
\label{eq:quark-mult}
\end{equation}
where the notation follows that of the fermion spinors i.e. the states
with a bar denote outgoing particles. An analogous on-shell multiplet
of states $Q^\pm$ and $\phi^\pm$ denoting outgoing antiparticles is
not displayed here.

\subsubsection{Massive vector multiplet}
\label{sec:susy-vector}

Apart from the quark multiplet, the other massive multiplets most relevant
for globally supersymmetric field theories are based on Clifford vacua with
spin one-half,
$\ket{\Omega_0,\pm\tfrac{1}{2}}=\ket{k,\tfrac{1}{2},\pm\tfrac{1}{2}}$.
Starting from the $s^0_n=\tfrac{1}{2}$ states one obtains the
multiplet containing two inequivalent massive ``wino'' states
$\chi$ with spin projection $+\tfrac{1}{2}$, a massive vector boson
$W$ with spin projection $+1$ and a linear combination of a longitudinal
massive vector boson and a scalar:
\begin{align}
\label{eq:massive-vec+}
\ket{\Omega_0,\tfrac{1}{2}}&
=\ket{k,\tfrac{1}{2},\tfrac{1}{2}}\equiv\ket{\chi_k^+}, &
\ket{\Omega_0',\tfrac{1}{2}}&
=\ket{k,\tfrac{1}{2}',\tfrac{1}{2}}\equiv\ket{\tilde\chi_k^{+}},\\
\ket{\Omega_+,1}&=\ket{k,1,1}\equiv\ket{W_k^+}, &
 \ket{\Omega_-,0}&=\frac{1}{\sqrt 2}(\ket{k,1,0}+\ket{k,0,0})\equiv
 \frac{1}{\sqrt 2}(\ket{W_k^0}+\ket{\Phi_k}).\nonumber
\end{align}
The states with the negative spin projections are obtained from the
Clifford vacuum with $s^0_n=-\tfrac{1}{2}$ and include the orthogonal
linear combination of the longitudinal vector and the scalar
\begin{equation}
\label{eq:massive-vec-}
\begin{aligned}
\ket{\Omega_0,-\tfrac{1}{2}}&
=\ket{k,\tfrac{1}{2},-\tfrac{1}{2}}\equiv \ket{\chi_k^-}, &
\ket{\Omega_0',-\tfrac{1}{2}}&=\ket{k,\tfrac{1}{2}',-\tfrac{1}{2}} 
\equiv\ket{\tilde\chi_k^{-}},
\\
\ket{\Omega_+,0}&
=\frac{1}{\sqrt 2}(\ket{W_k^0}-\ket{\Phi_k}), &
\ket{\Omega_-,-1}&=\ket{k,1,-1}=\ket{W_k^-}.
\end{aligned}
\end{equation}

\subsection{Extended supersymmetry and BPS multiplets}\label{subsec:explmassvsBPS}
Massive representations of extended supersymmetry can also be studied with the above methods. Without central charges, the analysis simply reduces to multiple copies of the $\mathcal{N}=1$ case. The most interesting example for a field theory in this class is the massive fundamental multiplet of $\mathcal{N}=2$, which contains a massive vector boson, $5$ real scalars and $2$ Dirac fermions. 

This multiplet can be reinterpreted \cite{Fayet:1978ig} as a BPS representation of $\mathcal{N}=4$. As pointed out in \cite{Fayet:1978ig}, there is a close link between the massive Dirac equation and BPS conditions. In the language and notation of this paper this link reads:
\begin{equation}\label{eq:relcentrcharg}
k^{\alpha \dalpha} Q^I_{\alpha}  = -\frac{1}{2} Z^{I}_J \bar{Q}^{J,\dalpha} \qquad \textrm{(BPS condition)}.
\end{equation}
The $Z^{I}_J$ can be identified with the central charges as follows: Multiplying both sides with $k_{\beta \dalpha}$ and summing gives,
\begin{equation}
Q^I_{\beta}  = -\frac{1}{2} k_{\beta \dalpha} \frac{Z^{I}_J}{m^2} \bar{Q}^{J,\dalpha}
\end{equation}
so that
\begin{equation}
\{Q^{I}_{\alpha}, Q^{J}_{\beta}\}  = -\frac{1}{2}k_{\alpha \dalpha} \frac{Z^{I}_L}{m^2} \{ \bar{Q}^{L,\dalpha}, Q^{J}_{\beta}\} 
 = Z^{I}_{L} \delta^{L,J} \frac{\left(  k_{\alpha \dalpha} k^{\dalpha}_{\beta}\right)}{m^2} = Z^{I, J} \epsilon_{\alpha \beta}.
\end{equation}
The conjugate equation follows easily as well. For comparison to the usual rest-frame analysis of BPS multiplets one can insert the expansion of the SUSY generators, equation \eqref{eq:covariantizationofsusy}, into equation \eqref{eq:relcentrcharg}. This gives
\begin{equation}
m \left( k^{\flat, \dalpha}  Q^I_+ - \tilde{q}^{\dalpha} Q^I_- \right) =  -\frac{1}{2} Z^{I}_J \left( \tilde q^{\dalpha}\bar{Q}^J_- + k^{\flat,\dalpha} \bar{Q}^J_+\right).
\end{equation}
Since the spinors form a complete basis of the spinor space, condition \eqref{eq:relcentrcharg} implies
\begin{equation}
m \, Q^I_+ = -\frac{1}{2} Z^{I}_J \, \bar{Q}^J_+\,
 \qquad m \, Q^I_- = \frac{1}{2} Z^{I}_J \, \bar{Q}^J_-
\end{equation}
for the rest-frame generators. This is the usual BPS condition: the supercharges are eigenvectors (up to complex conjugation)  under the central charges with eigenvalue equal to the mass.  The representation theory of these representations is seen to reduce to that of massive supersymmetry for a part of the algebra. Hence BPS representations of $\mathcal{N}= (2\textrm{ or } 4)$ SUSY for instance are isomorphic to complex massive representations of $\mathcal{N}= (1\textrm{ or } 2)$.  Of course, this can all be framed in terms of Dirac spinors with complex masses. Doing this yields immediately the 4D BPS analysis  of \cite{Boels:2010mj}. 

\section{On-shell superspaces for the masses}
\label{sec:coherent}
In this section on-shell superspaces for general massive representations will be constructed using the set-up introduced in section~\ref{sec:susyrepsmass}.  The main observation is that these on-shell superspaces can always be obtained as a fermionic coherent state representation since the above analysis reduces the supersymmetry algebra to copies of the fermionic harmonic oscillator. After a reminder on the properties of the coherent-state representation of massless simple supersymmetry the generalization to the massive case will be presented below. This parallels the construction in higher dimensions~\cite{Boels:2009bv} and $\mathcal{N}=4$ SUSY gauge theory on the Coulomb branch~\cite{Boels:2010mj}. Some general properties of the massive superamplitudes which follow will be obtained. In particular this includes  the vanishing of some simple amplitudes for appropriate choices of the spin axes of the massive particles, generalizing results for SQCD obtained from explicit calculations and SWIs~\cite{Schwinn:2006ca} and from diagrammatic arguments reviewed briefly below.

Impatient readers familiar with massless superspaces can skip ahead directly to a short-cut to on-shell superspaces for massive particles described in subsection \ref{subsec:repsupspacemass}. A quick reminder of massless on-shell four dimensional superspaces is contained in subsection \ref{subsec:repsupspacemassless}.

\subsection{Coherent states for massless $\mathcal{N}=1$ SUSY}
In order to set the stage, first recall the coherent states for massless, 4D $\mathcal{N}=1$ representations in a
notation suitable for generalization to the massive case. From the
decomposition of the SUSY generator~\eqref{eq:decompose-susy} one sees
that in the massless limit only the charges $Q_-$ and $\bar Q_+$ appear
and  the massive representation splits into  two massless
supermultiplets. The algebra of the $Q_-$ and $\bar Q_+$ is isomorphic to the creation
and annihilation operators of a fermionic oscillator, which motivates the
introduction of coherent states that are eigenstates of one of the supercharges.

For the supermultiplets containing the states of maximal and minimal helicity $s^+$ and $s^-$ one can define the coherent states by \footnote{In this section the in-states will be considered so that the helicity labels are reversed compared
   to~\cite{ArkaniHamed:2008gz}}
\begin{align}
\label{eq:coherent-massless}
\ket{s^+,\eta}&=
e^{-\frac{1}{\sqrt{2}}\eta  Q_-}
\ket{s^+}
= \ket{s^+}+\eta\ket{s^+-\tfrac{1}{2}} , \\
\ket{s^-,\bar\eta}&=
e^{-\frac{1}{\sqrt{2}}\bar\eta \bar Q_+}
\ket{s^-}
=\ket{s^-}+\bar\eta \ket{s^-+\tfrac{1}{2}}.
\end{align}
Here and in the following momentum labels have been suppressed. The individual spin-states in the multiplet can be obtained from the coherent state representation by appropriate integrations over the Grassmann parameters. The representation used for the maximal-helicity multiplet in~\eqref{eq:coherent-massless} will be denoted as the
$\eta$ representation and that for the minimal-helicity multiplet as the $\bar \eta$-representation. Alternatively to the description chosen above, one could e.g. use the $\eta$ description for both multiplets by defining the states
\begin{equation}
\ket{s^-+\tfrac{1}{2},\eta}=
e^{-\frac{1}{\sqrt{2}}\eta Q_-}
\ket{s^-+\tfrac{1}{2}}.
\end{equation}
These two equivalent representations are related by fermionic Fourier transform,
\begin{equation}\label{eq:fermfourtrafo}
\ket{\eta } = \int d\bar{\eta} \,e^{\bar{\eta} \eta} \,|\bar{\eta} \rangle.
\end{equation}

In complete analogy to the coherent states of a (fermionic) harmonic
oscillator, the two states defined above are eigenstates of $Q_-$ and
$\bar Q_+$, respectively,
\begin{equation}
\begin{aligned}
\frac{1}{\sqrt 2} \bar Q_+ \ket{\eta}&=-\eta\ket{\eta} ,&
\frac{1}{\sqrt 2}  Q_- \ket{\bar \eta}&
=-\bar\eta\ket{\bar\eta}.
\end{aligned}
\end{equation}
Therefore the $\eta$ representation diagonalizes the action of the
$\bar Q_+$ operator, while a SUSY transformation using the remaining
SUSY charge shifts the eigenvalue:
\begin{equation}
\begin{aligned}
e^{\frac{1}{\sqrt 2}\theta_{\dalpha} \bar Q^{\dalpha}} \ket{\eta}&=
e^{\frac{1}{\sqrt 2}\sbraket{\theta k}\bar Q_+} \ket{\eta}=
 e^{\eta\sbraket{\theta k}}\ket{\eta} ,\\
 e^{\frac{1}{\sqrt 2}\theta^\alpha  Q_\alpha} \ket{\eta}&=
e^{\frac{1}{\sqrt 2}\braket{\theta k}Q_-  }\ket{\eta}
=\ket{\eta-\braket{\theta k}} .
\end{aligned}
\end{equation}
The action on the $\bar\eta$ representation has the roles of the two
SUSY generators exchanged:
\begin{equation}
\begin{aligned}
e^{\frac{1}{\sqrt 2}\bar \theta_{\dalpha} \bar Q^{\dalpha}} \ket{\bar\eta}&=
 \ket{\bar\eta -\sbraket{\theta k} } ,\\
e^{\frac{1}{\sqrt 2}\theta^\alpha  Q_\alpha} \ket{\bar \eta}&=
e^{\bar\eta\braket{\theta k}} \ket{\bar \eta}.
\end{aligned}
\end{equation}

\subsubsection{Representing the SUSY generators on fields on on-shell superspace}\label{subsec:repsupspacemassless}
The above can be given a more geometrical interpretation. For this one maps the states to fields on the superspace spanned by the momentum $k_{\mu}$ and one fermionic variable $\eta$. By the above the action of the $\bar{Q}_+$ generator on the superfield is represented by fermionic multiplication, while the action of the $Q_-$ operator on the superfield is given by differentiation,
\begin{equation}
\bar{Q}_+ =  \sqrt{2} \eta \,,\qquad 
Q_-  = - \sqrt{2}  \frac{\partial}{\partial \eta}.
\end{equation}
It can be checked directly this forms a representation of the supersymmetry algebra since 
\begin{equation}
\{\bar{Q}_+, Q_-  \} =- 2 \{\eta, \frac{\partial}{\partial \eta} \} = -2.
\end{equation}
In other words, fermionic multiplication and differentiation form a natural representation of the fermionic harmonic oscillator. Since for massless states
\begin{equation}
Q_{\alpha} = k_{\alpha} Q_- \,,\qquad \bar{Q}_{\dalpha} = k_{\dalpha} \bar{Q}_+
\end{equation}
holds by equation ~\eqref{eq:decompose-susy}, we obtain 
\begin{equation}
Q_{\alpha} = -  \sqrt{2} k_{\alpha}  \frac{\partial}{\partial \eta}, \qquad  \bar{Q}_{\dalpha} =  \sqrt{2}  k_{\dalpha} \eta
\end{equation}
as a representation of the supersymmetry generators acting on fields on an on-shell superspace.  Note that for a choice of on-shell superspace spanned by variables $k$ and $\bar{\eta}$ the fermionic multiplication and differentiation are swapped between the chiral and anti-chiral generators. This is naturally encoded in the fermionic Fourier transform of equation \eqref{eq:fermfourtrafo}. Fields on the massless on-shell superspace have a finite expansion in $\eta$,
\begin{equation}
\phi(k_{\mu},\eta) = \phi^0(k) + \phi_{+} \eta,
\end{equation}
where the $+$ on the field $\phi$ serves as a reminder it has helicity $+\frac{1}{2}$ compared to the field $\phi$. The extension to multiple copies of the massless SUSY algebra needed to describe extended supersymmetry is obvious.

\subsection{Coherent states for massive  $\mathcal{N}=1$ SUSY}

In the massive $\mathcal{N}=1$ case in four dimensions, the $Q_+$ and $\bar Q_-$ components in the decompositions~\eqref{eq:covariantizationofsusy} do not decouple so two copies of the algebra of the fermionic oscillator are encountered, $\{Q_-,\bar Q_+\}$ and $\{\bar Q_-, Q_+\}$. 
Since any of the states of the supersymmetric multiplet can be used as a top state of a coherent state representation, there are four possible parameterizations:
\begin{description}
\item[$\eta$-representation]
Analogously to the massless case, the coherent state in the $\eta$ representation is obtained by acting
on the top-state $\Omega_+$ of the massive multiplet with an
exponential of the `lowering' operators $\bar Q_-$ and $Q_-$ and is
parameterized by two Grassmann valued eigenvalues $\eta$ and $\iota$
of the `raising' operators $\bar Q_+$ and $Q_+$:
\begin{equation}
\ket{k,s_n^+,\eta,\iota}\equiv
e^{-\frac{1}{\sqrt{2}}(\eta Q_-+\iota \bar Q_-)}
\ket{\Omega_+,s_n^+}. 
\end{equation}
Here $s_n^+=s_n+\tfrac{1}{2}$ is the maximal spin in the
supermultiplet based on the Clifford vacuum
$\ket{\Omega_0,s_n}=\ket{k,s,s_n}$. 
For the applications to scattering amplitudes only outgoing
states will be considered. For those the conjugate coherent states are relevant:
 \begin{equation}
\bra{k,s_n^+,\eta,\iota}\equiv
\bra{\Omega_+,s_n^+} e^{-\frac{1}{\sqrt{2}}(\bar Q_+\eta + Q_+\iota)}.
\end{equation}
\item[$\bar\eta$ representation]
Alternatively, a massive $\bar \eta$ representation can be introduced by acting with
the operators $Q_+$ and $\bar Q_+$ on the states with minimal spin
$s_n^-=s_n-\tfrac{1}{2}$:
\begin{equation}
\ket{k,s_n^-,\bar\eta,\bar\iota}\equiv
e^{-\frac{1}{\sqrt{2}}(\bar\eta \bar Q_++\bar \iota Q_+)}
\ket{\Omega_-,s_n^-}
\end{equation}
with the conjugate state given by
\begin{equation}
\label{eq:def-eta-bar}
\bra{k,s_n^-,\bar\eta,\bar\iota}\equiv
\bra{\Omega_-,s_n^-} e^{-\frac{1}{\sqrt{2}}(Q_-\bar\eta + \bar Q_-\bar\iota)}.
\end{equation}
In the following the spin- and momentum labels will be suppressed if no
confusion can arise.  As in the massless case, the two representations are
related by a Fourier transformation, 
\begin{equation}
  \ket{\eta,\iota}=\int d\bar\iota\; d\bar\eta\; e^{-\bar\iota\iota}\;
  e^{-\bar\eta\eta} \ket{\bar\eta,\bar\iota}\;,\quad
 \ket{\bar\eta,\bar\iota}=\int d\eta \;d\iota \; e^{\bar\iota\iota}\;
  e^{\bar\eta\eta} \ket{\eta,\iota}.
\end{equation}

\item[`mixed' representations]

The remaining two representations are constructed based on the  Clifford vacuum or its conjugate. These
`mixed' $\bar \eta \iota$ and  $\eta \bar\iota$ coherent state representations can be obtained e.g. from a partial Fourier transform of the $\eta$ representation:
\begin{equation}
  \ket{\bar\eta,\iota}=\int  d\eta\; 
  e^{\bar\eta\eta} \ket{\eta,\iota}\;,\quad
 \ket{\eta,\bar\iota}= \int\;d\iota \; 
 e^{\bar\iota\iota}\; \ket{\eta,\iota}.
\end{equation}
 Explicitly the in- and out states in the two mixed representations are given by
  \begin{align}
    \ket{k,s_n, \bar\eta,\iota}&= e^{-\frac{1}{\sqrt{2}}(\bar\eta \bar Q_+
      + \iota \bar Q_-)}
    \ket{\Omega_0,s_n} , &
    \ket{k,s_n, \eta,\bar\iota}&=
    -e^{-\frac{1}{\sqrt{2}}(\eta Q_- + \bar \iota Q_+)}
    \ket{\Omega_0',s_n},\\
   \bra{k,s_n, \bar\eta,\iota}&=  
    \bra{\Omega_0,s_n}e^{-\frac{1}{\sqrt{2}}(  Q_-\bar\eta
      +  Q_+\iota)}, &
     \bra{k,s_n, \eta,\bar\iota}&=
    - \bra{\Omega_0',s_n}e^{-\frac{1}{\sqrt{2}}(\bar Q_+\eta + \bar Q_-\bar \iota )}.
    \label{eq:mixed}
  \end{align}
\end{description}

The coherent states diagonalize the action of two of the four SUSY
generators: the raising operators $\bar Q_+$ and $Q_+$ in case of the
$\eta$-representation, the lowering operators $\bar Q_-$ and $Q_-$ in
the case of the $\bar\eta$-representation and the SUSY charge $\bar
Q^{\dalpha}$ ($Q_\alpha$) in case of the $\eta\bar\iota$ ($\bar\eta\iota$) 
representation. 
Acting with the $Q$ and $\bar Q$ SUSY transformations  using the
decomposition~\eqref{eq:covariantizationofsusy} the explicit form of the
transformations in the $\eta$ and $\bar\eta$ representations is found to read:
\begin{equation}
\label{eq:eta-susy}
  \begin{aligned}
 \bra{\eta,\iota} e^{ \frac{1}{\sqrt 2 }\theta^\alpha Q_\alpha }&=
    \bra{\eta,\iota'}e^{\eta\braket{\theta k^\flat}},&
    \bra{\eta,\iota}e^{\frac{1}{\sqrt 2 } \theta_{\dot\alpha}\bar Q^{\dot \alpha}}
    &=\bra{\eta',\iota}e^{- \iota m\frac{\sbraket{\theta q}}{\sbraket{k^\flat q}} } ,\\
  \bra{\bar\eta,\bar\iota} e^{ \frac{1}{\sqrt 2 }\theta^\alpha Q_\alpha }&=
\bra{\bar\eta',\bar\iota}
e^{ \bar\iota m\frac{\braket{\theta q}}{\braket{k^\flat q}} },&
\bra{\bar\eta,\bar\iota}e^{\frac{1}{\sqrt 2 } \theta_{\dot\alpha}\bar Q^{\dot \alpha}}
    &=\bra{\bar\eta,\bar\iota'}e^{\bar\eta\sbraket{\theta k^\flat}},
\end{aligned}
\end{equation}
where the shifted coherent-state parameters are given by
 \begin{equation}
   \begin{aligned}
      \eta'&=\eta +\sbraket{\theta k^\flat},&
    \iota'&=\iota+m\frac{\braket{\theta q}}{\braket{k^\flat q}},\\
  \bar\eta'&=\bar\eta +\braket{\theta k^\flat},&
  \bar\iota'&=\bar\iota-m\frac{\sbraket{\theta q}}{\sbraket{k^\flat q}}.
   \end{aligned}
 \end{equation}
The transformations of the parameters $\eta$ ($\bar\eta$) are
identical to those of the massless coherent states while those of
$\iota$ ($\bar\iota$)  vanish for SUSY transformation
parameters proportional to the reference spinors as in~\eqref{eq:susy-q}.

The analogous transformations in the two mixed representations are given by
\begin{equation}
\label{eq:mixed-susy}
  \begin{aligned}
 \bra{\eta,\bar\iota} e^{ \frac{1}{\sqrt 2 }\theta^\alpha Q_\alpha }&=
\bra{\eta,\bar\iota}
e^{ \eta \braket{\theta k^\flat}+\bar\iota m\frac{\braket{\theta q}}{\braket{k^\flat q}} },&
\bra{\eta,\bar\iota}e^{\frac{1}{\sqrt 2 } \theta_{\dot\alpha}\bar Q^{\dot \alpha}}
    &=\bra{\eta',\bar\iota'},\\
 \bra{\bar\eta,\iota} e^{ \frac{1}{\sqrt 2 }\theta^\alpha Q_\alpha }&=
    \bra{\bar\eta',\iota'},&
    \bra{\bar\eta,\iota}e^{\frac{1}{\sqrt 2 } \theta_{\dot\alpha}\bar Q^{\dot \alpha}}
    &=\bra{\bar\eta,\iota}
  e^{\bar\eta \sbraket{\theta k^\flat}-\iota m\frac{\sbraket{\theta q}}{\sbraket{k^\flat q}} } .
\end{aligned}
\end{equation}
The fact that the states in the mixed representations are
eigenstates of all components of the SUSY charges $Q_\alpha$ or $\bar Q^{\dot\alpha}$ will be
useful to implement supermomentum conservation in the construction of
superamplitudes below. A related observation for six dimensional massless particles has been made in  \cite{Dennen:2009vk}. 

\subsubsection{Representing the SUSY generators on fields on on-shell superspace}\label{subsec:repsupspacemass}
Just as was done in the massless case one can immediately obtain from the above a representation of the massive supersymmetry algebra acting on fields on a superspace whose coordinates are the momentum with two additional fermionic directions. The SUSY generators will correspond to fermionic multiplication and differentiation on this space. Just as above, four choices are possible here depending on which representation of the SUSY algebra one wishes to use. 

In the $\eta \bar{\iota}$ representation we obtain
\begin{equation}\label{eq:susygenonsuperspace}
Q_{\alpha} =   \sqrt{2} k^{\flat}_{\alpha} \eta + \sqrt{2} \frac{m}{\braket{k^{\flat}q}} q_{\alpha}  \bar{\iota}\,,  \qquad
  \bar{Q}_{\dalpha} = - \sqrt{2}  k^{\flat}_{\dalpha} \frac{\partial}{\partial \eta} - \sqrt{2} \frac{m}{\sbraket{q k^{\flat}}} q_{\dalpha} \frac{\partial}{\partial \bar{\iota}}\end{equation}
as a representation of the supersymmetry generators acting on fields on an on-shell superspace. Note that it can again be checked directly that this is a representation of the massive on-shell supersymmetry algebra. The other three possibilities for choices of fermionic multiplication and differentiation are obtained from the above by fermionic Fourier transform.

A field in the $\eta \bar{\iota}$ representation can be expanded into
components according to
\begin{equation}
\label{eq:superfield}
\Phi(k, \eta, \bar{\iota}) = -\phi'_0 + \phi_+ \bar{\iota} - \phi_- \eta 
+ \phi_0 \eta \bar{\iota},
\end{equation}
where the labels refer to the states in the massive multiplet discussed in section~\eqref{sec:mass-reps}.
 The phase factors in the definition of the states have been chosen in order to be compatible with the definition of the coherent state~\eqref{eq:mixed}.
Defining the SUSY transformation of the superfield by
\begin{equation}
\delta_\theta\Phi(k, \eta, \bar{\iota})= \frac{1}{\sqrt{2}}Q(\theta)\Phi(k, \eta, \bar{\iota})
\end{equation}
and evaluating the action of the differential operators,
the transformations of the component fields can be read off:
\begin{align}
 \delta_\theta\phi_0&= 
  \phi_+\braket{\theta k^\flat} + \phi_-m\frac{\braket{\theta
      q}}{\braket{k^\flat q}}, &
  \delta_\theta\phi_0'&=- \phi_-\sbraket{\theta k^\flat}
- \phi_+ m\frac{\sbraket{\theta q}}{\sbraket{k^\flat q}},\nonumber\\
 \delta_\theta\phi_-&=
 \phi_0' \braket{\theta k^\flat}
  + \phi_0m\frac{\sbraket{\theta q}}{\sbraket{k^\flat q}},&
  \delta_\theta\phi_+&=
 -\phi_0 \sbraket{\theta k^\flat} - \phi_0'm\frac{\braket{\theta
      q}}{\braket{ k^\flat q}}.
\label{eq:mass-multiplet}
\end{align}
These results can be shown to be consistent with the transformations obtained using the direct definitions of the states from section~\eqref{sec:susyrepsmass} or the expansion of the coherent states. These transformations generalize the result of~\cite{Schwinn:2006ca} to arbitrary massive representations.

\subsection{Superamplitudes}
Having defined the coherent states that are labeled by the particle
momentum $k$,the spin of the top state and two Grassmann valued
parameters, it is natural to define superamplitudes as functions of
these parameters,
\begin{equation}
A(\{k_i,s_{n_i}^+,\eta_{i},\iota_{i}\}, \{k_j,s_{n_j}^-,\bar\eta_{j},\bar\iota_{j}\},  \{k_k,s_{n_k}^0, \eta_{k},\bar\iota_{k}\},\{k_l,s_{n_l}^0, \bar\eta_{l},\iota_{l}\}).
\end{equation}
Here the particles with index $i$ are in the $\eta$-representation, the ones with index $j$ in the $\bar\eta$ representation, particles with index $k$ in the mixed $\eta\bar\iota$ representation and the index  $l$ denotes particles in the  $\bar \eta\iota$ representation. For the massive particles, the amplitude depends on the spin axes  $n_i$ which are defined through the reference vectors $q_i$.  For amplitudes including both massive and massless supermultiplets the same notation is used with the understanding that the variables $\iota_{i}$ and $\bar \iota_{j}$ vanish for the massless particles. Several properties of the massive superamplitudes will be demonstrated in this subsection.

\subsubsection{SUSY invariance}
First of all, scattering amplitudes must be invariant under the chiral and antichiral supersymmetry transformations. For amplitudes with particles in the $\eta$ and $\bar\eta$ representations, the transformations \eqref{eq:eta-susy}  imply the relations
\begin{equation}
\label{eq:coherent-swi}
\begin{aligned}
A(\{\eta_{i},\iota_{i}\},\{\bar\eta_{j},\bar\iota_{j}\})
&=e^{
\sum_i\eta_{i}\braket{\theta k_i^\flat}+
\sum_j  \bar\iota_{j}m_j\frac{\braket{\theta q_j}}{\braket{k_j^\flat q_j}} }
A(\{\eta_{i},\iota'_{i}\},\{\bar\eta'_{j},\bar\iota_{j}\}),\\
A(\{\eta_{i},\iota_{i}\},\{\bar\eta_{j},\bar\iota_{j}\})
&=e^{\sum_j \bar\eta_{j}\sbraket{\theta k_j^\flat}
-\sum_i\iota_{i}m_i\frac{ \sbraket{\theta q_i}}{\sbraket{k_i^\flat q_i}}}
A(\{\eta'_{i},\iota_{i}\},\{\bar\eta_{j},\bar\iota'_{j}\}),
\end{aligned}
\end{equation}
where the spin-and momentum labels are again suppressed. These identities encode the SUSY-Ward identities of helicity amplitudes
that can be extracted performing appropriate integrals over the Grassmann parameters.  This relation will be used in section~\ref{sec:vanish} to show the vanishing of the analog of the all-plus and one-minus amplitudes for appropriate choices of the spin axes. If the same spin axis defined by a reference vector $q$ is used for all massive particles and for a SUSY transformation aligned with the spin axis, the invariance relation resembles that in the massless case~\cite{ArkaniHamed:2008gz}:
\begin{equation}
\label{eq:coherent-swi-2}
A(\{\eta_{i},\iota_{i}\},\{\bar\eta_{j},\bar\iota_{j}\})|_{q_i=q}
=e^{
\sum_i\eta_{i}\theta\braket{q k_i^\flat}  +
\sum_j \bar\eta_{j} \theta\sbraket{q k_j^\flat} 
}
A(\{\eta'_{i},\iota_{i}\},\{\bar\eta'_{j},\bar\iota_{j}\})|_{q_i=q}\,.
\end{equation}

For amplitudes with all legs in one of the mixed representations, the SUSY transformations of  the coherent states are given by~\eqref{eq:mixed-susy} so the invariance of the superamplitudes takes the form
\begin{equation}
\label{eq:coherent-swi-3}
\begin{aligned}
A(\{\eta_{k},\bar\iota_{k}\},\{\bar\eta_{l},\iota_{l}\})
&=e^{
\sum_k\eta_{k}\braket{\theta k_k^\flat}+
  \bar\iota_{k}m_k\frac{\braket{\theta q_k}}{\braket{k_k^\flat q_k}} }
A(\{\eta_{l},\bar\iota_{l}\},\{\bar\eta'_{k},\iota_{k}'\}),\\
A(\{\eta_{k},\bar\iota_{k}\},\{\bar\eta_{l},\iota_{l}\})
&=e^{\sum_l \bar\eta_{l}\sbraket{\theta k_l^\flat}
-\iota_{l}m_l\frac{ \sbraket{\theta q_l}}{\sbraket{k_l^\flat q_l}}}
A(\{\eta'_{k},\bar\iota'_{k}\},\{\bar\eta_{l},\iota_{l}\}).
\end{aligned}
\end{equation}
Again, this identity simplifies to a form analogous to the massless case if the SUSY  transformation is aligned with the common reference spinors of the external legs.

\subsubsection{Supersymmetric momentum conservation}

In \cite{ArkaniHamed:2008gz} it was shown that for massless multiplets
in the coherent state formulation there is a supersymmetric partner of
the momentum conserving delta function which (for the case of extended
SUSY with $2\mathcal{N}$ supercharges) can be written as
\begin{equation}\label{eq:susymomcons}
\sim \delta^{2 \mathcal{N}}\left(\sum_{i=1}^n 
  \eta_i^I k^i_{\alpha} \right).
\end{equation}
Note the number of manifestly conserved charges here, which is twice the number of fermionic generators. This is related to the fact that
for massless fields the helicity equal and one-helicity unequal amplitudes vanish: one needs $2 \mathcal{N}$ integrations to get a
nonzero answer. Repeating the same reasoning as in \cite{ArkaniHamed:2008gz}, the question is under which supersymmetry
transformations a coherent state amplitude (for definiteness consider the case with all particles in the $\eta$ representation) only acquires a phase under a chiral SUSY transformation, e.g.
\begin{equation}
\label{eq:delta-condition}
A\left(\{\eta_{i}, \iota_{i}, k_i \} \right) 
\rightarrow e^{f_\alpha\left(\{\eta_{i},\iota_{i}, k_i\}\right)\theta^\alpha}
A\left(\{\eta_{i}, \iota_{i}, k_i \} \right)
\end{equation}
for some function $f$.
 In other words, the coherent state variables must remain unshifted. In the massless case this is true for all
chiral transformations if all particles are in the same coherent state representation. If the condition~\eqref{eq:delta-condition} is satisfied, one can infer that the amplitude must be proportional to a delta function $\delta^2\left(f_{\alpha}\left(\{\eta_{i},\iota_i,k_i\}\right)\right)$, in analogy to~ \eqref{eq:susymomcons}. In the massive case this is no longer true in general for all representations.

An inspection of the invariance conditions~\eqref{eq:coherent-swi} and~\eqref{eq:coherent-swi-3} shows that a condition of the form~\eqref{eq:delta-condition} is satisfied for the chiral transformation if all particles are in the $\eta\bar\iota$ representation and for the anti-chiral transformation  if all particles are in the $\bar\eta\iota$ representation. For the fundamental massive multiplet this means that the top state is one of the two scalars. In general with massive particles in the $\eta \bar{\iota}$ representation and massless fields in the $\eta$ representation one obtains a delta function of the type
\begin{equation}\label{eq:suppoincdelt}
\delta^{2 } \left( \mathcal{Q}_{\alpha} \right) \equiv\frac{1}{2} \braket{\mathcal{Q} \mathcal{Q}} =  \mathcal{Q}_1 \mathcal{Q}_2,
\end{equation}
where the conserved fermionic momentum reads
\begin{equation}
\label{eq:supermomentum}
\mathcal{Q}_\alpha = \sum_k \left( \eta_k k^\flat_{k,\alpha}  +\bar\iota_k m_k \frac{ q_{k,\alpha} }{\braket{k_k^\flat q_k}} \right).
\end{equation}
This naturally incorporates amplitudes with some massless particles in
the $\eta$ representation where  the corresponding $m_k$ vanish.
  The $\mathcal{Q}$-SUSY is
manifest through the delta-function while the remaining
SUSY-transformations  require that the vertex is annihilated by the
operator
\begin{equation}
  \bar{\mathcal{Q}}^{\dot\alpha}=\sum_i
  - k_i^{\flat,\dot\alpha}\frac{\partial}{\partial \eta_i}
  +q_i^{\dot\alpha}\frac{m_i}{\sbraket{k_i^\flat q_i}}
  \frac{\partial}{\partial \bar\iota_i}\,.
\end{equation}
 The invariance of the delta-function under the $\bar{\mathcal{Q}}$ operator is a consequence of momentum
  conservation. This is most easily seen using the SUSY algebra:
  \begin{equation}
    \lbrack \bar {\mathcal{Q}}_{\dot\alpha},\mathcal{Q}_\alpha
    \mathcal{Q}^\alpha\rbrack =-2 \sum_i k_{i,\dot\alpha\alpha} 
    \mathcal{Q}^\alpha=0\,.
  \end{equation}
Analogously, the $\bar \eta \iota$ representation will
lead to the naturally conjugate delta function 
\begin{equation}
\label{eq:supermomentum-con}
\delta^{2 } \left(\bar{\mathcal{Q}}^{\dot\alpha} \right)\;, \qquad
\bar{\mathcal{Q}}^{\dot\alpha} = 
\sum_k \left( \bar \eta_k k_{k}^{\flat,\dot\alpha}  
  -\iota_k m_k \frac{ q_{k}^{\dot\alpha} }{\sbraket{k_k^\flat q_k}} \right).
\end{equation}

The situation is more involved if all particles are in the $\eta \iota$ or $\bar\eta \bar{\iota}$ representation. In order to obtain a SUSY transformation where the superamplitudes only pick up a phase as required by~\eqref{eq:delta-condition}, it is necessary to use the same spin axis for all external particles and restrict to chiral SUSY transformations aligned to the spin quantization axis,
$\theta_\alpha=\theta q_\alpha$. In this case, the phase is given by the function
\begin{equation}
f_\alpha\left(\{\eta_{i},k_i\}\right)\theta^\alpha =  \sum_{i=1}^n 
\eta_{i} \theta \braket{q k_i^{\flat} }. 
\end{equation}
Hence the supersymmetric partner of the momentum conserving delta function for massive particles in the $\eta \iota$-representation reads
\begin{equation}\label{eq:susymassmomcons}
 \delta\left(
\sum_{i=1}^n \eta_{i} \braket{qk_i^{\flat}} \right).
\end{equation} 
In this case the delta function restricts only half of the SUSY
parameters. Moreover it is not invariant under the
full supersymmetry algebra.  An analogous argument using an anti-chiral transformation shows that
amplitudes in the $\bar\eta \bar{\iota}$ representation involve the delta function
\begin{equation}
 \delta\left(
  \sum_{i=1}^n \bar\eta_{i}\sbraket{ qk_i^{\flat} }\right).
\end{equation}
  From these results it can be seen immediately that the
$\eta$-and $\bar\eta$ representations are less suitable for the
construction of a manifestly supersymmetric formalism than the mixed
representations.

The arguments given here extend also to massive representations of
$\mathcal{N}$ extended SUSY where coherent states in a formalism
similar to the one used here have been constructed in~\cite{Boels:2010mj}.
In this case there are $2\mathcal{N}$ coherent state parameters $\eta^I$ and $\bar{\iota}^I$ with $I=1,\dots\mathcal{N}$, the conserved supermomentum~\eqref{eq:supermomentum} is generalized to $\mathcal{N}$ objects $\mathcal{Q}^I$ and the delta function is extended to 
\begin{equation}
\delta^{2\mathcal{N}} \left( \mathcal{Q}^I_{\alpha} \right) \equiv
\prod_I \frac{1}{2} \braket{\mathcal{Q}^I \mathcal{Q}^I} . 
\end{equation}
Similarly, in the $\eta \iota$ representation the delta function is given by
\begin{equation}
 \delta^{\mathcal{N}}\left(
\sum_{i=1}^n \eta^I_{i} \braket{qk_i^{\flat}} \right),
\end{equation} 
where it is seen that only half of the SUSY transformations are constrained. Again this delta function is not a solution to the SUSY Ward identities. 

\subsubsection{Superspin constraints}

The superamplitudes are further constrained by the Lorentz-invariance
condition~\eqref{eq:spin-constraint}. In analogy to the massless
case (e.g.~\cite{Drummond:2010ep}) it is useful to supersymmetrize this
constraint by assigning a superfield the spin of the top
state. This implies that the Grassmann variables $(\eta,\iota)$ and
$(\bar\eta,\bar\iota)$ are assigned the spin $\frac{1}{2}$ and
$-\frac{1}{2}$, respectively. The superamplitudes therefore satisfy
the identity (suppressing the Grassmann variables and the distinction among the various representations)
\begin{equation}
\label{eq:susy-spin}
  \mathcal{S}_iA(  \{k_k,s_{n_k}^0 \})
=-2s_{n_i}^0A(  \{k_k,s_{n_k}^0\}),
\end{equation}
where the
 superspin operator is given by
\begin{equation}
  \mathcal{S}_i=k_i^{\flat,\alpha}\frac{\partial}{\partial k_i^{\flat,\alpha}}
  -k_i^{\flat,\dot\alpha}\frac{\partial}{\partial k_i^{\flat,\dot \alpha}}
-\eta_i\frac{\partial}{\partial\eta_i}
+\bar\eta_i\frac{\partial}{\partial\bar\eta_i}
+\bar\iota_i\frac{\partial}{\partial\bar\iota_i}
-\iota_i\frac{\partial}{\partial\iota_i}\,.
\end{equation}
As mentioned before all amplitudes that will be
considered here are homogeneous of degree zero in the reference spinors.  Hence  the
$q$-dependent terms in~\eqref{eq:spin-constraint} always drop out and have not  been included in the expression above.

\subsubsection{Sums over supermultiplets}
One application of on-shell superspaces used often for massless particles is to simplify sums over spectra of on-shell particles. These sums appear for instance after taking unitarity cuts of scattering amplitudes at loop level or at a kinematic pole of a tree level amplitude. Typically this will take the form of
\begin{equation}
\sum_{s \in \textrm{species}, \textrm{spins}} A_L\left(\ldots, \{P, s\}\right) A_r\left(\{-P,-s \}, \ldots\right),
\end{equation}
where $A_L$ and $A_R$ appear on both sides of the tree level kinematic pole of an amplitude for instance. The momentum $P$ is on-shell, $P^2 = m^2$. The sum over $s$ ranges over all states which can appear in the theory with this particular mass. Hence it ranges over particle species as well as all values of the quantum number $s$ (helicity or spin) which can appear. The flip in sign occurs due to the difference in incoming and outcoming momentum. 

Using an on-shell superspace the discrete sum can be replaced by a continuous fermionic integral. Consider for simplicity first the fundamental massive multiplet. Expanding out the fermionic components of 
\begin{equation}\label{eq:sumtofermint}
\int d \eta d\bar{\iota} A_{L} (\ldots, \{P, \eta, \bar{\iota}\}) A_{R} (\{-P, \eta, \bar{\iota}\}, \ldots)
\end{equation}
for instance shows clearly all the different terms appearing. For more complicated representations than the fundamental one a sum over the different superfields has to be included. For the massive vector multiplet in $\mathcal{N}=1$ for instance which consists of two massive superfields one should write
\begin{equation}
\sum_{i=1}^2 \int d \eta d\bar{\iota} A_{L} (\ldots, \{P, \eta, \bar{\iota}\}_i) A_{R} (\{-P, \eta, \bar{\iota}\}_{i-1}, \ldots),
\end{equation}
where the sum over $i$ ranges over the two (naturally conjugate!) superfields.

\subsection{Vanishing all-multiplicity amplitudes}
\label{sec:vanish}

The compact form of the SWIs in the coherent state
formalism~\eqref{eq:coherent-swi} will now be applied to derive the
vanishing of some classes of massive amplitudes by generalizing the
discussion of the massless case in~\cite{ArkaniHamed:2008gz}. For
massive particles this will be seen in general to require special choices
of the spin-axes of the external particles.  For multi-gluon amplitudes with massive quarks or scalars some explicit results are
available~\cite{Forde:2005ue,Ferrario:2006np,Schwinn:2006ca} and for amplitudes
with massive vector bosons diagrammatic arguments
following~\cite{Dixon:1996wi,Boels:2009bv} can be used to cross-check
our findings.

\subsubsection{All particles in maximal spin-state}
As a generalization of massless amplitudes with all external particles
in the same helicity state, consider amplitudes where all particles
are in the state of either maximal or minimal spin of their SUSY
multiplet.  At this stage the Clifford vacuum of the representations
is kept arbitrary so this includes amplitudes with all particles in
the massive vector or quark multiplets as well as mixed
amplitudes. For definiteness consider the case were all particles are
in the state of maximal spin.  This amplitude can be extracted from
the superamplitude with all particles in the $\bar\eta$ representation
by the integral
\begin{equation}
  A((\{k_j,s_j^+\})
  =\int \prod_{j=1}^n (d\bar\iota_jd\bar\eta_j)\,
  A(\{k_j,\bar\eta_j,\bar\iota_j\}).
\end{equation}
For a common choice of spin axis the invariance under SUSY
transformations aligned with the spin axis implies the
identity~\eqref{eq:coherent-swi-2}.  Using a
transformation~\eqref{eq:eta-susy} with $\theta_{\alpha}=-\bar\eta_1
q_{\alpha}/\braket{qk_1^\flat}$ and $\theta^{\dot\alpha}=0$, the
variable $\bar\eta_{1}$ can be shifted to zero, resulting in the
following representation of the amplitude:
\begin{equation}
  A((\{k_j,s_j^+\})|_{q_j=q}
  =\int \prod_{j=1}^n (d\bar\iota_jd\bar\eta_j)
  A((k_1,0,\bar\iota_{1}),\{k_j,\bar\eta'_j,\bar\iota_j\})|_{q_j=q}
  =0.
\end{equation}
The amplitude vanishes upon performing the $\bar\eta_1$ integral since
the only dependence of the amplitude on $\bar\eta_{1}$ is through the
shifted $\bar\eta'_j$ variables and can be absorbed by a shift of the
integration variables.  The preceding argument is essentially the same
as in the massless example~\cite{ArkaniHamed:2008gz}, however for
massive particles it was necessary to pick the same spin axis and
choose a special SUSY transformation in order to avoid an
$\bar\eta_1$-dependent phase factor involving the $\iota$ variables
in~\eqref{eq:coherent-swi}.

These findings are in agreement with the explicit result that the
amplitudes with two massive 
quarks~\cite{Ferrario:2006np,Schwinn:2006ca} and an arbitrary number
of positive helicity gluons vanishes if the two quarks have positive
spin with respect to the same axes. For amplitudes with only massive
or massless gauge bosons the vanishing of the amplitudes with equal
spin labels can be checked at tree level by a diagrammatic argument
using the simple but powerful observation
\eqref{eq:simplebutpowerful}.  Since this argument is analogous to
that for unbroken Yang-Mills theory in four
dimensions~\cite{Dixon:1996wi} and higher
dimensions~\cite{Boels:2009bv} we will be very brief. Working in the
Feynman-'t Hooft gauge ($R_{\xi=1}$ gauge), momenta in the numerator of
Feynman diagrams can only arise through three-point vertices of gauge
bosons coupling to themselves and scalars that contain at most one
power of the momentum.  Since there are at most $n-2$ three-point
vertices in an $n$-point amplitude, each diagram must contain at least
one contraction of polarization vectors
$\epsilon_i^\pm\cdot\epsilon_j^\pm$ that vanishes according
to~\eqref{eq:simplebutpowerful}, provided the same spin axis is used
for all massive vector bosons.  The vanishing of the all-plus and
all-minus amplitudes for spontaneously broken gauge theories is
mentioned in \cite{Chalmers:2001cy}.

For the representation~\eqref{eq:massive-vec-} based on a spin one-half Clifford vacuum $\ket{\Omega_0,-\frac{1}{2}}$ the maximal representation is actually the state $\ket{\Omega_+,0}=\frac{1}{\sqrt 2}(\ket{W_k^0}-\ket{\Phi_k})$ so this yields an identity of an amplitude of a longitudinal vector and a massive scalar.

\subsubsection{One particle in minimal spin-state}
The next amplitudes to be considered are those with $n-1$ particles in the maximal spin
state of their multiplet and one particle (say particle $n$) in the minimal spin state. These amplitudes can be expressed as an integral over a superamplitude where particles one to $n-1$ are in the $\bar\eta$-representation and particle $n$ is in the $\eta$ representation:
\begin{equation}
  A((\{k_j,s_j^+\},(k_n,s_n^-))
  =\int \prod_{j=1}^{n-1} (d\bar\iota_jd\bar\eta_j)
  (d\eta_{n}d\iota_{n})
  A(\{k_j,\bar\eta_j,\bar\iota_j\},(k_n,\eta_{n},\iota_{n})).
\end{equation}
Both $\bar\eta_1$ and $\bar\eta_2$ can be transformed to zero using a
chiral SUSY transformation with
 \begin{equation}
\label{eq:eta-12}
\theta_{12,\alpha}=
\frac{k_{2,\alpha}^\flat\,\bar\eta_{1}-k_{1,\alpha}^\flat\,\bar\eta_{2}}{
  \braket{k_2^\flat k_1^\flat}}.
\end{equation}
Under this transformation, the amplitude becomes
\begin{align}
  A((\{k_j,s_j^+\},(k_n,s_n^-))
  =\int & \prod_{j=1}^{n-1} (d\bar\eta_jd\bar\iota_j) (d\eta_{n}d\iota_{n})
  e^{\sum_{j=1}^{n-1}\, \bar\iota_j m_j 
  \frac{\braket{\theta_{12} q_j}}{\braket{k_j^\flat q_j}} +
\eta_n \braket{\theta_{12} k_n^\flat}  } \nonumber\\
&\times  A((k_1,0,\bar\iota_1), (k_2,0,\bar\iota_2),
  \{k_j,\bar\eta'_j,\bar\iota_j\},(k_n,\eta_{n},\iota'_{n}))
\end{align}
with $\bar\eta_j'=\bar\eta_j+\braket{\theta_{12} k_j^\flat}$, 
$\iota_n'=\iota_n+m_n\frac{\braket{\theta_{12} q_n}}{\braket{k_n^\flat q_n}}$.

In the massless case considered in~\cite{ArkaniHamed:2008gz} the
amplitude depends on $\bar\eta_{1/2}$ only through the $\bar\eta_j'$
that are integrated over and through the phase involving
$\braket{\theta_{12}k_n^\flat}$. This spinor product is of the form
$a_1\bar\eta_{1}+a_2\bar\eta_{2}$ where the coefficients $a_{1,2}$
follow from~\eqref{eq:eta-12}.  Changing the $\bar\eta_1$ and
$\bar\eta_2$ integration variables to the linear combinations
$\bar\eta_\pm=a_1\bar\eta_1\pm a_2\bar\eta_2$, the amplitude vanishes
upon the $\bar\eta_-$ integration.  Once some particles are massive,
the dependence of the amplitude on $\bar\eta_1$ and $\bar\eta_2$ is
more involved.  However, the argument used in the massless case still
applies if a common spin quantization axis
$q_{i,\alpha}=k^\flat_{n,\alpha}$ is chosen for all massive particles,
except for the maximal-spin particle $n$. For this choice, as in the
massless case, the amplitude depends on $\bar\eta_{1/2}$ only through
$\braket{ k_n^\flat\theta_{12}}=\bar\eta_+$ and through the shifted
$\bar\eta_j'$ and $\iota_n'$ that are integrated over so that the
amplitude vanishes after integrating over $\bar\eta_-$:
\begin{equation}
\label{eq:vanish++-}
   A((\{k_j,s_j^+\},(k_n,s_n^-))|_{q_{i,\alpha}=k_{n,\alpha}^\flat}=0.
\end{equation}
This result generalizes observations based on SWIs or
diagrammatic arguments for specific multiplets.
In~\cite{Schwinn:2006ca} it was shown that the amplitude of a pair of
massive quarks with spin $+\tfrac{1}{2}$, one negative helicity gluon
and an arbitrary number of positive helicity gluons vanishes only if
the chiral reference spinor used in the definition of the massive
quarks is chosen as the momentum spinor of the negative helicity
gluon.  For amplitudes with external vector bosons only, the
diagrammatic argument given above for the all-plus amplitude can be
used in the present case as well since for the choice
$q_{i,\alpha}=k^\flat_{n,\alpha}$ as reference spinor for all
positive-helicity polarization vectors $\epsilon_j^+$, the identity
$\epsilon^+(k_j)\cdot\epsilon^-(k_n)$ is satisfied in addition
to~\eqref{eq:simplebutpowerful} (for the related higher-dimensional
case see~\cite{Boels:2009bv}).

For the choice of a common but arbitrary spin quantization axis for
all particles, the amplitude does not vanish, but a simple SWI can be
derived from the identity~\eqref{eq:coherent-swi-2} for SUSY
transformations aligned with the spin axis.  Projecting out the
auxiliary amplitude $A(s_j^+,\dots s_n^{0'})$ with
$\bra{s_n^{0'}}\equiv \bra{\Omega_0',s^0_n}$ from the superamplitude and
using  SUSY invariance results in:
\begin{equation}
  \begin{aligned}
  A(\{k_j,s_j^+\},(k_n, s_n^{0'}))=\int \prod_{j=1}^{n-1}
(d\bar\eta_jd\bar\iota_j)  (d\iota_{n}d\eta_{n})\; \eta_n \;
  A(\{k_j,\bar\eta_j,\bar\iota_j\},(k_n,\eta_{n},\iota_{n}))\\
    =\int \prod_{j=1}^{n-1} (d\bar\eta_jd\bar\iota_j)
  (d\iota_{n}d\eta_{n}) \;\eta_n\;
  e^{\sum_j \bar\eta_j \theta\sbraket{qk_j^\flat}}\;
  A(\{k_j,\bar\eta_j,\bar\iota_j\},(k_n,\eta'_{n},\iota_{n})).
  \end{aligned}
\end{equation}
Expanding the exponential and performing the integrals generates a sum of amplitudes where one of the negative-spin states is replaced  by $\bra{s_i^{0}}$:
\begin{equation}
  \int \dots (d\bar\eta_id\bar\iota_i)\dots \bar\eta_i\dots 
  A(\dots,\{k_i,\bar\eta_i,\bar\iota_i\}\dots)=-
(-1)^{\chi_i} A(\dots (k_i,s_i^{0})\dots)
\end{equation}
as can be seen from the definition of the coherent
states~\eqref{eq:def-eta-bar}. 
Here $\chi_j$ counts the \emph{fermionic} states $s_j^+$ with $j<i$.
  Shifting the $\eta_n$ integral to
$\eta_n'=\eta_n-\theta\sbraket{qk_n^\flat}$ results in the SWI
\begin{equation}
\label{eq:swi++-}
\sbraket{qk_n^\flat}A(\{k_j,s_j^+\},(k_n, s_n^-)) =\sum_{j=1}^{n-1}(-1)^{\chi_j}
 \sbraket{qk_j^\flat}\,A((k_1,s_1^+),\dots (k_j,s_j^{0}),\dots (k_n,s_n^{0'})).
\end{equation}


\section{Applications to SUSY models with massive particles}
\label{sec:exampamplis}
In this section the on-shell superspace techniques developed in the previous section are applied to a series of examples. Particular attention is paid to the three point vertices as they are crucial in the application of supersymmetric on-shell recursion relations to be presented in the next section.

\subsection{SQCD with massive matter}
\label{sec:sqcd}
As a first example, consider  SQCD with a massive matter multiplet in the fundamental representation. The SUSY Ward identities of helicity amplitudes in this model have been studied previously in~\cite{Schwinn:2006ca}. The Lagrangian and the on-shell three-point vertices are summarized in appendix~\ref{app:sqcd}. In the on-shell superspace formulation, the ingredients are a massive superfield $\Phi$ in the fundamental of $SU(N)$, a massive superfield  $\bar\Phi$ in the anti-fundamental and  massless superfields including the positive and negative helicity gluon and gluinos. In the $\eta\bar\iota$ representation for the massive superfields and the $\eta$ representation for the massless superfield they read
\begin{equation}
\label{eq:sqcd-fields}
  \begin{aligned}
   \Phi(\eta,\bar\iota)&=-\phi^-+\eta Q^--\bar\iota Q^++\eta\bar\iota\; \phi^+,&
   \bar\Phi(\eta,\bar\iota)&=-\bar\phi^-+\eta \bar Q^--\bar\iota \bar Q^+
   +\eta\bar\iota\; \bar \phi^+,\\
    G^-(\eta)&=\Lambda^-+\eta g^- ,&
    G^+(\eta)&= g^+ +\eta \Lambda^+.
  \end{aligned}
\end{equation}
From the representation of the component fields in terms of polarization vectors and spinors it is seen that the superfield $G^+$ has superspin $1$ while the field $G^-$ has superspin $-\frac{1}{2}$.

\subsubsection{Three-point superamplitudes}
There are two three-point supervertices of matter fields and the gluon multiplet that can be constructed out of the superfields~\eqref{eq:sqcd-fields}: those with field content $\bar\Phi G^-\Phi$ and $\bar\Phi G^+\Phi$. For definiteness, let us express the superamplitudes in the  $\eta\bar\iota$ representation. From the general argument in the previous section, both these amplitudes should be proportional to the supermomentum-conserving delta function\footnote{As will be shown explicitly further on, a subtlety which arises for massless fields in the case of three-point kinematics does not arise in the massive case.} of equation ~\eqref{eq:suppoincdelt}.
Therefore the three-point superamplitudes take the form\footnote{
  In the discussion of SQCD, we will consider colour ordered vertices with the colour structures and gauge coupling stripped off. All the reference spinors will be chosen equal.
}: 
\begin{equation}
\label{eq:v3-fund}
  A_3(\bar \Phi_1, G_2^\pm,\Phi_3)=\delta^2(\mathcal{Q}_\alpha)\,
  F^\pm\left(\eta_i,\bar\iota_i, \braket{ij},\braket{iq}\right)
\end{equation}
with the explicit form of the Grassmann delta-function
\begin{equation}
\label{eq:delta-Q}
  \delta^{2 } \left( \mathcal{Q}_{\alpha} \right)=
\sum_{i,j}\left(\frac{1}{2}\braket{i j}\eta_i\eta_j 
  +m  \frac{\braket{i q}}{\braket{j q}}\,\eta_i\bar\iota_j 
\right).
\end{equation}
The functions $F$ can be expressed entirely in terms of angular spinor
brackets as can be derived by exploiting the on-shell three-point
kinematics. As a result of momentum conservation one can obtain the relation 
\begin{equation}
\label{eq:3-pt-mom-con}
\frac{\sbraket{23}}{\braket{1q}} = \frac{\sbraket{31}}{\braket{2q}} = \frac{\sbraket{12}}{\braket{3 q}}.
\end{equation}
Furthermore the on-shell conditions
\begin{equation}
2 k_2 \cdot k_3 = (k_2 + k_3)^2 - m^2 = (k_1)^2 - m^2 = 0
\end{equation}
imply the identity
\begin{equation}
\label{eq:swap-square}
\braket{2 3} \sbraket{2 3} = - m^2 \frac{\braket{q 2} \sbraket{q
    2}}{\braket{q 3} \sbraket{q 3}} = - m^2 \frac{\braket{3
    1}\sbraket{3 1}}{\braket{2 1} \sbraket{2 1}}
\end{equation}
and an analogous relation for $\braket{21}\sbraket{21}$.
Since the vertex should be homogeneous of degree zero in the reference spinors, the conjugate equation of~\eqref{eq:3-pt-mom-con} allows to eliminate ratios of square spinor brackets involving auxiliary spinors.
Taken together, these results therefore can be used to express the three point amplitudes entirely in terms of angular spinor brackets.

A similar representation of the vertices can be written down in the
conjugate superspace (the $\bar{\eta} \iota$ representation)
 where they are
conventionally expressed through square brakets:
\begin{equation}
\label{eq:v3-fund-conj}
  A_3(\bar \Phi_1, G_2^\pm,\Phi_3)|_{\bar\eta\iota}
=\delta^2(\bar{\mathcal{Q}}^{\dot\alpha})\,
\bar F^\pm\left(\bar\eta_i,\iota_i, \sbraket{ij},\sbraket{iq}\right).
\end{equation}
The vertices in the $\eta\bar\iota$-representation are related to 
that in the conjugate superspace by  a Grassmann-Fourier-transform:
\begin{equation}
  A(\bar \Phi_1, G_2^\pm,\Phi_3)|_{\eta\bar \iota}=  
 \bar F^\pm\left(\left\{\frac{\partial}{\partial \eta_i},
      \frac{\partial}{\partial \bar\iota_j}\right\}\right)
\int\prod_{i=1}^3 d\bar\eta_i\, \prod_{j=1,3}d\iota_j \;
e^{\eta_i\bar\eta_i} e^{-\iota_j\bar\iota_j}
\delta^{2 } \left( \bar{\mathcal{Q}}^{\dot\alpha} \right).
\end{equation}
Hence fixing one three point amplitude also fixes the conjugate amplitude. 
The functions $F$ and $\bar F$ are subject to certain requirements.  In a renormalizable theory, the
vertex must have mass dimension one, since the coupling constant is
dimensionless and the vertices are either proportional to a momentum
and dimensionless external wavefunctions or involve two external quark
spinors with mass dimension one half.  Since the Grassmann delta
function has mass dimension one, the function $F$ must be
dimensionless. The three-point supervertices must also satisfy constraints implied by
the Lorentz-invariance condition~\eqref{eq:susy-spin}.
Since $ \mathcal{S}_i\mathcal{Q}_\alpha=0$ the
Grassmann-delta function is annihilated by the superspin operator and
the spin information is carried by the function $F$ alone.
 Finally, as argued before, the vertex should be homogeneous of degree
 zero in the auxiliary spinors $q$.

 The relation between the representations~\eqref{eq:v3-fund}
 and~\eqref{eq:v3-fund-conj} can be used to show that the functions
 $F$ and $\bar F$ are at most linear in the Grassmann variables. To
 see this, note that the fermionic Fourier transform of the $\bar
 {\mathcal{Q}}^{\dot\alpha}$-delta function has fermionic weight $3$
 in the $\eta \bar{\iota}$ representation, while the
 $\mathcal{Q}_\alpha$-delta function has degree two. Since a
 Grassmann-parameter dependent function $\bar F$ can only lower the
 degree in the $\eta \bar{\iota}$ representation, this implies that
 the maximum fermionic weight for the functions $F$ and $\bar F$ is
 one.  The most general Ansatz is therefore of the form
 $F=F_0+\sum_i(c_i\eta_i+d_i\bar\iota_i)$. Due to the supersymmetric delta
 function it is possible to eliminate two of these terms by adding a
 term proportional to $\braket{\xi \mathcal{Q}}$ for some spinor
 $\xi$. For the case of two massive legs, this leaves a possible
 dependence on three Grassmann parameters.  There are then three
 possible non-trivial solutions to the condition $\bar{\mathcal
   Q}F=0$:
\begin{subequations}
\begin{align}
\label{eq:qbar-susy}
  F_1(\{\bar\iota_i\})&=\sum_{i=1,3} c_i \bar\iota_i\qquad\text{with}\qquad
\sum_{i=1,3} \frac{m_i}{\sbraket{k_i^\flat q}} c_i=0 ,\\
 F_2(\{\eta_i\})&=\eta_1\sbraket{23}+\eta_2\sbraket{31}+\eta_3\sbraket{12}
\label{eq:qbar-susy-2}, \\[7pt]
 F_3 & \neq F_3(\eta, \bar{\iota}).
\end{align}
\end{subequations}
However, due to the identity~\eqref{eq:3-pt-mom-con} the case $F_2$ is
actually proportional to $\braket{q \mathcal{Q}}$ which annihilates the
fermionic delta function by construction. The solution to the SUSY
constraints for different reference spinors of the three legs is
discussed in appendix~\ref{app:3-pt}. From this analysis it also
follows that an in principle possible solution involving the two
$\bar\iota$ variables and one of the $\eta$-variables is excluded if
all reference spinors are chosen equal.

Let us now take the spin-constraint on the solutions $F$ into account.
These functions can depend on six spinor products so they are not
completely determined by the three spin constraints $\mathcal{S}_i
F=-2s_i F$. Taking also the requirement of equal powers of the
reference spinors in numerator an denominator into account, there are
two free degrees of freedom left.  Consider the function $F_3$ first.
Taking the two independent parameters to be the exponents of
$\braket{q1}$ and $\braket{q3}$ the solution of the superspin
constraints can be written as
\begin{equation}
\label{eq:f3-spin}
F_3=g_{123}  \braket{12}^{\alpha_3+\beta_3}
\braket{23}^{\alpha_1+\beta_1}
\braket{31}^{\alpha_2-(\beta_1+\beta_3)}
\frac{\braket{q1}^{\beta_1}\braket{q3}^{\beta_3}}{\braket{q2}^{\beta_1+\beta_3}},
\end{equation}
where $\alpha_1=s_1-s_2-s_3$, $\alpha_2=s_2-s_3-s_1$ and  $\alpha_3=s_3-s_1-s_2$ are the solutions in the massless case~\cite{Benincasa:2007xk} and the $g_{ijk}$ are constants of mass dimension $-\sum_i\alpha_i=-\sum_i s_i$.
At tree level, only natural numbers can arise for the coefficients $\alpha_i$ and $\beta_i$ since the vertices must be equivalent to expressions obtained by contracting the Feynman rules with polarization vectors and spinors.
Since it is not possible to fix the exponents $\beta_i$ from this symmetry argument, the requirement to reproduce the correct massless results must be imposed to uniquely determine the form of the vertices.
Turning to the solutions $F_1$ of the SWIs, for the vertices considered here with  $m_1=m_3\equiv
m$, $m_2=0$ the solution is proportional to $F_1\sim \bar\iota_1\sbraket{q1}-\bar\iota_3\sbraket{q3}$. Taking the spin constraints into account, the  general form is
\begin{equation}
\label{eq:f1-spin}
F_1=h_{123} \braket{12}^{\alpha_3+\beta_3+\frac{1}{2}}
\braket{23}^{\alpha_1+\beta_1+\frac{1}{2}}
\braket{31}^{\alpha_2-(\beta_1+\beta_3)-\frac{1}{2}}
\frac{\braket{q1}^{\beta_1}\braket{q3}^{\beta_3}}{\braket{q2}^{\beta_1+\beta_3}}
\left(\frac{\bar\iota_1}{\braket{12}}-\frac{\bar\iota_3}{\braket{23}}\right),
\end{equation}
where  the $h_{ijk}$ are constants of mass dimension $\frac{1}{2}-\sum_i s_i$. Here the conjugate equation of~\eqref{eq:3-pt-mom-con} has been used to write the result entirely in terms of angular brackets.

The above constraints can now be applied  to the vertex
$A_3(\bar\Phi_1,G_2^-,\Phi_3)$ where $s_1=s_3=0$ and $s_2=-1/2$. It is seen that a vertex of the form~\eqref{eq:f3-spin} requires half-integer
exponents of the spinor brackets that have been excluded above while 
the solution~\eqref{eq:f3-spin} leads to integer exponents.
It can be checked that the solution with $\beta_i=0$, 
\begin{equation}\label{eq:mhv-fund}
\begin{aligned}
     A_3(\bar \Phi_1, G_2^-,\Phi_3)&=\delta^2(\mathcal{Q}_\alpha)
  \frac{\bar\iota_1\braket{23}-\bar\iota_3\braket{12}}{\braket{31}} \\
&=\frac{1}{2}\sum_{i,j}\braket{i j}\eta_i\eta_j 
\frac{\bar\iota_1\braket{23}-\bar\iota_3\braket{12}}{\braket{31}}
  +m\sum_i  \frac{\braket{i q}\braket{2q}}{\braket{1q}\braket{3q}}\,
\eta_i\bar\iota_1 \bar\iota_3
  \end{aligned}
\end{equation}
leads to the correct massless limit for the scalar-gluon vertex:
\begin{equation}
 A_3(\bar \phi_1^-, g_2^-,\phi^+_3)= 
-A_3(\bar \Phi_1, G_2^-,\Phi_3)|_{\eta_2\eta_3\bar\iota_3}
=\frac{\braket{12}\braket{23}}{\braket{31}}.
\end{equation}
In the second line of~\eqref{eq:mhv-fund}, momentum conservation has
been used to simplify the $\bar\iota$-dependent terms. This vertex
reproduces the three-point amplitudes for the component fields~\cite{Schwinn:2006ca} collected in appendix~\ref{app:sqcd}.

The interaction with the positive-helicity gluon multiplet can be
either obtained from symmetry arguments as above or by a
Grassmann-Fourier transform from the
$\bar\eta\iota$ representation. It is instructive to follow the second
approach here.  The three-point amplitude with the positive helicity
gluon multiplet $\bar\eta\iota$-representation is obtained by taking
the complex conjugate of the result~\eqref{eq:mhv-fund}:
\begin{equation}
  A_3(\bar \Phi_1,
  G_2^+,\Phi_3)|_{\bar\eta\iota}=\delta^2(\bar{\mathcal{Q}}^{\dot\alpha})
  \frac{\iota_1\sbraket{23}-\iota_3\sbraket{12}}{\sbraket{31}},
\end{equation}
where the positive helicity gluon superfield in the $\bar\eta$ representation is
\begin{equation}
  G^+(\bar\eta)=\Lambda^++\bar\eta g^+.
\end{equation}
The vertex in the $\eta\bar\iota$-representation is obtained by a Fourier transformation
\begin{equation}
\label{eq:sv-fourier}
  \begin{aligned}
    A_3(\bar \Phi_1, G_2^+,\Phi_3)|_{\eta\bar\iota}
&=\int \prod_i d\bar\eta_i\; d\iota_i e^{\eta_i\bar\eta_i} e^{-\iota_i\bar\iota_i}
A_3(\bar \Phi_1, G_2^+,\Phi_3)|_{\bar\eta\iota}\\
&=\epsilon^{ijk}\left[\frac{1}{2}\sbraket{i j}\eta_k
\frac{\bar\iota_3\sbraket{23}-\bar\iota_1\sbraket{12}}{\sbraket{31}}
  +m  \frac{\sbraket{i q}\sbraket{2q}}{\sbraket{1q}\sbraket{3q}}\,
\eta_j\eta_k\right].
  \end{aligned}
\end{equation}
It can be checked from this form easily that the component  vertices obtained from this vertex reproduce the component vertices in appendix~\ref{app:sqcd}.

According to the general discussion, it should be possible to write
the last line of equation \eqref{eq:sv-fourier} in the general
form~\eqref{eq:v3-fund} where the SUSY-invariance is manifest. Indeed,
using~\eqref{eq:swap-square} to extract a common prefactor between the
`$\eta^2$' and `$\eta \bar{\iota}$' type terms the vertex can be
written as
\begin{equation}\label{eq:barmhv-fund}
A_3(\bar \Phi_1, G_2^+,\Phi_3)|_{\eta\bar\iota} = 
\delta^2(\mathcal{Q}_{\alpha}) m \frac{\braket{31}}{\braket{12} \braket{23}}.
\end{equation}
This result is in agreement with~\eqref{eq:f3-spin} for $s_1=s_3=0$
and $s_2=1$.  Equations \eqref{eq:mhv-fund} and \eqref{eq:barmhv-fund}
contain the two supersymmetric three point amplitudes in massive SQCD,
in addition to the two known three point amplitudes which involve only
gluons.

Note that in the formulation~\eqref{eq:barmhv-fund} the non-vanishing
of the $\overline{\text{MHV}}$-type vertices 
\begin{equation}
A_3(\bar Q_1^+, g_2^+,Q^-_3) \textrm{  and  } A_3(\bar \phi_1^+, g_2^+,\phi^-_3)
\end{equation} 
in the massless limit is not manifest, but holds due
to~\eqref{eq:swap-square}. In contrast, the form~\eqref{eq:sv-fourier}
has a massless limit in agreement with the form of the
$\overline{\text{MHV}}$-three point supervertices in massless
maximally supersymmetric theories \cite{Brandhuber:2008pf,
  ArkaniHamed:2008gz} where the degenerate on-shell three-point
kinematics only allows to extract a reduced SUSY-conserving delta
function
$\delta(\eta_1\sbraket{23}+\eta_2\sbraket{31}+\eta_3\sbraket{12})$.

\subsubsection{All-multiplicity amplitudes with positive helicity gluons}

Using the result that the amplitudes in the $\eta\bar\iota$
representation are proportional to the Grassmann delta
function~\eqref{eq:suppoincdelt} it is easy to write down the
supersymmetric version of the massive scalar amplitude coupled to gluons
given in equation \eqref{eq:rodrigo}:
\begin{equation}
\label{eq:sqcd-all-plus}
A(\bar \Phi_1, G_2^+,\ldots G_{n-1}^+, \Phi_n)
 = \delta^{2 } \left( \mathcal{Q}_{\alpha} \right)  
\frac{A(\bar \phi_1^+,g_2^+,\dots,\phi_n^-)}{m}.
\end{equation}
Integrating out $\eta_1$ and $\bar \iota_1$ gives back the original
amplitude. Integrating out $\bar\iota_1$ and $\eta_n$ gives the
amplitude with massive quarks:
\begin{equation}\label{eq:simplesqcdwasid}
A(\bar Q_1^{+}, g_2^+,\ldots g_{n-1}^+ ,Q_n^{-}) =
 \frac{\braket{ n q}}{\braket{1 q}} A(\bar \phi_1^+,g_2^+,\dots,\phi_n^-)
\end{equation}
This reproduces the result in \cite{Schwinn:2006ca} obtained from a
more conventional use of the Ward identities. Integrating out two
$\eta$ variables or $\bar\iota_1$ and $\eta_i$ yields the remaining
non-vanishing component amplitudes:
\begin{equation}
  \begin{aligned}
A(\bar Q_1^{-}, g_2^+,\ldots g_{n-1}^+ ,Q_n^{-}) &=
 \frac{\braket{ 1n}}{m} A(\bar \phi_1^+,g_2^+,\dots,\phi_n^-),\\
A(\bar Q_1^{-}, g_2^+,\ldots\Lambda_i^+, g_{i+1}^+,\ldots ,\phi_n^{-}) &=
 \frac{\braket{ 1i}}{m} A(\bar \phi_1^+,g_2^+,\dots,\phi_n^-),\\
A(\bar Q_1^{+}, g_2^+,\ldots\Lambda_i^+, g_{i+1}^+,\ldots ,\phi_n^{-}) &=
 \frac{\braket{iq }}{\braket{1q}} A(\bar \phi_1^+,g_2^+,\dots,\phi_n^-).
  \end{aligned}
\end{equation}

\subsection{Three point vertices with massive and massless vector bosons}
The symmetry analysis in section~\ref{sec:sqcd} can also be applied to
vertices with two massive vector bosons and a massless one. This
provides a template for the $W^+W^-\gamma$ vertex in (a
supersymmetrized extension of) the standard model. Here, however, the concrete model will be left unspecified. The ingredients are two superfields containing the positive and negative helicity ``photons''
\begin{equation}
  \mathcal{A}^+(\eta)=A^++\eta\Lambda^+\;,\quad  
  \mathcal{A}^-(\eta)=\Lambda^-+\eta A^-
\end{equation}
and two massive superfields for charged vector fields $W$, two Dirac
fermions $\chi$ and $\tilde\chi$ and scalars $H$, as contained in the supermultiplets~\eqref{eq:massive-vec+} and~\eqref{eq:massive-vec-}:
\begin{equation}
  \begin{aligned}
    \mathcal{W}^+(\eta,\bar\iota)&=-\chi^+
  +\eta \frac{1}{\sqrt 2}(W^0+H)- \bar\iota\; W^++\eta\bar\iota\;
 \tilde \chi^{+},\\
   \mathcal{W}^-(\eta,\bar\iota)&=-\chi^-
  +\eta W^- - \bar\iota\; \frac{1}{\sqrt 2}(W^0-H)+\eta\bar\iota\;
 \tilde \chi^{-}.
  \end{aligned}
\end{equation}
There are analogous superfields in the conjugate representation of the unbroken gauge group denoted by a bar. Note that the scalars are charged under the unbroken gauge group with massless gauge bosons $A$ and should not be confused with the Higgs bosons responsible for symmetry breaking. The corresponding Higgs-superfields  would have to be added in a complete model.

Since the two superfields $\mathcal{W}^\pm$ are needed to describe all
spin states of the massive vector bosons, there are two MHV-type
vertices with the field content $ \overline{\mathcal{W}}{}_1^+
\mathcal{A}_2^-\mathcal{W}_3^-$ (and the analog with the $\mathcal{W}$
plus-minus labels exchanged) and $ \overline{\mathcal{W}}{}_1^-
\mathcal{A}_2^+\mathcal{W}_3^- $.  For the first vertex, the
superspin constraint~\eqref{eq:susy-spin} holds with $s_{1/3}=\pm
\frac{1}{2}$ for the $\mathcal{W}^\pm$-legs, and $s_2=-1/2$ for the
$\mathcal{A}^-$ leg. As in the case of the matter supermultiplet there
is no solution involving a Grassmann-parameter independent function
$F_3$ with integer coefficients $\alpha_i$ and one is led to an
expression similar to~\eqref{eq:mhv-fund}:
\begin{equation}
  A_3(\overline{\mathcal{W}}_1^+, \mathcal{A}_2^-,\mathcal{W}_3^-)
=  \delta^2(\mathcal{Q}_\alpha)
\left(\bar\iota_1 \braket{23}
  -\bar\iota_3 \braket{12} \right)
 \frac{\braket{23}}{\braket{12}\braket{31}}
\end{equation}
For the second MHV-type vertex with $s_1=s_3=-\frac{1}{2}$, $s_2=1$ a
Grassmann-parameter independent solution is available
\begin{equation}
\label{eq:waw}
  A_3(\overline{\mathcal{W}}_1{}^-, \mathcal{A}_2^+,\mathcal{W}_3^-)=
  \delta^2(\mathcal{Q}_\alpha)
\frac{\braket{13}^2}{\braket{12}\braket{23}}.
\end{equation} 
It can be seen that these two vertices include the correct MHV-type
triple gauge boson vertices that are identical to the massless case.
It is also seen that the $(\eta_1\bar\iota_1)$ and
$(\eta_3\bar\iota_3)$ coefficients of~\eqref{eq:waw} yield the correct
helicity-flip vertices for massive fermions
\begin{equation}
  A_3(\overline\chi{}^-_1,A_2^+,\tilde\chi^-_3)= 
  A_3(\overline{\tilde\chi}{}^-_1,g_2^+,\chi^-_3)=-m 
  \frac{\braket{13}^2}{\braket{12}\braket{23}}
\end{equation}
and a vertex
involving a longitudinal gauge boson
\begin{equation}
  A_3(\overline {W}{}_1^-,A_2^+, W_3^0)=2
  m \frac{\braket{1q}}{\braket{3q}} \frac{\braket{13}^2}{\braket{12}\braket{23}}
\end{equation}
that agrees with an explicit computation.

The conjugate supervertices can either be obtained by a Fourier
transformation from the $\bar\eta\iota$ representation or from
symmetry considerations and matching to the massless limit. The results are
\begin{align}
  A_3(\overline{\mathcal{W}}_1^+, \mathcal{A}_2^+,\mathcal{W}_3^-)
&= \delta^2(\mathcal{Q}_{\alpha}) m\frac{\braket{q3}}{\braket{q1}}
 \frac{\braket{31}}{\braket{12} \braket{23}},\\
 A_3(\overline{\mathcal{W}}_1^+, \mathcal{A}_2^-,\mathcal{W}_3^+)&=
 \delta^2(\mathcal{Q}_{\alpha}) m\frac{\braket{q2}^2}{\braket{q1}\braket{q3}}
 \left(\frac{\bar\iota_1}{\braket{12}}-\frac{\bar\iota_3}{\braket{23}}\right).
\end{align}

\subsection{Abelian Higgs model}\label{sec:AbelianHiggs}
As the simplest toy model for a theory with massive vector bosons, consider the
supersymmetric Abelian Higgs model of a vector multiplet
$(A_\mu,\lambda,D)$ and a chiral multiplet $(\phi,\psi,F)$ with a
Fayet-Illopolus term. This model was first constructed
in~\cite{Fayet:1975yh}. The Lagrangian of the model is discussed in
appendix~\ref{app:AbelianHiggs}. In the broken phase, the physical
spectrum of the model contains a Dirac fermion
$\Psi\equiv(\psi_-,\psi_+)^T=(\psi,-i\lambda^\dagger)^T$, a vector
boson $A$ and a scalar $H$ with a common mass $m$. In this minimal form the model is not anomaly free, which does, however, not affect the tree amplitudes discussed in the following.
 
In order to match the physical spectrum of the model to the
massive vector multiplet described in section~\ref{sec:susy-vector},
the antifermion degrees of freedom have to be treated as independent
from the fermion states.  From~\eqref{eq:mass-multiplet} it is
seen that the spin one-half state $\bra{\Omega^0,\tfrac{1}{2}}$ is
the superpartner of the maximal spin-state $\bra{\Omega^{+},1}$ in the
massless limit. Therefore it has to be identified with the
right-handed anti-particle state created by the field
$\psi_+=-i\lambda^\dagger$:
$\bra{\psi^+}=\bra{\Omega_0,\tfrac{1}{2}}$. Similarly the left-handed
particle state created by $\bar\psi_-=i\lambda$ is identified with the
superpartner of the minimal spin state,
i.e. $\bra{\bar\psi^-}=\bra{\Omega_0',-\tfrac{1}{2}}$.  In superfield
notation, the physical states of the model are contained in the two
supermultiplets in the $\eta\bar\iota$-representation:
\begin{equation}
  \begin{aligned}
    \mathcal{A}^+(\eta,\bar\iota)&=-\bar\psi^+
  +\eta \frac{1}{\sqrt 2}(A^0+H)- \bar\iota\; A^++\eta\bar\iota\,\psi^+,
 \\
   \mathcal{A}^-(\eta,\bar\iota)&=-\bar\psi^-
  +\eta A^- - \bar\iota\; \frac{1}{\sqrt 2}(A^0-H)+\eta\bar\iota\;
  \psi^{-}.
  \end{aligned}
\end{equation}

\subsubsection{Three-point superamplitudes}
The three point amplitudes of the model are contained in two
supervertices with the field content
$\mathcal{A}_1^-\mathcal{A}_2^+\mathcal{A}_3^-$ and
$\mathcal{A}_1^+\mathcal{A}_2^-\mathcal{A}_3^+$.  For the first
vertex, the superspin constraint~\eqref{eq:susy-spin} applies with
$s_{1/3}=-1/2$ and $s_2=1/2$.  Demanding integer exponents of the
spinor brackets leads to an expression similar to~\eqref{eq:mhv-fund}:
\begin{equation}
   A_3(\mathcal{A}_1^-, \mathcal{A}_2^+,\mathcal{A}_3^-)
=  \delta^2(\mathcal{Q}_\alpha)
\left(\bar\iota_1 \braket{23}
  -\bar\iota_3 \braket{12} \right)
 \frac{\braket{13}}{\braket{12}\braket{23}}.
\end{equation}
The precise form of the solution to the $\bar{\mathcal{Q}}$-SUSY
invariance condition has been fixed by the requirement of the absence
of a triple gauge boson interaction in this Abelian model, which
implies that the vertex must not depend on $\bar\iota_2$. This results also in the absence of
vertices involving $\psi^+$ and two fields with a minus label, which
agrees with the explicit vertices obtained in
appendix~\ref{app:AbelianHiggs}.

The second vertex takes its natural form in the $\bar\eta\iota$
representation where it is given by
\begin{equation}
   A_3(\mathcal{A}_1^+, \mathcal{A}_2^-,\mathcal{A}_3^+)|_{\bar\eta\iota}
=  \delta^2(\bar{\mathcal{Q}}_\alpha)
\left(\iota_1 \sbraket{23}
  -\iota_3 \sbraket{12} \right)
 \frac{\sbraket{13}}{\sbraket{12}\sbraket{23}}
\end{equation}
Fourier-transforming to the $\eta\bar\iota$-representation results in
\begin{equation}
\begin{aligned}
   A_3(\mathcal{A}_1^+, \mathcal{A}_2^-,\mathcal{A}_3^+)|_{\eta\bar\iota}
&= \frac{\sbraket{31}}{\sbraket{12}\sbraket{23}}
\epsilon^{ijk}\left[\frac{1}{2}\sbraket{i j}\eta_k
\left(\bar\iota_1\sbraket{12}
+\bar\iota_3\sbraket{23} \right)\bar\iota_2\right.\\ 
 & \qquad  \left. -m   \frac{\sbraket{i q}}{\sbraket{2q}}\,
\eta_j\eta_k\left[\sbraket{31}
\frac{\sbraket{2q}^2}{\sbraket{1q}\sbraket{3q}}\bar\iota_2+\sbraket{23}\bar\iota_3+\sbraket{12}\bar\iota_1\right]\right]\\
&= \delta^2(\mathcal{Q}_{\alpha}) \left(
\frac{\eta_1\braket{12} - \eta_3 \braket{23}}{\braket{31}}
+\frac{\sbraket{31}}{m} \bar{\iota}_2 \right).
\end{aligned}
\end{equation}
As in the case of~\eqref{eq:barmhv-fund}, in the last line the
$\mathcal{Q}$-supersymmetry has been made manifest. The
$\bar{\mathcal{Q}}$-supersymmetry can be checked to hold as a result
of the Schouten identity and~\eqref{eq:3-pt-mom-con}.  Note
that~\eqref{eq:swap-square} does not hold for three massive legs, so
it is not possible to argue that the three point vertices can be
expressed entirely in terms of angular brackets. Instead, the on-shell
conditions and momentum conservation imply the identity
\begin{equation}
\begin{aligned}
0=(p_1+p_3)^2-m^2&=\braket{13}\sbraket{31}+m^2\left[1+
\frac{\braket{q3}\sbraket{q3}}{\braket{q1}\sbraket{q1}}
+\frac{\braket{q1}\sbraket{q1}}{\braket{q3}\sbraket{q3}}\right]\\
\Rightarrow\quad
\braket{13}\sbraket{31}-m^2&=m^2
\frac{\sbraket{q2}^2}{\sbraket{q3}\sbraket{q1}}
\frac{\sbraket{31}^2}{\sbraket{12}\sbraket{23}}
\end{aligned}
\end{equation}
that can be used to verify the equivalence of the two expressions for the vertex.

Note that both three point vertices are antisymmetric under interchange of the two legs with equal quantum numbers. This reflects the simple fact that the lowest components of the $\mathcal{A}$-superfields are fermions. This should be contrasted with the usual massless three point vertices in Yang-Mills: due to the structure constant the color ordered three point amplitudes are anti-symmetric under interchange of bosonic legs with equal quantum numbers.

\subsubsection{Results for scattering amplitudes}
As a result of the general discussion of section~\ref{sec:vanish} the
amplitudes with only vector bosons in the maximal or minimal spin
state vanish, as in the massless case. For amplitudes with one vector
boson in the opposite spin state, the general discussion allows to
conclude that they vanish only for a special choice of the spin axes,
c.f.~\eqref{eq:vanish++-}. In the Abelian Higgs model, however, it is
possible to show diagrammatically that these amplitudes vanish even
for arbitrary choices of the spin axes. This is discussed in
appendix~\ref{app:AbelianHiggs}. Therefore the simplest non-vanishing ``maximally spin violating'' amplitudes are, in analogy to massless non-Abelian gauge theories, those with two opposite spin labels. Concretely, picking particles one and two to be polarized opposite to the rest for instance the following $n$-point amplitude is obtained from the Feynman graphs,
\begin{equation}
\label{eq:msv}
A_{\textrm{MSV}/\overline{\textrm{MSV}}} =  i(2ig)^{n-2} m^2\!\!\!\!\!\!\!
\sum_{\textrm{perm}(3,\ldots, n)}\!\!\!
 \left(  \left(\prod_{i=2}^{n/2-1} \frac{ (\epsilon^{\pm}_{2i}\cdot k_{1,2i-1}) 
      (\epsilon^{\pm}_{2i+1}\cdot k_{1,2i})}{
      (k_{1,2i-1}^2 - m^2)(k_{1,2i}^2 - m^2)}  \right)
  \frac{ \left(\epsilon^{\mp}_2 \cdot\epsilon^{\pm}_3\right)
    \left(\epsilon^{\pm}_n\cdot \epsilon^{\mp}_1\right)}{
    (k_n+k_1)^2 -m^2 } \right) .
\end{equation}
Details can be found in the appendix. This amplitude should be generated by a solution to the SWIs which on the superspace in $\eta \iota$ representation should have weight $4$. The above amplitude for instance is the coefficient of $\eta_1 \iota_1 \eta_2 \iota_2$. We have been unable however to find the general superamplitude on this space which reproduces the above amplitude. 

The amplitudes~\eqref{eq:msv} are related to amplitudes with fermions by the SWI
\begin{multline}
\label{eq:swi--+}
 \braket{qk_n^\flat}\,A_n(A_1^-,A_2^+,\dots,A_i^- ,\dots,A_n^+)
=\braket{q k_1^\flat}\,
A_n(\bar\Psi_1^-,A_2^+,\dots A_i^-,\dots,\Psi_n^+)\\
+\braket{q k_i^\flat}\,A_n(A_1^-,A_2^+,\dots,\bar\Psi_i^- ,\dots,\Psi_n^+)\,.
\end{multline}
Amplitudes with two outgoing antifermions $\Psi^+$ generated by the SWI obviously vanish due to fermion number conservation and have been dropped. 
Using the results from appendix~\ref{app:AbelianHiggs} this identity can be checked at the four-point level. The four-point vector boson amplitude is found to be
\begin{equation}
\label{eq:4ptMSV}
    \mathcal{A}_4(A^-_{k_1}, A^+_{k_2}, A^-_{k_3}, A^+_{k_4}) =  i^3(2gm)^2 
    \frac{\braket{q1}\sbraket{2q}\braket{q3}\sbraket{4q}}{
        \braket{q2}\sbraket{1q} \braket{q4}\sbraket{3q}}
      \left(\frac{1}{(k_{1,2}^2-m^2)}+\frac{1}{(k_{2,3}^2-m^2)}
      \right)\,
\end{equation}
where $k_{i,j}=k_i+\cdots k_j$. The result for the fermion amplitude is
\begin{equation}
\label{eq:4ptMSVfermi}
   \mathcal{A}_4(\bar\Psi^-_{k_1}, A^+_{k_2}, A^-_{k_3}, \Psi^+_{k_4})
   =\frac{i^3(2gm)^2\sbraket{q2}\braket{3q}}{\sbraket{q1}
\braket{q2}\sbraket{3q}\braket{4q} }\left[
\frac{\braket{q-|\fmslash k_{3,4}|q-}}{
(k_{1,2}^2-m^2)}+
\frac{\braket{q+|\fmslash k_4-\fmslash k_1|q+} }{(k_{2,3}^2-m^2)}
\right].
\end{equation}
Adding the contribution from the second fermionic amplitude that is
obtained by exchanging $k_1\leftrightarrow k_3$, the
SWI~\eqref{eq:swi--+} is manifestly satisfied.


\subsection{Massive $\mathcal{N}=2$ superfields on the Coulomb branch of $\mathcal{N}=4$}\label{subsec:massivecoulombbranch}
The minimal massive multiplet in four dimensions with $\mathcal{N}=2$ supersymmetry is the massive vector multiplet. Hence any theory which contains this type of multiplet must be a spontaneously broken gauge theory. These fall in two classes: those which became massive by giving a vacuum expectation value to a scalar in a massless gauge multiplet (Coulomb branch), or by giving a vev to a scalar in a matter multiplet (Higgs branch). In this subsection we briefly consider the former. Typically, these theories arise as a generalized dimensional reduction of a six dimensional theory (see \cite{Boels:2010mj} for an explicit description of this map). However, the multiplets which arise this way are BPS multiplets. As described in section \ref{subsec:explmassvsBPS}, BPS representations of $\mathcal{N}=4$ can be treated as massive $\mathcal{N}=2$ multiplets. In the following only this $\mathcal{N}=2$ will be made explicit. 

In an $\mathcal{N}=2$ framework one can introduce an on-shell massive vector boson superfield $V$
\begin{align}\label{eq:susyfieldcontentN2}
V = \phi +   \psi_I  \, \eta^{I}+  \psi^I \, \bar{\iota}_{I} + & \nonumber \\ 
 V^-  \,  \eta^1 \eta^2 + & V^+  \,  \bar{\iota}^1 \bar{\iota}^2 +   \phi_I^{\,J}   \,  \bar{\iota}_{J} \eta^I + \\ 
 & \bar{\psi}^{I}  \,  \bar{\iota}_{I}  \eta^1 \eta^2 + \bar{\psi}_{K}  \,   \eta^K  \bar{\iota}^1 \bar{\iota}^2 +  \bar{\phi} \,   \bar{\iota}^1\bar\iota^2 \eta^1 \eta^2\nonumber 
\end{align}
and  the $\mathcal{N}=2$ positive helicity gluonic superfield
\begin{equation}
G^+ = g^+ + \nu^+_I  \,  \eta^I + s  \,  \eta^1 \eta^2
\end{equation}
as well as a negative helicity superfield,
\begin{equation}
G^- = \bar{s}  + \bar{\nu}^+_I   \, \eta^I + g^-  \, \eta^1 \eta^2.
\end{equation}
All capital roman letters run from $1$ to $2$. These superfields are in a different phase convention than the previous ones. All fields are in the $\eta \bar{\iota}$ representation. On this superspace the minimal solution to the SUSY Ward identities is the supermomentum conserving delta function and hence
\begin{equation}
A \sim \delta^{4} \left(\mathcal{Q}^I_{\alpha} \right) 
\end{equation}
for every amplitude in this representation. 

\subsubsection{Three-point superamplitudes}
In addition to the known three point amplitudes for the Yang-Mills fields there are three superamplitudes which involve the massive vectors. Note that two massive vectors are needed by six dimensional momentum conservation. The simplest of these is simply proportional to the delta function,
\begin{equation}
A(V, G^+, V) = \frac{\braket{31}}{\braket{12}\braket{23}}
\delta^{4} \left(\mathcal{Q}^I_{\alpha} \right), 
\end{equation}
which can be easily checked by integrating out both $\eta \bar{\iota}$ on the last leg for instance and using the identity \eqref{eq:swap-square}. The conjugate to this amplitude reads
\begin{equation}
A(V, G^-, V) = \frac{\sbraket{31}}{\sbraket{12}\sbraket{23}} 
\delta^{4} \left(\bar{\mathcal{Q}}^I_{\dalpha} \right) 
\end{equation}
on the conjugate superspace. The triple massive vector boson vertex (which has weight $6$ and is self-dual under fermionic Fourier transform) will not be needed here as it only arises for symmetry breaking patterns which involve more than $2$ unbroken subgroups. Note though that with one supersymmetry in the $\eta \bar{\iota}$ and the other in the $\bar{\eta} \iota$ representation
\begin{equation}\label{eq:guess3massvecs}
A(V,V,V) \stackrel{?}{\sim} \sum_{i=1}^3 \left(\eta_i^1 \bar{\eta}^2_i 
+ \bar{\iota}^1_i \iota^2_i\right) \delta^{2} \left(\mathcal{Q}^1_{\alpha} \right) \delta^{2} \left(\bar{\mathcal{Q}}^2_{\dalpha} \right) 
\end{equation}
is a solution to the Ward identities. Integrating out all fermionic variables on the first leg and $\eta^1 \iota^2$ on the second for instance from this guess yields after a short calculation
\begin{equation} 
A(\phi^{1}_1 V^- \phi^{2}_2) \sim \frac{m_2 \sbraket{1 q} \braket{1 2}}{\sbraket{q 2}}.
\end{equation}
Since one expects this to be the usual amplitude of a complex scalar coupled to a gluon, one has to divide the proposed three point function in equation \eqref{eq:guess3massvecs} by $m_2$. The field identification follows from a double fermionic Fourier transform on the second supersymmetry of the muliplet in equation \eqref{eq:susyfieldcontentN2}. The explicit appearance of the mass $m_2$ however makes it difficult to turn this three point amplitude into a generic amplitude on the Coulomb branch of $\mathcal{N}=4$ as only for equal masses one can make a symmetric amplitude out of equation \eqref{eq:guess3massvecs}. This does however make it a prime candidate for the $\mathcal{N}=2$ three point coupling in the Abelian Higgs model. Note that as a special feature \eqref{eq:guess3massvecs} is a solution to the SUSY Ward identities for arbitrary spin axes on the three different legs.

\subsubsection{All multiplicity  superamplitudes}
In the following we will study the symmetry breaking pattern $U(N) \rightarrow U(N_1) \otimes U(N_2)$. The naturally massive vector bosons are charged in the bi-fundamental while the gluons are taken to transform in one of the two $U(N_i)$ factors. The simplest amplitude follows by almost trivial extension of the above results in SQCD to $\mathcal{N}=2$ supersymmetry,
\begin{equation}\label{eq:massvecbos}
A(\bar V_1, G_2^+,\ldots G_{n-1}^+, V_n)
= \delta^{4 } \left( \mathcal{Q}^I_{\alpha} \right)  
\frac{A(\bar \phi_1^+,g_2^+,\dots,\phi_n^-)}{m^2}.
\end{equation}
Note that the fields in the massive multiplets remain charged in the fundamental and anti-fundamental of the $U(N_1)$ gauge group, while the multiplet now incorporates a massive vector boson. An interesting component amplitude follows by integrating out the $\bar\iota^1_1, \bar\iota^2_1$ and $\eta_n^1 \eta_n^2$ coordinates:
\begin{equation}
A(\bar V_1^{+}, g_2^+,\ldots g_{n-1}^+ ,V_n^{-}) =
\left( \frac{\braket{ n q}}{\braket{1 q}} \right)^2 A(\bar \phi_1^+,g_2^+,\dots,\phi_n^-)
\end{equation}
which is the natural extension of equation \eqref{eq:simplesqcdwasid}. It contains two massive vector boson components, one with positive and one with negative spin w.r.t the common spin-axis defined by $q$.  

To verify that \eqref{eq:massvecbos} is correct, one can consider one component amplitude which involves the massive scalar pair. By dimensional reduction from six dimensions it is seen that the part of the action of the symmetry broken field theory relevant for this amplitude agrees with the action used to compute the all-multiplicity amplitude of massive scalars in QCD~\eqref{eq:rodrigo}. Since equation \eqref{eq:massvecbos} is invariant under on-shell $\mathcal{N}=2$ supersymmetry and reproduces one component amplitude, it is indeed the correct amplitude on the Coulomb branch. Note that the other superamplitudes where some or all of the massless matter is charged under the other gauge group are related to the above one by the usual Kleiss-Kuijf relations \cite{Kleiss:1988ne}, applied in the six dimensional parent theory.  For instance,
\begin{multline}
A(G^+_\alpha, \bar \Phi_1, G_2^+,\ldots G_{n-1}^+, \Phi_n) = A( \bar \Phi_1, G^+_\alpha, G_2^+,\ldots G_{n-1}^+, \Phi_n) +\\  A( \bar \Phi_1, G_2^+, G^+_\alpha, \ldots G_{n-1}^+, \Phi_n)  + \ldots  + A( \bar \Phi_1, G_2^+, \ldots G_{n-1}^+, G^+_\alpha,  \Phi_n) 
\end{multline}
follows. Note that the supermomentum conserving delta function always factors out of these sums as it is completely symmetric.

\subsection{Effective Higgs-gluon couplings}
In the standard model one can consider effective Higgs-gluon couplings obtained by integrating out a top quark loop in the limit of large top mass. In this limit the effective interaction is given by the following local mass dimension 5 operators added to the Yang-Mills action~\cite{Wilczek:1977zn, Shifman:1979eb},
\begin{align}\label{eq:QCDhiggscoupl}
S_{Hgg}&=  \frac{\alpha_s}{6 \pi m_r} \tr \int d^4x \left[ H F^2 + A F \tilde{F} \right]
\end{align}
which can be rewritten as
\begin{align}
S_{Hgg}& = \frac{\alpha_s}{6 \pi v} \tr \int d^4x \left[ (\phi + \bar{\phi}) F^2 + (\phi - \bar{\phi}) F \tilde{F} \right] \\ 
& =  \frac{\alpha_s}{6 \pi v}  \tr \int d^4x \left[ \phi F_{+}^2 +  \bar{\phi} F_-^2 \right].\label{eq:chiralQCDhiggscoupl}
\end{align}
Here $H = \phi + \bar{\phi}$ is the real Higgs field decomposed into two chiral scalar fields, $A = \phi - \bar{\phi}$ the would-be axion field, $F_+$ and $F_-$ denote the self-dual and anti-self-dual field strengths respectively and $v \sim 246 GeV$\cite{Dixon:2004za}. Of course, the Higgs and axion fields are uncharged under the strong gauge group. As shown in \cite{Dixon:2004za}, the scattering amplitudes of either $\phi$ or $\bar{\phi}$ fields display an 'almost' MHV or $\overline{\textrm{MHV}}$ form, e.g.
\begin{equation}\label{eq:comphigsslgueMHV}
A(\phi, g_2^+ \ldots g^-_i \ldots g^-_j \ldots g_n^+)
 = \frac{\alpha_s^{n-2}}{6 \pi v} \frac{\braket{ij}^4}{\braket{2 3} \ldots \braket{n 2}}
\end{equation}
for an amplitude with $n-1$ gluons. In addition to the MHV-type amplitude, the model under study has one more simple amplitude: the amplitude with all negative helicity gluons,
\begin{equation}\label{eq:comphigsslgueNMAXMHV}
A(\phi,g^-_2 \ldots g^-_n) = \frac{\alpha_s^{n-2}}{6 \pi v} \frac{m_{\phi}^4}{\sbraket{2 3} \ldots \sbraket{n 2}}.
\end{equation}
The difference to the actual MHV amplitude is in the momentum conserving delta function which now includes the momentum of the chiral Higgs field $\phi$. For the field $\bar{\phi}$ there is a corresponding natural anti-MHV type amplitude. As shown in \cite{Boels:2008ef} there is actually a tower of these amplitudes with an arbitrary number of insertions of either $\phi$ or $\bar{\phi}$ fields (but none of mixed type), as well as a natural generalization to higher dimension operators. Here we will focus on the example of a single $\phi$ or $\bar{\phi}$ field and the operator in equation \eqref{eq:QCDhiggscoupl} for clarity, the generalization should be obvious. 

The above model can be embedded into a supersymmetric field theory \cite{Dixon:2004za}. This will generically turn the Higgs field into a massive multiplet of the $\mathcal{N}=1$ algebra 
\begin{equation}
\Phi(\eta,\bar\iota)=-\phi + \eta\,\psi^+ -\bar\iota\,\psi^-
+\eta\bar\iota\, \bar\phi   
\end{equation}
 Hence the field content of this model is very similar to that studied for supersymmetric QCD but the action and symmetry properties are different. The Ward identities of the massless Higgs case were discussed in \cite{Dixon:2004za}. Using the technology developed above it is easy to generalize to massive Higgs particles. 
In particular,  a SUSY argument for the vanishing of amplitudes of the form $A(\phi, g_2^+ \ldots g^-_i  \ldots g_n^+)$ for a massive Higgs can be constructed along the lines of  section~\ref{sec:vanish}.

\subsubsection{Three point amplitudes}
There are two non-vanishing three point amplitudes which involve the massive Higgs field. For instance in the $\eta \bar{\iota}$ representation we have
\begin{equation}
A(\Phi_1, G_2^-, G_3^-) =  \frac{\alpha_s}{6 \pi v} \braket{2 3}^2 \delta^2(\mathcal{Q}_\alpha) \delta^4(k_1+k_2+k_3)
\end{equation}
for a $\phi F_+^2$ type coupling as well as 
\begin{equation}
A(\overline\Phi_1, G_2^+, G_3^+) = \frac{\alpha_s}{6 \pi v} \sbraket{2 3}^2 \delta^2(\bar{\mathcal{Q}}_{\dot\alpha}) \delta^4(k_1 + k_2 + k_3)
\end{equation}
in the $\bar{\eta}\iota$ representation for a $\bar{\phi} F_-^2$ type coupling.  
Here the scalar superfield in the  $\bar{\eta}\iota$ representation is denoted by $\overline\Phi(\bar\eta,\iota)=\bar\phi+\bar\eta \,\psi^++\iota\,\psi^- -\bar\eta\iota\,\phi$. It is straightforward to check that these solve the supersymmetric Ward identities. Note that $\mathcal{Q}$ now also includes the super-momentum of the chiral Higgs fields.

\subsubsection{All multiplicity superamplitudes of MHV type}
From the three point amplitudes it is easily guessed that the $n$-point amplitudes with MHV type configuration reads
\begin{equation}
A(\Phi, G^+ \ldots  G_i^- \ldots  G_j^- \ldots G^+) =  \frac{\alpha_s^{n-2}}{6 \pi v}\frac{\braket{ij}^3}{\braket{2 3} \ldots \braket{n 2}} \delta^2(\mathcal{Q}_\alpha).
\end{equation}
Integrating out $\eta_i$ and $\eta_j$ gives back the known component amplitude of equation \eqref{eq:comphigsslgueMHV}. Similarly, one can immediately write down the conjugate amplitude on the conjugate superspace
\begin{equation}
A(\overline\Phi, G^- \ldots  G_i^+ \ldots  G_j^+ \ldots G^-) =  \frac{\alpha_s^{n-2}}{6 \pi v} \frac{\sbraket{ij}^3}{\sbraket{2 3} \ldots \sbraket{n 2}} \delta^2(\bar{\mathcal{Q}}^{\dot\alpha}).
\end{equation}
These are both solutions to the supersymmetric Ward identities and reproduce one component amplitude. Hence they must be the correct superamplitudes. From these expression explicit forms of amplitudes which involve a Higgs-ino and a gaugino can be easily calculated.

\subsubsection{All multiplicity superamplitudes of $\textrm{N}^{\textrm{max}}$MHV type}
The other simple amplitude, i.e. the one from equation \eqref{eq:comphigsslgueNMAXMHV} can also easily be written as a superamplitude. Since the amplitude in equation \eqref{eq:comphigsslgueNMAXMHV} has an arbitrary number of minus helicity gluons it is natural to work on the conjugate superspace. The simplest guess reads
 \begin{equation}\label{eq:comphigsslguesuperNMAXMHV}
A(\overline\Phi, G^-, \ldots, G^-)|_{\bar\eta\iota} = \frac{\alpha_s^{n-2}}{6 \pi v} \frac{m_{\phi}^3}{\sbraket{2 3} \ldots \sbraket{n 2}} \delta^2(\bar{\mathcal{Q}}^{\dot\alpha}).
\end{equation}
The component amplitude in \eqref{eq:comphigsslgueNMAXMHV} follows by integrating out the fermionic $\iota_1$ and $\bar{\eta}_1$ coordinates. Of course, there is a conjugate superamplitude involving large numbers of $G^+$ fields in the conjugate representation as well, 
 \begin{equation}\label{eq:comphigsslguesuperNMAXMHVbar}
A(\Phi, G^+, \ldots, G^+) = \frac{\alpha_s^{n-2}}{6 \pi v} \frac{m_{\phi}^3}{\braket{2 3} \ldots \braket{n 2}} \delta^2(\mathcal{Q}_\alpha).
\end{equation}
These superamplitudes solve the supersymmetric Ward identities by construction. Note that by applying a fermionic Fourier transform to all legs the conjugate representation can be obtained: this will have a total fermionic weight $n$.  

\subsubsection{Extension to $\mathcal{N}=4$ and its string theory interpretation}
It is easy to extend the above amplitudes at least to theories which involve more supersymmetry, leading for instance to an $\mathcal{N}=2$ amplitude with a massive vector boson. Adding even more supersymmetry will lead to multiplets which involve massive states with higher spins. In the $\eta \bar{\iota}$ representation $\mathcal{N}=4$ supersymmetry leads to perhaps the simplest generalization of the MHV amplitude,
\begin{equation}\label{eq:n4stringamp}
A(\Phi, G, \ldots, G) =  \frac{\alpha_s^{n-2}}{6 \pi v} \frac{1}{\braket{2 3} \ldots \braket{n 2}} \delta^8(\mathcal{Q}^I_\alpha) \qquad\qquad I=1,\ldots 4
\end{equation} 
This displays explicit $\mathcal{N}=4$ supersymmetry. Integrating out the obvious fermionic coordinates yields back the amplitudes equations \eqref{eq:comphigsslgueMHV} and (the conjugate of) \eqref{eq:comphigsslgueNMAXMHV}.  A conjugate amplitude to this (in the $\bar{\eta} \iota$ representation) also exists. The $\mathcal{N}=4$ multiplet stretches all the way to massive spin $2$ fields which are physically problematic in field theory. 
On-shell methods could be, however,  useful here to evade complications of a Lagrangian description, as advocated in~\cite{Benincasa:2007xk}.
There might also be an application to correlation functions, see also the next subsection. 

In string theory a massive $\mathcal{N}=4$ scalar multiplet in four dimensions arises by restricting the momentum of the first Regge excitation to lie in a chosen four dimensions. One can choose all momenta of the fields in an amplitude in the superstring to have this type of four dimensional kinematics. The chiral couplings written in \eqref{eq:chiralQCDhiggscoupl} appear for instance in a certain rewriting \cite{Tseytlin:1999dj} of the Dirac-Born-Infeld action in four dimensions inspired by string field theory. Hence one can map the Higgs-gluon coupling amplitudes to amplitudes in the superstring. 

The three point version of the amplitude in equation \eqref{eq:n4stringamp} for instance can neatly be compared to the complete string amplitude in four dimensional kinematics written in equation (V.107) of \cite{Feng:2010yx} after application of the appropriate fermionic integrations. The dimensionful constant $v$ is exchanged for $ \sqrt{ \alpha'}$, while the mass scale is also set by this constant, $m^2 \sim \frac{1}{\alpha'}$. From one of the explicit four point results in \cite{Feng:2010yx} it follows that the full four point string superamplitude with one massive state in four dimensional kinematics is a Veneziano-type factor times the above amplitude, i.e.
\begin{equation}\label{eq:n4stringampcompl}
A(\Phi, G, G, G)|_{\eta \bar{\iota}} = \sqrt{ \alpha'}  \frac{1}{\braket{2 3} \braket{34} \braket{4 2}} \delta^8(\mathcal{Q}^I_\alpha) \frac{\Gamma(1-\alpha' s)\Gamma(1-\alpha' t)}{\Gamma(1-\alpha' s - \alpha' t)}
\end{equation} 
up to some numerical constant and $I$ still runs from $1$ to $4$. This form reproduces one color ordered component amplitude and is invariant under the on-shell SUSY algebra, so must be the full answer. Note that in the case of one massive leg the four point MHV-like and $\overline{\textrm{MHV}}$-like amplitude, e.g.
\begin{equation}\label{eq:n4stringampcomplbar}
A(\overline{\Phi}, G, G, G)|_{\bar{\eta} \iota}  =  \sqrt{ \alpha'}   \frac{1}{\sbraket{2 3} \sbraket{34} \sbraket{4 2}} \delta^8(\bar{\mathcal{Q}}^I_{\dot\alpha}) \frac{\Gamma(1-\alpha' s)\Gamma(1-\alpha' t)}{\Gamma(1-\alpha' s - \alpha' t)}
\end{equation} 
are \emph{not} dual to each other under fermionic Fourier transform as in the massless case. One of these amplitudes is natural in the $\eta \bar{\iota}$ and the other in the $\bar{\eta} \iota$ representation. Simply counting fermionic weight shows that the fermionic Fourier transform of one of these four point amplitudes to the conjugate representation will yield a function with fermionic weight $12$ instead of $8$. This can therefore not be the other superamplitude. Hence the four point amplitude with four dimensional kinematics has both an MHV and a  $\overline{\textrm{MHV}}$ like configuration with one massive leg. We strongly suspect the three and four point superamplitudes reproduce all amplitudes in \cite{Feng:2010yx} of this type by integrating out appropriate fermionic variables. 
To make a precise comparison would require to properly identify the lower-spin states in the $\mathcal{N}=4$ supermultiplet with maximal spin $2$ that will be in general superpositions of states with equal spin projection but different total spin (c.f.~\eqref{eq:clebsch-states}), analogous to the spin zero states in the vector multiplet~\eqref{eq:massive-vec+}.  For higher multiplicities the string theory amplitudes with four dimensional kinematics and one massive leg should reduce to equation \eqref{eq:n4stringamp} in the $\alpha' \rightarrow 0$ limit. We leave many interesting questions and obvious guesses which are raised by these short paragraphs to future work.

\subsection{Incorporating off-shell elements: vector boson currents}
Although the main focus of this article is on on-shell states, it is an interesting question how the technology developed here can be extended to incorporate elements of off-shell states as well. After all the difference between on-shell and off-shell for massive states is not as big as for massless states, see for instance the parallel between the space considered in  \cite{Raju:2011ed, Raju:2011mp} for the $\mathcal{N}=4$ stress-energy multiplet and the $\mathcal{N}=4$ massive multiplet considered above. Another prime example of this are vector boson currents, discussed in the context of CSW rules in~\cite{Bern:2004ba} for instance. These are the currents for an electroweak vector boson coupled to quarks which in turn couple to glue. All states apart from the electroweak vector boson are put on-shell. 

As calculated by the ancients~\cite{Berends:1988yn},  this type of current has a particularly nice form if the gluons are all of positive helicity while the quarks are of positive and negative type,
\begin{equation}\label{eq:simplevecboscur}
J^{\alpha \dalpha}(f_1^+, g_2^+, \ldots, g_{n-1}^+, f_n^- | P_V) = C \sqrt{2} \frac{n_{\beta} P_V^{ \beta \dalpha} n^{\alpha}}{\braket{1 2} \ldots \braket{(n-1) n}}.
\end{equation}
Here $C$ is an unimportant numerical constant. In this color-ordered current particles $1$ and $n$ correspond to the fundamental quarks with indicated helicities and $P_V$ is the momentum flowing through the off-shell vector boson leg. Note that the current is transverse $P_V^{\mu} J_{\mu} =0$, as it should. In four dimensions this implies the current can be expanded into a basis of the transverse space. A natural basis for us is given in equation \eqref{eq:massivepolarization} and consists of the polarization vectors for the massive vector boson with mass $P_V^2$. In particular, the mass of the vector boson is \emph{not} set to its physical value. This results in 
\begin{equation}\label{eq:simplevecboscurcomp}
J^{\alpha \dalpha}(f_1^+, g_2^+, \ldots, g_{n-1}^+, f_n^- | P_V) = e^{\alpha \dalpha}_- c_1 +  e^{\alpha \dalpha}_+ c_2 + e^{\alpha \dalpha}_0 c_3
\end{equation}
with 
\begin{align}
c_1 & = C  \frac{P_V^2}{\braket{q p_V}^2} \frac{ \braket{q n }^2 }{\braket{1 2} \ldots \braket{(n-1) n}} ,\\
c_2 & = C \frac{\braket{p_V n }^2 }{\braket{1 2} \ldots \braket{(n-1) n}}, \\
c_3 & = \sqrt{2} \frac{\braket{p_V n } \braket{n q} \sqrt{P_V^2} }{\braket{p_V q}\braket{1 2} \ldots \braket{(n-1) n}} .
\end{align}
What is important for us is that the $c_i$ coefficients can be interpreted as specific scattering amplitudes for massive vector bosons. It is this interpretation which can be incorporated neatly into the on-shell framework advocated above. 

Since the problem under study contains a massive vector boson it is natural to try to use a $\mathcal{N}=2$ massive on-shell superspace as in subsection \ref{subsec:massivecoulombbranch}. We will in addition need a massless quark superfield $M$ charged in the fundamental and a similar superfield $\bar{M}$ charged in the anti-fundamental,
\begin{equation}
M = \xi^+ + m_I \eta^I + \xi^- \eta^{1}\eta^2, \qquad \bar{M} = \bar{\xi}^+ + \bar{m}_I \eta^I + \bar{\xi}^- \eta^{1}\eta^2.
\end{equation}
By analogy to the MHV amplitude case it is easy to suspect that the components of the current \eqref{eq:simplevecboscurcomp} will arise as the simplest solution to the SUSY Ward identities: i.e. the ones which have no fermionic weight beyond the delta function. This leads to 
\begin{equation}\label{eq:supercurrent}
A(M_1, G_2^+, \ldots, G^+, \bar{M}_n | V) = C \frac{1}{\braket{1 2} \ldots \braket{(n-1) n}}  \delta^{4} \left(\mathcal{Q}^I_{\alpha} \right) .
\end{equation}
Integrating out both $\eta$ variables on the massive vector boson leg and both $\eta$ variables on the anti-fundamental scalar leg gives,
\begin{equation}
A(f_1^+, g_2^+, \ldots g_{n-1}^+, f_n^- | P^-_V) = C \frac{\braket{p_V n }^2 }{\braket{1 2} \ldots \braket{(n-1) n}},
\end{equation}
which is exactly equivalent to coefficient $c_2$. Instead integrating out both $\iota$ variables on the vector boson leg and both $\eta$ variables on the anti-fundamental scalar leg gives,
\begin{equation}
A(f_1^+, g_2^+,\ldots g_{n-1}^+, f_n^- | P^+_V) = C \frac{m^2}{\braket{q p_V}^2} \frac{ \braket{q n }^2 }{\braket{1 2} \ldots \braket{(n-1) n}} ,
\end{equation}
which is exactly equivalent to coefficient $c_1$, when taking into account the relation $m^2 = P_V^2$. The last coefficient follows from
\begin{align}
A(f_1^+, g_2^+ \ldots, g_{n-1}^+, f_n^- | P^0_V) & = \int d\eta_n^2 \frac{1}{\sqrt{2}} \left(\left[ d\eta^1_V d\bar{\iota}^2_V + d\eta^2_V d\bar{\iota}^1_V \right] \right) A(M, G^+, \ldots, G^+, \bar{M} | V) \\
& =  \sqrt{2} \frac{\braket{p_V n } \braket{n q} m }{\braket{p_V q}\braket{1 2} \ldots \braket{(n-1) n}} ,
\end{align}
which is the final coefficient, $c_3$.  

With these observations a supercurrent can be constructed:
\begin{equation}\label{eq:expractcurr}
J^{\alpha \dalpha}(M G^+ \ldots G^+ \bar{M} | P_V) = e^{\alpha \dalpha}_+ \tilde{c}_1 +  e^{\alpha \dalpha}_- \tilde{c}_2 + e^{\alpha \dalpha}_0 \tilde{c}_3
\end{equation}
with 
\begin{align}
c_1 & = \int d\eta^1_V d\eta^2_V  A(M, G^+, \ldots, G^+, \bar{M} | V),  \\
c_2 & = \int d\bar{\iota}^1_V d\bar{\iota}^2_V  A(M, G^+, \ldots, G^+, \bar{M} | V),  \\
c_3 & = \frac{1}{\sqrt{2}} \left(\left[\int d\eta^1_V d\bar{\iota}^2_V + \int d\eta^2_V d\bar{\iota}^1_V \right] \right) A(M, G^+, \ldots, G^+, \bar{M} | V) .
\end{align}
Note that this supercurrent is more like a component part of a natural function on superspace. As such it is not completely annihilated by the on-shell superspace supersymmetry generators. Similar supercurrents can be constructed for the scalar and fermionic fields in the massive vector boson multiplet. 

The calculation in this subsection shows that off-shell elements like currents can be encoded into an essentially on-shell approach using the massive on-shell superspace. The prescription is to expand the current into a complete set of solution to the free field equations. The coefficients of this expansion can then be treated as amplitudes with massive legs whose $P^2$ does not equal their physical mass. However, when expressed into the original currents along the lines of equation \eqref{eq:expractcurr} the explicit supersymmetry becomes obscured. It would be very interesting to find a formulation which yields a more supersymmetric form of the supercurrents. This should arise from blending an off-shell superfield formalism with the above on-shell framework. Although interesting, this is far beyond the scope of this article. 

\subsubsection{Three-point ``supercurrent''}
For completeness let us list the three point supercurrent,
\begin{equation}
A(M_1, \bar{M}_2 | V) = C \frac{1}{\braket{1 2}}  \delta^{4} \left(\mathcal{Q}^I_{\alpha} \right), 
\end{equation}
which is just the three point version of equation \eqref{eq:supercurrent}.


\section{Solving the SUSY Ward identities through supersymmetric on-shell recursion}
\label{sec:susyrecur}
A natural question from the previous section is how one should calculate generic superamplitudes at tree level without resorting to off-shell techniques. In the massless case the answer to this question is the existence of supersymmetric on-shell recursion relations \cite{Brandhuber:2008pf} which are a supersymmetric version of Britto-Cachazo-Feng-(Witten)  (BCFW) \cite{Britto:2004ap,Britto:2005fq} on-shell recursion. These relations allow one to calculate a superamplitude at tree level with any number of legs  from lower point amplitudes. Hence this reduces the problem down to calculating three point superamplitudes, which were found explicitly in the previous section for various different theories. Recursion relations which involve BCFW shifts of massive legs have been discussed in~\cite{Badger:2005zh,Badger:2005jv,Schwinn:2007ee}. 

\subsection{Constructing BCFW supershifts for massive particles}
A key ingredient in the BCFW relations is the concept of a shift of two in principle arbitrarily chosen legs which turns an amplitude of interest into a function of a single complex parameter $z$, 
\begin{equation}
k_1 \rightarrow \hat k_1\equiv k_1 + z  \, n \,,\qquad  
k_2 \rightarrow \hat k_2\equiv k_2 - z \, n\,.
\end{equation}
If the vector $n$ satisfies
\begin{equation}\label{eq:bcfwconstr}
n \cdot k_1 = n \cdot k_2 = n \cdot n = 0
\end{equation}
this shift keeps the masses of particles one and two invariant. This shift is known as a BCFW shift. To construct this shift for massive particles in terms of spinors, note that two massive momenta $k_1$ and $k_2$ can always be decomposed into
\begin{equation}
k_1 = k_1^{\flat} + \frac{m^2_1}{2 k_1^{\flat} \cdot k_2^{\flat}}  k_2^{\flat},  \qquad k_2=  k_2^{\flat} + \frac{m^2_2}{2 k_1^{\flat} \cdot k_2^{\flat}} k_1^{\flat} ,
\end{equation}
where $k_1^{\flat}$ and $k_2^{\flat}$ are massless. A BCFW shift vector $n$ which satisfies equation \eqref{eq:bcfwconstr} can be written in terms of the $\flat$-spinors as
\begin{equation}\label{eq:bcfwshiftsol}
n_{\alpha\dalpha} =   \, k_{1,\alpha} k_{2,\dalpha} \qquad \textrm{or} \qquad 
n_{\alpha\dalpha} =  \, k_{2,\alpha} k_{1,\dalpha}
\end{equation}
up to a proportionality constant. Here and below in this section for notational convenience the spinors have been written without $\flat$. Picking the first solution in equation \eqref{eq:bcfwshiftsol} the shifted momenta can be written as
\begin{equation}\label{eq:bcfwshift}
\begin{aligned}
\hat k_{1,\alpha\dalpha} &= k_{1,\alpha} (k_{1,\dalpha} + z k_{2,\dalpha}) 
+ \frac{m^2_1}{\braket{1 2}\sbraket{21}} k_{2\alpha} k_{2\dalpha} , \\
\hat k_{2,\alpha\dalpha}&=  (k_{2,\alpha} -z k_{1,\alpha}) k_{2,\dalpha} +
\frac{m^2_2}{\braket{1 2}\sbraket{21}} k_{1,\alpha}k_{1,\dalpha} .
\end{aligned}
\end{equation}
This way of writing the momenta is manifestly consistent with the massless limits. Since for the chosen shift 
\begin{equation}
\braket{1 2}\sbraket{1 2}  = \braket{\hat{1} 2}\sbraket{1 \hat{2}} 
\end{equation}
holds, these spinors can be used just as above to construct supersymmetric amplitudes: the BCFW shift reduces again to studying shifted spinors. Care should be taken though that the reference vectors of legs $1$ and $2$ are not shifted, but kept fixed at $k_2^{\flat}$ and $k_1^{\flat}$ respectively. We will employ an on-shell superspace for the two massive legs for which the SUSY generators can be written as 
\begin{align}
-\frac{1}{\sqrt 2}(Q_{1,\alpha} + Q_{2,\alpha}) & 
=  k_{1,\alpha} \eta_1 + \frac{m_1}{\braket{1 2}} k_{2,\alpha} \bar{\iota}_1 + (1 \leftrightarrow 2),\\
\frac{1}{\sqrt 2}(\bar{Q}_{1,\dalpha} + \bar{Q}_{2,\dalpha})
 & = k_{1,\dalpha} \frac{\partial}{\partial \eta_1} + 
 \frac{\overline{m}_1}{\sbraket{ 21}}k_{2,\dalpha}  
\frac{\partial}{\partial \bar{\iota}_1} + (1 \leftrightarrow 2).
\end{align}
Note that in this step we have chosen a spin polarization axis for both of the shifted legs. As explained above, if the same type of superspace is used for all the legs throughout an amplitude that amplitude is proportional to a super momentum conserving delta function,
\begin{align}
A(1,2,X) \sim \delta^{2}(\mathcal{Q}_\alpha) & =
\delta^{2}\left(\eta_1 k_{1,\alpha}  + \bar{\iota}_1 \frac{m_1}{\braket{12}} k_{2,\alpha}  + \eta_2 k_{2,\alpha}  + \bar{\iota}_2 \frac{m_2}{\braket{21}} k_{1,\alpha} + \mathcal{Q}_{X,\alpha}\right),
\end{align}
where $\mathcal{Q}_{X,\alpha}$ stands for the appropriate supermomentum of the rest of the diagram. Note that this does not imply a choice of spin-polarization axis for the remaining particles, but only a choice of representation of the SUSY algebra.

Under the shift of the momenta given in equation \eqref{eq:bcfwshift} the supermomentum $\mathcal{Q}_{\alpha}$ is not invariant but shifts linearly\footnote{Recall that the terms $\sim\bar\iota$ in the supermomentum arise from the reference spinors and are not shifted }
\begin{equation}
\mathcal{Q}_{\alpha} \rightarrow \mathcal{Q}_{\alpha} -  z\, \eta_2 \, k_{1,\alpha}.
\end{equation}
The supermomentum can be made invariant under the BCFW shift by an additional shift of the fermionic variables $\eta_1$,
\begin{equation}\label{eq:supershift}
\eta_1 \rightarrow \eta_1 + z \, \eta_2.
\end{equation} 
More general possibilities to cancel the $z$-dependence of $Q$ involving a shift of $\bar{\iota}_2$ would in general not permit a massless limit and will therefore not be considered here. The resulting combined shift of equation \eqref{eq:bcfwshift} together with \eqref{eq:supershift} will be referred to as a supershift in the following. 

\subsection{Supersymmetric on-shell recursion relations}
As BCFW observed, the original amplitude one wishes to study arises as
\begin{equation}
A(0) = \oint_{z=0} \frac{A(z)}{z},
\end{equation}
where the contour integral is around a small contour circling zero and will include a normalization factor of $\frac{1}{2 \pi i}$ by convention. Pulling the contour to infinity gives a sum over simple poles at finite values of $z$ whose residue is the product of tree level amplitudes summed over species and spins, with in addition a possible pole at infinity,
\begin{multline}
A(0) = - \sum_{\textrm{finite }z \textrm{ poles}} \left(\sum_{s \in {\textrm{species} \atop \textrm{spins}}} A_L\left(\hat{1} \ldots, \{\hat{P}, s\}\right) \frac{1}{P_L^2-m_P^2}A_R\left(\{-\hat{P},-s \}, \ldots, \hat{2}\right)
 \right)\\ + \textrm{Res}(z=\infty).
\end{multline}
Here $P_L$ is the sum over undeformed momenta of all known particles into the
$A_L$ amplitude and $M$ the mass of the intermediate particle with momentum
$P_L$. Hence if the residue at infinity vanishes then an on-shell recursion
relation is obtained. From the discussion around equation
\eqref{eq:sumtofermint} it follows that the sums over spins and to some extend
the sum over species can be replaced by fermionic integrations on
superamplitudes,
\begin{multline}\label{eq:susyonsherec}
A(0) = - \!\!\!\!\! \sum_{\textrm{finite }z \textrm{ poles}} \left(\sum_{s}\!\! \int d\eta_P d\bar{\iota}_P A_L\left(\hat{1} \ldots, \{\hat{P}, \eta_P, \bar{\iota}_P\}_s \right) \frac{1}{P_L^2-m_P^2} \right. \\ \left. A_R\left(\{-\hat{P}, \eta_P, \bar{\iota}_P \}_{-s}, \ldots, \hat{2}\right)
 \right) + \textrm{Res}(z=\infty),
\end{multline}
where the sum $s$ runs over the possible superfields. Here a choice of spin axis for the 'cut' leg can still be made. 

It is instructive to verify that the relation in equation \eqref{eq:susyonsherec} produces a solution to the on-shell SUSY Ward identities. In fact, it does so term-by-term, assuming the lower order terms satisfy the Ward identities\footnote{For the theories under study in this article the necessary ``three point amplitude'' base step for this proof by induction has been shown explicitly in section \ref{sec:exampamplis}.}. To see this, pick one particular term of the sum over finite poles in equation  \eqref{eq:susyonsherec} . For this term we would like to verify
\begin{equation}
\mathcal{Q}_{\alpha} (\int d\eta_P d\bar{\iota}_P  A_L A_R) \,  = \,0 \,= \,
\bar{\mathcal{Q}}_{\dalpha} (\int d\eta_P d\bar{\iota}_P  A_L A_R).
\end{equation}
Since both these operators leave the factor $\frac{1}{P_L^2-m_P^2}$ invariant, this has been stripped off. Now by construction of the supershift,
\begin{equation}
\mathcal{Q}_{\alpha} =
 \mathcal{Q}^{L}_{\alpha} + \mathcal{Q}^{R}_{\alpha} 
=  \hat{\mathcal{Q}}^{L}_{\alpha} + \hat{\mathcal{Q}}^{R}_{\alpha}.
\end{equation}
The shifted $\hat{\mathcal{Q}}^{L/R}$ almost annihilate the amplitudes, apart from the missing `cut'-leg term, 
\begin{equation}
 \hat{\mathcal{Q}}^{L}_{\alpha} A_L = 
\left(  \eta_P p^{\flat}_{\alpha} + \bar{\iota}_P \frac{m_P q^P_{\alpha}}{\braket{ p^{\flat}q^P}} \right) A_L.
 \end{equation}
Choosing the spinor momentum of the leg with momentum $-P$ to be $-p^{\flat}_{\alpha}$ the other amplitude yields
\begin{equation}
 \hat{\mathcal{Q}}^{R}_{\alpha} A_R = 
 \left( -\eta_P p^{\flat}_{\alpha} - \bar{\iota}_P \frac{m_P q^P_{\alpha}}{\braket{ p^{\flat}q^P}} \right) A_R
 \end{equation}
and hence
\begin{equation}
\mathcal{Q}_{\alpha} (\int d\eta_P d\bar{\iota}_P  A_L A_R) = 0.
\end{equation}

The other operator can be written
\begin{equation}
 \bar{\mathcal{Q}}_{\dalpha}  =  \bar{\mathcal{Q}}^L_{\dalpha} + 
 \bar{\mathcal{Q}}^R_{\dalpha} .
\end{equation}
From the explicit form of the above shift:
\begin{equation}
\eta_1 \rightarrow \eta_1 + \eta_2 z \equiv \hat{\eta}_1\, ,\quad
 \eta_2\equiv\hat\eta_2, \qquad 
k_{1,\dalpha} \rightarrow k_{1,\dalpha} + z k_{2,\dalpha} \equiv \hat{k}_{1,\dalpha}
\end{equation}
holds, which leads  to 
\begin{equation}
\frac{\partial}{\partial \eta_2} = \left(\frac{\partial}{\partial \hat \eta_2}
  + z \frac{\partial}{\partial \hat{\eta}_1} \right), \qquad \frac{\partial}{\partial \eta_1}  = \frac{\partial}{\partial \hat{\eta}_1}
\end{equation}
so that 
\begin{align}
k_{1,\dalpha}  \frac{\partial}{\partial \eta_1}  
+ k_{2,\dalpha} \frac{\partial}{\partial \eta_2}
 & = k_{1,\dalpha} \frac{\partial}{\partial \hat{\eta}_1} +
 k_{2,\dalpha}  \left( \frac{\partial}{\partial \hat \eta_2} 
 + z \frac{\partial}{\partial \hat{\eta}_1} \right) \nonumber\\ 
& = \hat{k}_{1,\dalpha}  \frac{\partial}{\partial \hat{\eta}_1} + k_{2,\dalpha} \frac{\partial}{\partial \hat \eta_2} 
\end{align}
and hence
\begin{equation}
\bar{\mathcal{Q}}^L_{\dalpha} +  \bar{\mathcal{Q}}^R_{\dalpha}  =
 \hat{\bar{\mathcal{Q}}}^L_{\dalpha} +  \hat{\bar{\mathcal{Q}}}^R_{\dalpha} .
\end{equation}
Now the operators on the right hand side of this equation act on $A_L$ and $A_R$ respectively. The Ward identities of both these superamplitudes then can be used to show that one obtains a fermionic derivative on the cut leg in both cases. This total derivative under the fermionic integral vanishes trivially. This completes the proof that the finite $z$ residues in \eqref{eq:susyonsherec} satisfy the supersymmetric Ward identities term-by-term. In passing we note that this implies also that the residue at infinity satisfies the supersymmetric Ward identities even if it is not zero.

\subsection{Residues at infinity: general case} 
To have a constructive method of solving the Ward identities the residue at
infinity has to be understood properly. This residue will first be studied in
the generic case of the supershift contained in equations \eqref{eq:bcfwshift}
and \eqref{eq:supershift}. In particular, we allow these legs to be either
massive or massless. Since the BCFW shift involves a momentum which will
become large compared to any fixed scale, one might naively expect that masses
do not change the analysis of large BCFW shifts in field theory compared to
the massless case. However, the shift involves complex momenta which may
invalidate this hand-waving argument. Also, for massive particles new helicity
combinations are allowed which can lead to complications. This will be checked explicitly below in the example of SQCD.  

We study the shifts for a general superfield in the
$\eta\bar\iota$-representation~\eqref{eq:superfield}.  The $z$-dependence of
the amplitude is contained in both the momentum as well as the coherent state
parameter $\hat\eta_1$,
\begin{equation}
A(z) = A(\{\hat{k}_1, \hat{\eta}_1, \bar{\iota}_1 \}, \{\hat{k}_2, \eta_2, \bar{\iota}_2 \} , X)
\end{equation}
The point of this representation is that just as in the gluonic massless case discussed in \cite{ArkaniHamed:2008gz} there is a supersymmetry transformation which shifts $\hat{\eta}_1$ and $\eta_2$ to zero, while it itself is independent of $z$. Any spinor $\xi_{\dalpha}$ can be expanded,
\begin{equation}\label{eq:susytrafobcfw}
\xi_{\dalpha}  = \frac{\hat{k}_{1\dalpha} \xi_1 - k_{2,\dalpha} \xi_2}{\sbraket{1 2}}  = \frac{k_{1,\dalpha} \xi_1 - k_{2,\dalpha} \tilde{\xi}_2}{\sbraket{1 2}}
\end{equation}
where we have defined 
\begin{equation}
\tilde{\xi}_2 \equiv \xi_2 - z \xi_1
\end{equation}
Consider now the supersymmetry transformations generated by $\sbraket{\xi \bar Q}$. These shift the fermionic variables on legs $1$ and $2$ as
\begin{equation}
\begin{array}{cccccc}
\hat{\eta}_1 & \rightarrow & \hat{\eta}_1 + \sbraket{\xi \hat{1}} & = &\hat{\eta}_1 + \xi_2 &= \eta_1 + z (\eta_2+  \xi_1)  + \tilde{\xi}_2\,, \\
\eta_2 & \rightarrow & \eta_2 + \sbraket{\xi 2} & = & \eta_2 + \xi_1\,, & \\
\bar{\iota}_1 & \rightarrow & \bar{\iota}_1 - \frac{m_1}{\sbraket{21}} \sbraket{\xi 2} & = &  \bar{\iota}_1 -  \frac{m_1}{\sbraket{21}} \xi_1\,, & \\
\bar{\iota}_2 & \rightarrow & \bar{\iota}_2 -  \frac{m_2}{\sbraket{1 2}} \sbraket{\xi 1} & = &  \bar{\iota}_2 -  \frac{m_2}{\sbraket{1 2}} \tilde{\xi}_2 \,.&
\end{array}
\end{equation}
Note that not all spinors are shifted, since some are part of the definition of the spin axis of the other leg (see equation \eqref{eq:bcfwshift}). To shift $\hat\eta_1$ and $\eta_2$ to zero, set
\begin{equation}
\xi_1 = - \eta_2 \,,\qquad \tilde{\xi}_2 = -\eta_1\,.
\end{equation}
This SUSY transformation is manifestly independent of $z$ from the second expression in equation \eqref{eq:susytrafobcfw}. 

Schematically the amplitude now reads
\begin{equation}
A(z) \sim A(\{\hat{k}_1, 0, \tilde{\bar{\iota}}_1 \}, \{\hat{k}_2, 0, \tilde{\bar{\iota}}_2 \} , \tilde{X}),
\end{equation}
where the proportionality factor and the transformed field content $\tilde{X}$ are independent of $z$ (but dependent on $\eta_1, \eta_2$). Expanding out this superamplitude over $\tilde{\iota}_1$ and $\tilde{\iota}_2$ then yields the result that the supershifted amplitude is proportional to a sum over four amplitudes, 
\begin{multline}\label{eq:shiftamptostudy}
A(\{\hat{k}_1, 0, \tilde{\bar{\iota}}_1 \}, \{\hat{k}_2, 0, \tilde{\bar{\iota}}_2 \} , \tilde{X}) = A(\{\phi_0' \}, \{\phi_0'\} , \tilde{X}) +\\  \tilde{\bar{\iota}}_2 A(\{\phi_0' \}, \{\phi_+\} , \tilde{X})  +  \tilde{\bar{\iota}}_1 A(\{\phi_+ \}, \{\phi_0'\} , \tilde{X})  +  \tilde{\bar{\iota}}_1 \tilde{\bar{\iota}}_2 A(\{\phi_+ \}, \{\phi_+\} , \tilde{X}) 
\end{multline}
with all the $z$-dependence in the momenta of the first two particles in each of the component amplitudes. Hence the large-$z$ behavior of the super-shifted amplitudes can be obtained from studying the ordinary BCFW shifts of the component amplitudes of the states with spin $s_n^0$ and  $s_n^0+\tfrac{1}{2}$ . The latter can be done through various means and the outcome depends on the field content of the theory under study. Note that there can be no cancellations between the four different amplitudes in the above equation since they are multiplied by the $\bar{\iota}_i$ variables. These are the only places $\bar{\iota}_1$ and $\bar{\iota}_2$ occur in the full superamplitude. 
One can take a massless limit of the above shifts without problems: in
practice this means that one sets either $\tilde{\bar{\iota}}_1$ or
$\tilde{\bar{\iota}}_2$ or both to zero in the equation just derived. 


\subsection{Residues at infinity: SQCD example}
The shifts of tree level amplitudes in \eqref{eq:shiftamptostudy} can be studied in any theory which one might be interested in. BCFW shifts at tree level with at least one massless leg in quite a large class of theories have been studied in \cite{Cheung:2008dn}. Here we will confine ourselves to an initial study of the example of supersymmetric QCD, already considered above as the first example theory in section \ref{sec:exampamplis}. While the generalization of the results of this subsection to other minimally coupled supersymmetric theories with massive matter is immediate, a systematic study of supersymmetric theories with massive gauge bosons (or higher spins) is beyond the scope of this article.

First consider the case of two shifted and color-adjacent massless legs. For the fields in the fundamental color-adjacent is taken to mean on opposite ends of the same color structure, while for mixed adjoint-fundamental type shifts this means that the adjoint field appears next to the fundamental in color-ordering. In this case there are four possibilities depending on the choice of which massless superfields are shifted:
\begin{equation}
A(\hat{G}^+, \hat{G}^+, \ldots), \qquad A(\hat{G}^-, \hat{G}^+, \ldots), \qquad A(\hat{G}^+, \hat{G}^-, \ldots), \qquad A(\hat{G}^-, \hat{G}^-, \ldots) .
\end{equation}
By the analysis just presented, this implies we need to study 
\begin{equation}
A(g^+, g^+, \ldots), \qquad A(\Lambda^-, g^+, \ldots),\qquad A(g^+, \Lambda^-, \ldots), \qquad A(\Lambda^-, \Lambda^-, \ldots), 
\end{equation}
respectively. An analysis of the shifts of these amplitudes can be found for
instance in appendix B of  \cite{Boels:2010nw}, see also \cite{Cheung:2008dn}
for a slightly different tree level analysis \footnote{It can be checked that
  the analysis in \cite{Boels:2010nw} goes through basically unchanged if
  massive fermions charged in the fundamental are added. QCD with massive
  fermions has also been analyzed in~\cite{Schwinn:2007ee}.}. This leads to the massless sector of table \ref{tab:largezmassless} whose structure in this sector might  also have been guessed by familiarity with the usual result. By the results in the table, supershifts exist such that there are supersymmetric on-shell recursion relations without boundary contributions.

\begin{table}
\begin{center}
\begin{tabular}{c|c c c}
$1 \;\backslash \;2  $ & $G^+$           & $G^-$    & $\overline{\Phi}$             \\
\hline
$G^+$                   & $ +1$ & $ -2$     & $-1$ \\
$G^-$ 		   & $ +1$ & $ +1$    & $-1$ \\
$\Phi$                   & $ +1$ & $ -1$     & $-1$
\end{tabular}

\caption{Estimate of the leading asymptotic power in $z^{-\kappa}$ of the adjacent BCFW supershift of all superfields in a tree amplitude in SQCD.}
 \label{tab:largezmassless}
\end{center}
\end{table}

There are four possible supershifts for cases with one massive and one massless leg. 
\begin{equation}
A(\hat{G}^+, \hat\Phi, \ldots), \qquad 
A(\hat{G}^-, \hat\Phi, \ldots), \qquad 
A(\hat\Phi, \hat{G}^-, \ldots), \qquad A(\hat\Phi, \hat{G}^+, \ldots) .
\end{equation}
By the analysis above this reduces to the study of shifts of
\begin{eqnarray}
\left\{ A(g^+, \phi, \ldots)\, ,\, A(g^+, Q^+, \ldots) \right\} &, & \left\{ A(\Lambda^-, \phi, \ldots)\, ,\, A(\Lambda^-, Q^+, \ldots) \right\} , \\
 \left\{ A(\phi, \Lambda^-, \ldots)\, ,\, A(Q^+, \Lambda^-, \ldots) \right\}&,& 
\left\{ A(\phi, g^+, \ldots)\, ,\, A(Q^+, g^+,  \ldots) \right\} .
\end{eqnarray}
Since these amplitudes multiply different fermionic weight functions in \eqref{eq:shiftamptostudy} the total shift is determined by the 'worst behaving' shift of the two component amplitudes. As observed in \cite{ArkaniHamed:2008yf} there is a natural gauge to study shifts in: the lightcone gauge with gauge vector $q$ from the BCFW shift. In this gauge one can perform a quick analysis which graphs contribute at leading order in the shift parameter $z$. The leading graphs for instance invariably include those where the shifted legs end on the same vertex. Here a difference arises between the massive and massless cases: in the massive case the Yukawa couplings behave differently since a helicity violating $\Lambda^-Q^+\phi$-vertex is allowed (c.f.~\eqref{eq:gluino-yukawa}). 
This analysis results in the entries involving $G$ and $\Phi$-fields in table~\ref{tab:largezmassless}. A more refined analyis might improve the behaviour for some of these shifts from $z^1$ to $z^0$.

The shift of two massive legs is easier to analyze as there is only one possibility for a shift. By equation \eqref{eq:shiftamptostudy} we need to study the component amplitudes
\begin{equation} 
\left\{  A(\phi, \phi, \ldots)\, ,\, A(\phi, Q^+, \ldots) \, ,\, A(Q^+, \phi, \ldots)\, ,\, A(Q^+, Q^+, \ldots)\right\}.
\end{equation}
From diagrams which involve the Yukawa couplings between matter and vector multiplet, the third amplitude in this list for instance scales as $z$. 

Hence on-shell recursion relations work with massive multiplets in SQCD in general, provided one shifts at least one massless leg with
the shifts indicated on the first column of table \ref{tab:largezmassless}. This coincides with the conclusion of \cite{Cheung:2008dn} for tree level amplitudes. Of course, in cases there is a residue at infinity one could still use on-shell recursion relations to calculate as long as the residue is known. Alternatively, one could also pursue the use of shifts of more than two legs, for instance  amplitudes with only massive quark legs can be constructed with three-line shifts~\cite{Schwinn:2007ee}. Note also that all-leg shifts have been considered recently in~\cite{Cohen:2010mi}. There is one interesting exceptional case of this type which is interesting in its own right. This is the case of a massive fundamental scalar/anti-scalar pair coupled to all-plus glue which will be discussed below.

It should be noted that the results of table \ref{tab:largezmassless} can be improved if other shifts than adjacent ones are considered. For the adjoint-valued fields this means non-color-adjacent shifts which generically scale one order of $\frac{1}{z}$ better than the adjacent shift. For the mixed case, i.e. a shifted fundamental and adjoint valued field, a similar result is expected to hold, as well as for two shifted fundamental/anti-fundamental pairs which appear on different color structures. The latter possibility only arises of course when more than one fundamental/anti-fundamental pair is involved in the amplitude.

\subsubsection{Massive scalar pair with only positive helicity gluons}
In this particular case the amplitude is known in a particularly nice form as was mentioned in the introduction in equation \eqref{eq:rodrigo}. In this case it can be checked explicitly that the residue at infinity vanishes under a BCFW shift of both massive legs. This can also be understood from powercounting in the special lightcone gauge: the leading diagram for a shift of the massive legs involves the scalar pair coupled directly to a single gluonic current which involves only plus gluons. This current has been calculated long ago in \cite{Berends:1987me} and reads
\begin{equation}
J_{+}^{\alpha\dalpha}(3^+,4^+,\ldots, n^+)  \sim  \frac{\xi^{\dalpha} \xi^{\dbeta} (k_{3,n})_{\dbeta}^{\,\alpha}}{\braket{\xi 3} \braket{34}\ldots \braket{n \xi}} \ ,
\end{equation}
where the spinor $\xi$ is a lightcone gauge choice. For the case at
hand with the gauge choice that
$\xi^{\dot\alpha}\xi^{\alpha}=n^{\dot\alpha\alpha}=k_2^{\dot\alpha}k_1^\alpha$
it is easy to check that the diagram with two off-shell scalars
contracted into this current through a three vertex vanishes. This
follows as from the explicit form of the first solution for the BCFW
shift from equation \eqref{eq:bcfwshiftsol}
\begin{equation}
\begin{aligned}
\xi^{\dalpha} \xi^{\dbeta} (k_{3,n})_{\dbeta}^{\,\alpha}  &  =- (1+\frac{m^2}{2 k_1 \cdot k_2}) k_2^{\dalpha} k_2^{\dbeta} \left(k_2^{\alpha} k_{2,\dbeta} + k_1^{\alpha} k_{1,\dbeta}\right) \\
& =- (1+\frac{m^2}{2 k_1 \cdot k_2}) k_2^{\dalpha} k_1^{\alpha} \sbraket{12}\\
& \sim n^{\alpha \dalpha}
\end{aligned}
\end{equation}
where on the last line we have identified the vector as the first solution from equation \eqref{eq:bcfwshiftsol}. Note that the same result would have been obtained for the other solution in equation \eqref{eq:bcfwshiftsol}. Hence this current is orthogonal to all momenta appearing in the three vertex coupling. Moreover, the sub-leading graph with two scalars attached to the four vertex also vanishes as it involves contracting two all-plus currents together with a metric. Therefore the large BCFW shift of the two massive legs of a scalar pair amplitude with all plus glue vanishes. In particular, these amplitudes can be calculated through on-shell recursion. 

By simple extension of the above, the $\mathcal{N}=1$ superamplitude version of the above scalar pair amplitude~\eqref{eq:sqcd-all-plus}
also scales as $\frac{1}{z}$ under a large BCFW shift on the massive legs, simply because the supermomentum conserving delta function is invariant under the supershift by construction. The same holds for the $\mathcal{N}=2$ extension of this amplitude discussed above in equation \eqref{eq:massvecbos}.

\section{Discussion and conclusions}
This article should be viewed as another step in extending recent analytic results on scattering amplitudes which involve massless particles to particles with mass in four dimensions. This step consist of a fully covariant treatment of massive representations of the supersymmetry algebra which is a natural extension of \cite{Grisaru:1977px} to the massive case. The Ward identities which follow from this should hold to any loop order as their derivation is based on algebra alone and in particular does not depend on any coupling constants. It is particularly illuminating to see the Ward identities derived in \cite{Schwinn:2006ca} from an off-shell point of view reappear here from an on-shell point of view.  Our on-shell treatment is completely general and can be applied to any theory with unbroken supersymmetry and particles in any massive or massless representation of the four dimensional supersymmetry algebra. 

A particularly useful tool in recent developments in the study of amplitudes with massless particles has been the on-shell massless superspaces originally pioneered in \cite{Nair:1988bq}. As explained in \cite{ArkaniHamed:2008gz} these can be understood as covariant coherent state representations of the on-shell supersymmetry algebra. This point of view was used above to construct on-shell superspaces for the massive representations. A prime application of on-shell superspaces in the massless case is to provide a method for obtaining solutions to the supersymmetric Ward identities, usually implemented through BCFW on-shell recursion. In several example amplitudes in several example theories with massive particles we have shown how this works in practice. A formal solution to the Ward identities in terms of amplitudes alone was also obtained in terms of on-shell recursion. To turn this into an actual solution one needs an analysis of allowed BCFW shifts for the theory one is interested in. For shifts of one or more massless particles on-shell recursion is expected to work generically. The question of other shifts was studied for SQCD above, a corresponding analysis for other theories with massive particles is left to future work. 

In general it would be interesting to see if the methods described in this paper can be applied to yield more examples of supersymmetric scattering amplitudes. The maximally spin violating amplitudes in the Abelian Higgs model show for instance that there certainly is scope for the appearance of simple results for amplitudes in seemingly complicated theories. Usually these results are indicative of an unappreciated underlying symmetry which would be interesting to find. A further worthwhile avenue to explore is the interpretation of massive representations of the SUSY algebra as off-shell states. Some baby-steps along these lines were taken above in the example of off-shell vector boson currents but there is clearly a lot more which may be done here, especially considering recent work on form factors at weak \cite{Brandhuber:2010ad, Bork:2010wf} and strong \cite{Maldacena:2010kp} coupling. 

One interesting research direction leading closer to experiment would be to investigate what happens to the SUSY Ward identities in scenario's where supersymmetry is (spontaneously) broken. The basic derivation of the transformations will hold there as well, but the Ward identity will change as the vacuum is not invariant under the supersymmetry any more. Note that the analog of an Adler zero has already been utilized in the context of spontaneously broken supersymmetry in \cite{deWit:1975th}, and it would be interesting to explore this further. A further direction which stands out is the study of non-perturbative corrections to the Ward identities. The vanishing of the all-plus amplitude for the massless sector for instance depends on the existence of a $U(1)_R$ symmetry which is known to be broken by instanton effects. In favorable circumstances, these corrections can be calculated and presumably also analyzed in an amplitude type of approach.

\section*{Acknowledgements}
We would like to thank Nathaniel Craig, Henriette Elvang, Michael Kiermaier and Tracy Slatyer for discussions and coordination of their related work \cite{Craig:2011ws}. In addition, we would like to thank Gang Yang for some useful discussions and the Niels Bohr International Academy for hospitality while this work was initiated. This work has been supported by the German Science Foundation (DFG) within the Collaborative Research Center 676 ``Particles, Strings and the Early Universe". 

\appendix

\section{Notation and conventions}
\label{app:notation}

\subsection{Spinor conventions and collection of useful results}
\label{app:spinors}
In this article the conventions of~\cite{Schwinn:2006ca} have been used which will be summarized here for convenience.  The sigma matrices are defined as
$\sigma_{\alpha\dot{\beta}}^{\mu} = \left( 1 , - \vec{\sigma}
\right)$, $\bar{\sigma}^{\mu \dot{\alpha} \beta} = \left( 1 ,
  \vec{\sigma} \right)$ where $\vec{\sigma} = \left( \sigma_x,
  \sigma_y, \sigma_z \right)$ are the Pauli matrices.  Four-vectors
$x^\mu$ are mapped to bi-spinors according to
\begin{equation}
\begin{aligned}
x_{\alpha\dot{\alpha}}&=x_\mu \sigma_{\alpha\dot{\alpha}}^\mu 
=
\begin{pmatrix}
  x_0-x_3 &, -x_1+i x_2\\ -x_1- i x_2& x_0+x_3
\end{pmatrix}
,\\
x^{\dot{\alpha} \alpha}&=x_\mu \bar\sigma^{\mu;\dot{\alpha} \alpha}
=
\begin{pmatrix}
  x_0+x_3 &, x_1-i x_2\\ x_1+ i x_2& x_0-x_3
\end{pmatrix}
\end{aligned}
\label{eq:bi-spinor}
\end{equation}
 so that 
$2 x^\mu y_\mu = x_{\alpha\dalpha}y^{\dalpha\alpha}$.
The two-dimensional antisymmetric tensor is defined by
\begin{equation}
\varepsilon^{\alpha\beta} = \varepsilon^{\dot{\alpha}\dot{\beta}} =
\varepsilon_{\alpha\beta} = \varepsilon_{\dot{\alpha}\dot{\beta}} =
\left(\begin{array}{cc} 0 & 1\\ -1 & 0 \\
\end{array} \right).
\end{equation}
Indices of two-component Weyl spinors are raised and lowered as follows:
\begin{equation}
 k^\alpha = \varepsilon^{\alpha\beta} k_\beta,
  \;\;\;
 k^{\dot{\alpha}} = \varepsilon^{\dot{\alpha}\dot{\beta}}k_{\dot{\beta}},
  \;\;\;
 k_{\dot{\beta}} =
 k^{\dot{\alpha}} \varepsilon_{\dot{\alpha}\dot{\beta}},
  \;\;\;
 k_\beta = k^\alpha \varepsilon_{\alpha\beta}.
\
\end{equation}
In the bra-ket notation spinor products are denoted as
\begin{eqnarray}
 \braket{p q} = \braket{p - | q + } = p^\alpha q_\alpha,
 &\qquad &
 \sbraket{q p} = \braket{q + | p - } = q_{\dot{\alpha}} p^{\dot{\alpha}}.
\end{eqnarray}

We also make use of the matrices
\begin{equation}
  \sigma^{\mu\nu}=\frac{1}{4}
  (\sigma^\mu\bar\sigma^\nu - \sigma^\nu\bar\sigma^\mu)\;,\quad
 \bar\sigma^{\mu\nu}=\frac{1}{4}
  (\bar\sigma^\mu\sigma^\nu-\bar\sigma^\nu\sigma^\mu)
\end{equation}
that satisfy the relations (using the convention $\epsilon^{0123}=1$)
\begin{equation}
\label{eq:sd-sigma}
  \sigma^{\mu\nu}=\frac{i}{2}\epsilon^{\mu\nu\rho\sigma}\sigma_{\rho\sigma}
  \,,\quad
 \bar \sigma^{\mu\nu}=-\frac{i}{2}\epsilon^{\mu\nu\rho\sigma}
 \bar\sigma_{\rho\sigma}.
\end{equation}
It is useful to recall the translation of the totally antisymmetric
tensor to spinor notation:
\begin{equation}
\label{eq:eps-spinor}
  \epsilon^{\mu\nu\rho\sigma}\Leftrightarrow
  \epsilon^{\alpha\dot\alpha\beta\dot\beta\gamma\dot\gamma\delta\dot\delta}
  =4i\left(\varepsilon^{\alpha\gamma}\varepsilon^{\beta\delta}
    \varepsilon^{\dot\alpha\dot\delta}\varepsilon^{\dot\beta\dot\gamma}-
    \varepsilon^{\alpha\delta}\varepsilon^{\beta\gamma}
    \varepsilon^{\dot\alpha\dot\gamma}\varepsilon^{\dot\beta\dot\delta}
  \right).
\end{equation}

For the Dirac matrices the representation
\begin{equation}
  \gamma^\mu =
  \begin{pmatrix}
    0&\sigma^\mu\\\bar\sigma^\mu &0
  \end{pmatrix}\,,\quad
  \gamma^5=i\gamma^0\gamma^1\gamma^2\gamma^3=
  \begin{pmatrix}
    1&0\\0&-1
  \end{pmatrix}
\end{equation}
will be used.

\subsection{Super Poincar\'e algebra}
\label{app:susy}
The Pauli-Lubanski vector is given by
\begin{equation}
\label{eq:pauli}
W^\mu=-\frac{1}{2}\epsilon^{\mu\nu\rho\sigma} P_\nu M_{\rho\sigma},
\end{equation}
where $M_{\rho\sigma}$ are the generators of the Lorentz
transformations. For a massive particle in the restframe, the
Pauli-Lubanski vector reduces to $W^0=0$, $W^i=m
J^i\equiv\frac{m}{2}\epsilon^{ijk}M_{jk}$. The generators of Lorentz-transformations in the Dirac representation
are given by $M_{\mu\nu}=\frac{1}{2}\Sigma_{\mu\nu}$ with
\begin{equation}
\label{eq:gen-onehalf}
  \Sigma^{\mu\nu}=\frac{i}{2}[\gamma^\mu,\gamma^\nu]=
  \begin{pmatrix}
    2i \sigma^{\mu\nu}&0\\ 0& 2i\bar\sigma^{\mu\nu}
  \end{pmatrix}.
\end{equation}
The relations~\eqref{eq:sd-sigma} imply the identity
\begin{equation}
\label{eq:SD-sigma}
 \gamma^5 \Sigma^{\mu\nu}=\frac{i}{2}\epsilon^{\mu\nu\rho\sigma}\Sigma_{\rho\sigma}.
\end{equation}

The SUSY algebra in four dimensions without central charges reads in the conventions of~\cite{Schwinn:2006ca}
\begin{align}\label{eq:susyalg}
\{Q_{\alpha}, \bar{Q}_{\dalpha}  \} &= -2 \sigma^{\mu}_{\alpha \dalpha} k_{\mu},&
\{\bar{Q}^{\dalpha}, Q^{\alpha}  \} &= -2 \bar{\sigma}^{\mu,\dalpha \alpha} k_{\mu}.
\end{align}
The commutation relations of the Lorentz-generators with the supercharges are
\begin{equation}
\label{eq:m-q-comm}
\begin{aligned}
  \lbrack M^{\mu\nu},Q_\alpha\rbrack&=-i (\sigma^{\mu\nu})_\alpha^\beta Q_\beta,\\
 \lbrack M^{\mu\nu},\bar Q^{\dot\alpha} \rbrack&=-i
 (\bar \sigma^{\mu\nu})^{\dot \alpha}_{\dot \beta} \bar Q^{\dot \beta}.
\end{aligned}
\end{equation}

\section{SUSY models with massive particles}\label{app:details}
This appendix contains some details on the models of SUSY gauge theories with
massive particles considered in the main text.
\subsection{SQCD with massive matter} 
\label{app:sqcd}
A supersymmetrized model of QCD with a massive quark can be obtained by
coupling  super-Yang Mills multiplet $(A_\mu^a,\lambda^a, D^a)$
 to a chiral multiplet $\Phi=(\phi,\psi,F)$ in the fundamental 
 and 
$\bar\Phi=(\bar\phi,\bar\psi,\bar F)$ in the anti-fundamental representation of
the gauge group. The Lagrangian  with a superpotential
$m\bar\Phi\Phi$ reads after elimination of the $D$ and
$F$ terms
\begin{eqnarray}
\label{eq:sqcd-mass}
 {\cal L} & = & - \frac{1}{4} \left( F^a_{\mu\nu} \right)^2 +
 \frac{i}{2} \bar{\Lambda}^a \gamma^\mu D^{ab}_\mu \Lambda^b + i
 \bar{\Psi} \gamma^\mu D_\mu  \Psi +
  \left( D_\mu \phi_+ \right)^\dagger \left( D_+^\mu \phi \right) +
\left( D_\mu \phi_- \right)^\dagger \left(D^\mu  \phi_- \right)
\nonumber \\ & &
 -\sqrt{2}g\left[ \bar\phi_+ \bar\Lambda^a_- T^a \psi_- +
   \bar\psi_+ \Lambda^{a}_+ T^a  \phi_-
   -\bar\phi_- \bar\Lambda^{a}_+ T^a \psi_+
   - \bar\psi_- \Lambda^a_- T^a \phi_+ \right]
 \nonumber \\ & & - \frac{1}{2} g^2 \left( \bar{\phi}_+ T^a  \phi_- -
 \bar\phi_- T^a \phi_+ \right)^2, \nonumber
\end{eqnarray}
where the covariant derivative in the fundamental and adjoint representation is given by
\begin{eqnarray}
 D_\mu = \partial_\mu - i g T^a A^a_\mu,
 &\qquad &
 D^{ab}_\mu = \partial_\mu - g f^{abc} A^c_\mu.
\end{eqnarray}
Here the following notation for the relation of two-component and four-component spinors is used,
\begin{equation}
\begin{aligned}
  \Psi&=
  \begin{pmatrix}
    \psi\\ \bar \psi^\dagger
  \end{pmatrix}
  \equiv 
  \begin{pmatrix}
    \psi_-\\  \psi_+
  \end{pmatrix}
\,,&
 \bar\Psi&=
  \begin{pmatrix}
    \bar\psi ,&  \psi^\dagger
  \end{pmatrix}
  \equiv
  \begin{pmatrix}
    \bar\psi_- ,&  \bar\psi_+
  \end{pmatrix}
\,,\\
  \Lambda&=
  \begin{pmatrix}
    i\lambda\\ -i\lambda^\dagger
  \end{pmatrix}
\equiv \begin{pmatrix}
    \Lambda_-\\ \Lambda_+
  \end{pmatrix}
\,,&
 \bar\Lambda&=
  \begin{pmatrix}
    i\lambda ,& -i \lambda^\dagger
  \end{pmatrix}
\equiv 
  \begin{pmatrix}
    \bar\Lambda_-,& \bar\Lambda_+
  \end{pmatrix}
\end{aligned}.
\end{equation}
The plus and minus labels  are chosen such that e.g.
$\psi_+$ creates an antiquark with positive helicity and thus differ from the
notation in~\cite{Schwinn:2006ca}.
The scalar fields have been relabeled according to
\begin{equation}
  \phi=\phi_-\, ,\quad  \bar\phi=\bar\phi_-
\end{equation}
with the convention $(\phi_\pm)^\dagger=\bar\phi_\mp$. 

The colour ordered three-point amplitudes obtained from~\eqref{eq:sqcd-mass}
can be found in~\cite{Schwinn:2006ca}. In addition to the usual vertices of
massive scalars and quarks with gluons there are the Yukawa couplings of
scalars, quarks and gluinos that read
\begin{align}
\label{eq:gluino-yukawa}
 V(\bar{Q}_1^+,\phi_2^-,\Lambda_3^+) = - i \sqrt{2} \sbraket{13},
 &&
 V(\bar{Q}_1^-,\phi_2^+,\Lambda_3^-) =  i \sqrt{2} \braket{13},
 \nonumber \\
 V(\bar{Q}_1^-,\phi_2^-,\Lambda_3^+) = - i \sqrt{2} m \frac{\braket{12}}{\braket{23}},
 &&
 V(\bar{Q}_1^+,\phi_2^+,\Lambda_3^-) =  i \sqrt{2} m \frac{\sbraket{12}}{\sbraket{23}},
 \nonumber \\
 V(\bar{\phi}_1^+,Q_2^-,\bar{\Lambda}_3^-) =  i \sqrt{2} \braket{23},
 &&
 V(\bar{\phi}_1^-,Q_2^+,\bar{\Lambda}_3^+) =  -i \sqrt{2} \sbraket{23},
 \nonumber \\
 V(\bar{\phi}_1^+,Q_2^+,\bar{\Lambda}_3^-) =   i \sqrt{2} m \frac{\sbraket{12}}{\sbraket{31}},
 &&
 V(\bar{\phi}_1^-,Q_2^-,\bar{\Lambda}_3^+) =  i \sqrt{2} m \frac{\braket{12}}{\braket{13}}.
\end{align}

\subsection{Abelian Higgs model}\label{app:AbelianHiggs}
The simplest supersymmetric version of the Abelian Higgs model was first
constructed in~\cite{Fayet:1975yh}. After elimination of the auxiliary
fields the Lagrangian reads
\begin{eqnarray}
 {\cal L}_{\text{AHM}} & = & - \frac{1}{4} \left( F_{\mu\nu} \right)^2 +
 i \lambda^\dagger \bar\sigma^\mu \partial_\mu \lambda + i
 \psi^\dagger \bar\sigma^\mu D_\mu  \psi +
  \left( D_\mu \phi \right)^\dagger \left( D^\mu \phi \right)
\nonumber \\ & &
 -\sqrt{2}ig\left[ \phi^\dagger \lambda  \psi -
   \psi^\dagger \lambda^{\dagger}   \phi\right]
- \frac{1}{2}  \left(g \phi^\dagger  \phi +\xi\right)^2, \nonumber
\end{eqnarray}
with $D_\mu=\partial_\mu -ig A_\mu$.  Because of the axial anomaly, this model is not a consistent quantum field theory but it is adequate as a simple example for the study of amplitudes at tree level.
For $\xi<0$ the scalar develops a vev, $\braket{\phi}=\frac{v}{\sqrt{2}}=\sqrt{|\xi|/g}$. The mass eigenstates can be introduced by decomposing the scalar as 
\begin{equation}
  \phi=\frac{1}{\sqrt 2}(v+H+i\varphi)
\end{equation}
and introducing the Dirac fermion
\begin{equation}
\label{eq:def-psi}
  \Psi=
  \begin{pmatrix}
    \psi\\ -i\lambda^\dagger
  \end{pmatrix}
\equiv 
  \begin{pmatrix}
    \psi_-\\ \psi_+
  \end{pmatrix}\,,\quad
\bar\Psi=
  \begin{pmatrix}
    i\lambda ,&  \psi^\dagger
  \end{pmatrix}
  \equiv
  \begin{pmatrix}
    \bar\psi_- ,&  \bar\psi_+
  \end{pmatrix}.
\end{equation}
In terms of the mass eigenstates and adding a gauge fixing term
$\mathcal{L}_{\text{GF}}=-\frac{1}{2\alpha}(\partial_\mu A^\mu-\alpha m\varphi)^2$ with $m= gv$,
 the Lagrangian is given by
\begin{align}
\label{eq:Lagrangian-sahm}
 {\cal L}_{\text{AHM}+\text{GF}}
  = & - \frac{1}{4} \left( F_{\mu\nu} \right)^2 
-\frac{1}{\alpha}(\partial_\mu A^\mu)^2+
\frac{1}{2}\left[(m+gH)^2+g^2\varphi^2\right]A_\mu A^\mu\nonumber\\
&+\frac{1}{2}  \partial_\mu H^2+\frac{1}{2}
\partial_\mu \varphi^2 -\alpha m^2\varphi^2 
+ g A^\mu(\varphi\overleftrightarrow{\partial_\mu} H) - \frac{g}{8}
\left(2vH+H^2+\varphi^2\right)^2
\nonumber \\
& + i \bar\psi_- \bar\sigma^\mu\partial_\mu \psi_+ 
+ i \bar\psi_+\sigma^\mu D_\mu \psi_-
-(m + gH)\bar\Psi\Psi+ig\varphi \bar\Psi\gamma^5\Psi.
\end{align}
The physical particles $A,H,\Psi$ all acquire the common mass $m$ so they form a massive supermultiplet as expected, while the field $\varphi$ is the unphysical Goldstone boson that cancels the gauge dependent pole at $\alpha m^2$ in the massive vector propagator.
\subsubsection{Three-point amplitudes}
The non-vanishing three-point amplitudes with external physical states
obtained from the
Lagrangian~\eqref{eq:Lagrangian-sahm} using the external massive
spinors and polarization vectors of
section~\ref{sec:massivespinorhelictyform} are given by
\begin{equation}
  \begin{aligned}
   A_3(A^+_1,H_2,A^-_3)
	&=
        igm_W\frac{\braket{q3}\braket{23}}{\braket{q1}\braket{12}} ,&
  A_3(A_1^0,H_2,A^+_3)
	&=
        i\frac{g}{\sqrt 2}\sbraket{13}\frac{\braket{q1}}{\braket{q3}},\\
 A_3(A_1^0,H_2,A^-_3)
	&=i\frac{g}{\sqrt 2}\braket{13}\frac{\sbraket{q1}}{\sbraket{q3}} ,&
 A_3(A_1^0,H_2,A^0_3)
	&= 2igm \left(\epsilon^0_1 \cdot \epsilon^0_3 \right).
  \end{aligned}
\end{equation}
The nonvanishing vertices involving fermions and vector bosons read
\begin{equation}
  \begin{aligned} 
    A_3(\bar Q^+_1,A_2^-,Q^-_3)&=-i\sqrt 2 g \frac{\braket{23}^2}{\braket{31}}
    ,&
     A_3(\bar Q^+_1,A_2^-,Q^+_3)&=-i\sqrt 2 g m \frac{\braket{23}}{\braket{31}}
     \frac{\braket{2q}}{\braket{3q}},\\
     A_3(\bar Q^+_1,A_2^0,Q^-_3)&=-i \frac{g}{m}
     \left(\sbraket{12}\braket{23}-
       m^2\frac{\sbraket{1q}\braket{q3}}{\sbraket{2q}\braket{q2}}\right)  ,&
       A_3(\bar Q^+_1,A_2^0,Q^+_3)&=i g \, \sbraket{13} \\
 A_3(\bar Q^+_1,A_2^+,Q^-_3)&=-i\sqrt 2 g  \frac{\sbraket{12}^2}{\sbraket{31}}
    ,&
     A_3(\bar Q^-_1,A_2^+,Q^-_3)&=-i\sqrt 2 g m \frac{\sbraket{12}}{\sbraket{31}}
     \frac{\sbraket{2q}}{\sbraket{1q}}.
  \end{aligned}
\end{equation}
Note that the chiral interaction of the fermions with the vector bosons
results in the  absence of the amplitudes $  A_3(\bar Q^-_1,A_2^-,Q^+_3)$ and
$  A_3(\bar Q^-_1,A_2^+,Q^+_3)$. Also the form of the 'helicity flip' vertices
differs from those in a non-chiral theory. The coupling of fermions to the
Higgs boson are given by
\begin{equation}
  \begin{aligned} 
    A_3(\bar Q^+_1,H_2,Q^+_3)&=ig \sbraket{31}
    ,&
     A_3(\bar Q^+_1,H_2,Q^-_3)&=i mg\left[
   \frac{\braket{q3}}{\braket{q1}}-  \frac{\braket{23}}{\braket{12}} \right],
 \\
 A_3(\bar Q^-_1,H_2,Q^-_3)&=ig \braket{31}
    ,&
     A_3(\bar Q^-_1,H_2,Q^+_3)&=i mg\left[
   \frac{\braket{12}}{\braket{23}}-  \frac{\braket{q1}}{\braket{q3}} \right].
\end{aligned}
\end{equation}

\subsubsection{Maximally spin violating amplitudes}
Let us now examine some examples of scattering amplitudes with only external massive vectors and their relation to fermionic amplitudes through SUSY.  Throughout this section the same reference spinors $\ket{q\pm}$ for all the external massive states will be used. Furthermore,  unless stated otherwise the Feynman- 't Hooft gauge (with $\alpha=1$) will be employed so that all poles of the propagators appear at $p^2=m^2$.  From the vertices obtained from~\eqref{eq:Lagrangian-sahm} it is seen that
$A$-lines in the massive vector-boson amplitudes must couple either to a $HA^2$ or $A(\varphi\overleftrightarrow\partial H)$ cubic vertex, where only the former does not conserve the number of scalars, or trough quartic vertices.

The only contributions to the four vector-boson amplitude arise from the exchange of an $H$-boson:
\begin{equation}\label{eq:4ptMSVabhiggs}
\mathcal{A}_4(A_{k_1}, A_{k_2}, A_{k_3}, A_{k_4}) =  (2igm)^2 \sum_{\sigma\in S_4} 
\left( (\epsilon_{k_{\sigma(1)}}\cdot \epsilon_{k_{\sigma(2)}}) 
\frac{i}{(k_{\sigma(1)}+k_{\sigma(2)})^2 - m^2} 
(\epsilon_{k_{\sigma(3)}}\cdot \epsilon_{k_{\sigma(4)}}) \right),
\end{equation}
where one sums over all the permutations $\sigma$ of the external
momenta. This amplitude vanishes in the limit $m \rightarrow 0$, as
it should. From the expression it is straightforward to derive
scattering amplitudes for all the states of the massive vector
boson. From equation \eqref{eq:simplebutpowerful} it follows that the
following amplitudes (and the ones with plus and minus labels
exchanged) simply vanish,
\begin{align}
  \mathcal{A}(A^+,A^+,A^+,A^+)=\mathcal{A}(A^-,A^+,A^+,A^+)
=\mathcal{A}(A^0,A^+,A^+,A^+)=0.
\end{align}
The vanishing of the amplitudes with one unequal helicity for
arbitrary spin axes is a special property of the Abelian Higgs model
and does not follow from the general analysis of
section~\ref{sec:vanish}. 
The only non-vanishing four-point amplitude that does not involve
longitudinal polarizations is the analog of the MHV amplitude,
 $\mathcal{A}_4(A_1^-,A_2^+,A_3^- ,A_4^+)$ and permutations thereof.
 For this amplitude there are only two distinct non-zero
contributions in the permutation sum in equation
\eqref{eq:4ptMSVabhiggs}, leading to the explicit
result quoted in~\eqref{eq:4ptMSV}.

The SWI~\eqref{eq:swi--+} relates the four point Maximally Spin Violating (MSV)
 amplitude to a sum of two fermionic amplitudes.
It is seen that each of the fermionic amplitudes receives a
non-vanishing contribution from a Higgs exchange diagram and a diagram
with a fermion propagator:
\begin{equation}
  \begin{aligned}
     \mathcal{A}_4(\bar\Psi_{k_1}^-, A^+_{k_2}, A^-_{k_3}, \Psi^+_{k_4})
   =&(ig)^2\bar u(k_1,-)\left[\fmslash \epsilon(k_2,+)
\left(\tfrac{1+\gamma_5}{2}\right)
\frac{i(\fmslash k_{3,4}+m)}{(k_{3,4}^2-m^2)}\fmslash \epsilon(k_3,-)
\left(\tfrac{1+\gamma_5}{2}\right)\right.\\
&\left.-\frac{i  \epsilon(k_2,+)\cdot\epsilon(k_3,-) }{(k_{2,3}^2-m^2)}
\right]v(k_4,+),
  \end{aligned}
\end{equation}
where the diagram with exchanged photon attachments vanishes for the
given spin assignments.  Inserting the external wave-functions results
in the expression quoted in~\eqref{eq:4ptMSVfermi}.

\begin{figure}[t]
 \begin{center}
\includegraphics[width=0.6\textwidth]{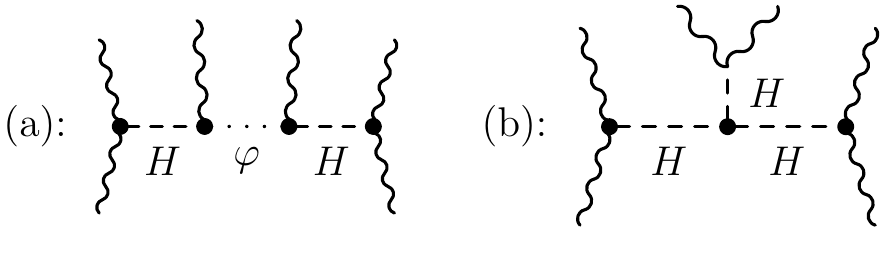}
\end{center}
  \caption{Example topologies contributing to the six point amplitudes of massive vectors in the Abelian Higgs model.}
  \label{fig:six-point}
\end{figure}
The above observation on the vanishing of vector boson amplitudes with
one unequal spin label can be extended to an arbitrary number of legs.
As a first example, consider the six-point amplitude of massive vector
bosons where diagrams of two different topologies with only cubic
vertices contribute, as shown in figure~\ref{fig:six-point}, and
additional diagrams with quartic vertices not shown. It can be seen
that diagrams of class (b) only contribute to amplitudes with at most
three different external spin states since otherwise they involve at
least one product $\epsilon^+_{k_i}\cdot\epsilon^+_{k_j}=0$. The same
conclusion holds for diagrams with quartic couplings.  The only
contributions to diagrams with one or two negative helicity labels
therefore arise from class (a) and read explicitly
\begin{equation}
  A^{(a)}_6(A_{k_1},\cdots,A_{k_6})=\sum_{\sigma\in S_6}
a^{(a)}_6(A_{k_\sigma(1)},\cdots,A_{k_{\sigma(6)}})
\end{equation}
with
\begin{equation}
\label{eq:msv6}
a^{(a)}_6(A_{k_1},\cdots,A_{k_6})
 =   i (2g)^4 m^2  (\epsilon_{k_1} \cdot \epsilon_{k_2 })
\frac{ \left(\epsilon_{k_3}\cdot k_{1,2}\right)}{k_{1,2}^2 - m^2} 
\frac{\left( \epsilon_{k_4}\cdot k_{1,3} \right) }{k_{1,3}^2 - m^2} 
  \frac{(\epsilon_{k_5}\cdot \epsilon_{k_6})}{k_{1,4}^2 - m^2} .
\end{equation}
Again at least two polarization vectors with a different polarization
than the remaining ones are required for a non-vanishing result so
that the vector boson amplitudes with one negative helicity
vanish. 

The topologies contributing to $n$-point amplitudes with at most two
unequal spins are the direct generalization of
fig.~\ref{fig:six-point}~(a), i.e. involve one line of alternating $H$
and $\varphi$ propagators.  Again they are non-vanishing only for the
`maximally spin violating' amplitudes with two unequal spins. This is
based on the fact that at least two $A A H$ vertices are needed to
soak up the scalar degrees of freedom at tree level. This will require
at least $2$ vectors having a different polarization from the
others. In particular, the all-plus and all-minus amplitudes vanish,
just as the one-minus and one-plus amplitudes. Furthermore, all
vertices except $A H \varphi$ will lead to an additional contraction
of polarization vectors: in other words, these contribute to NMSV and
onward. With the particular set of vertices in the Abelian Higgs model
there is just one (sum over permutations of one) diagram for MSV
amplitudes that generalizes~\eqref{eq:msv6}, as quoted in~\eqref{eq:msv}.

\section{Three particle vertices in SQCD with arbitrary polarization axes}
\label{app:3-pt}

The results for three point vertices obtained in subsection \ref{sec:sqcd} in supersymmetric QCD can be generalized to the case of  arbitrary polarization axes. The vertices for the case of the equal polarization axes can be found in equations \eqref{eq:mhv-fund} and \eqref{eq:barmhv-fund}, reproduced here for the readers' convenience:
\begin{align}
A_3(\bar \Phi_1, G_2^-,\Phi_3) & =\delta^2(\mathcal{Q}_\alpha) \frac{\bar\iota_1\sbraket{q1}-\bar\iota_3\sbraket{q3}}{\sbraket{q2}} , \nonumber \\
A_3(\bar \Phi_1, G_2^+,\Phi_3)_{\eta\bar\iota} & = \delta^2(\mathcal{Q}_{\alpha}) m \frac{\braket{31}}{\braket{12} \braket{23}} .\nonumber
\end{align}
In the case of unequal polarization axes, these expressions change slightly. Let us first classify the solutions to the Ward identities. These either have fermionic weight two or three and should reduce to the above if the axes are chosen to coincide. Of course, both of these are proportional to the delta function, 
\begin{equation}
A_3 = \delta^2(\mathcal{Q}_{\alpha}) F(1,2,3).
\end{equation}
Hence in the weight two case only one component amplitude needs to be calculated. It is convenient to let this amplitude be $\phi g^+ \bar{\phi}$ as in this case
\begin{equation}
A(\phi, g^+, \bar{\phi})  = \int d\eta_3 d\bar{\iota}_3 A_3(\bar \Phi_1, G_2^+,\Phi_3)_{\eta\bar\iota}  =m F(1,2,3)
\end{equation}
holds. Hence in the case of unequal polarization axes
\begin{align}
A_3(\bar \Phi_1, G_2^+,\Phi_3)_{\eta\bar\iota}  & =
\delta^2(\mathcal{Q}_{\alpha})
\frac{  A(\phi, g^+, \bar{\phi})}{m}\nonumber \\
& =  \delta^2(\mathcal{Q}_{\alpha})  \frac{\xi_{\alpha} 2_{\dalpha} K_1^{\alpha \dalpha}}{m\braket{\xi 2}}
\end{align}
holds for one of the vertices. The other one can in general be written as
\begin{equation}
A_3(\bar \Phi_1, G_2^-,\Phi_3)=\delta^2(\mathcal{Q}_\alpha) \left(c_1 \eta_1 + c_2 \eta_2 + c_3 \eta_3 + c_4 \bar{\iota}_1 + c_5 \bar{\iota}_3 \right)
\end{equation}
for some coefficients $c_i$. To this linear polynomial one can always add multiples of $\braket{\xi \mathcal{Q}}$ for some spinor $\xi$. This allows one to eliminate two of the five coefficients, leading to 
\begin{equation}
A_3(\bar \Phi_1, G_2^-,\Phi_3)=\delta^2(\mathcal{Q}_\alpha) \left( \tilde{c}_2 \eta_2 + \tilde{c}_4 \bar{\iota}_1 + \tilde{c}_5 \bar{\iota}_3 \right)
\end{equation}
for some new coefficients $\tilde{c}$. Now we demand invariance under $\bar{\mathcal{Q}}$. This leads to,
\begin{equation}
\tilde{c}_2 k_2^{\dalpha} + \tilde{c}_4  \frac{m}{\sbraket{q_1 1}} q_1^{\dalpha} + \tilde{c}_5  \frac{m}{\sbraket{q_3 3}} q_3^{\dalpha} =  0
\end{equation}
which constitutes two linear equations. These can be isolated by contracting once with $q_1$ and once with $q_3$,
\begin{align}
\tilde{c}_2 \sbraket{2 q_3}  + \tilde{c}_4  \frac{m}{\sbraket{q_1 1}} \sbraket{q_1 q_3 } & =  0,\\
\tilde{c}_2 \sbraket{2 q_1}  + \tilde{c}_5  \frac{m}{\sbraket{q_3 3}} \sbraket{q_3 q_1}  & =  0.
\end{align}
Hence the superamplitude has to be proportional to
\begin{equation}
A_3(\bar \Phi_1, G_2^-,\Phi_3) \sim \delta^2(\mathcal{Q}_\alpha)  \left( m \sbraket{q_1 q_3} \eta_2 - \sbraket{2 q_3}\sbraket{ q_1 1} \bar{\iota}_1 +  \sbraket{2 q_1}\sbraket{ q_3 3}  \bar{\iota}_3 \right)
\end{equation}
to solve the SUSY Ward identity constraints. The answer here was rescaled to display a smooth $q_1 \rightarrow q_3$ limit. This leaves the determination of the prefactor. This can be obtained by comparing to the $\phi g^- \bar{\phi}$ amplitude,
\begin{equation}
A(\phi, g^-, \bar{\phi})  = \int d\eta_2 d\eta_3 d\bar{\iota}_3 A_3(\bar \Phi_1, G_2^-,\Phi_3) =  \frac{\xi_{\dalpha} 2_{\alpha} K_1^{\alpha \dalpha}}{\sbraket{\xi 2}}.
\end{equation}
This yields
\begin{multline}
A_3(\bar \Phi_1, G_2^-,\Phi_3) = \frac{1}{\left( \sbraket{q_1 1} \braket{13} \sbraket{3 q_3}\right)} \left(\frac{\xi_{\dalpha} 2_{\alpha} K_1^{\alpha \dalpha}}{\sbraket{\xi 2}} \right) \\ \delta^2(\mathcal{Q}_\alpha)  \left( m \sbraket{q_1 q_3} \eta_2 - \sbraket{2 q_3}\sbraket{ q_1 1} \bar{\iota}_1 +  \sbraket{2 q_1}\sbraket{ q_3 3}  \bar{\iota}_3 \right).
\end{multline}
Unequal masses can be incorporated into the above calculation without essential difficulty.

\bibliographystyle{apsrev4-1}

\bibliography{susbiblio}
\end{document}